\documentclass[aps,prc,twocolumn,longbibliography,reprint,superscriptaddress,amsmath,nofootinbib]{revtex4-1}

\usepackage{verbatim}
\usepackage{amsmath,amssymb,amsfonts}
\usepackage{graphicx}
\usepackage[dvipsnames]{xcolor}
\usepackage[colorlinks=true,linkcolor=black,urlcolor=blue,citecolor=blue]{hyperref}
\usepackage[normalem]{ulem}

\graphicspath{{figs_PRC/}}

\usepackage{bm} 

\begin{document}

\title{Far-off-equilibrium expansion trajectories in the QCD phase diagram}

\author{Chandrodoy Chattopadhyay}
\affiliation{Department of Physics, North Carolina State University, Raleigh, NC 27695, USA}
\author{Ulrich Heinz}
\affiliation{Department of Physics, The Ohio State University, Columbus, OH 43210, USA}
\author{Thomas Sch\"afer}
\affiliation{Department of Physics, North Carolina State University, Raleigh, NC 27695, USA}

\date{\normalsize \today}

\begin{abstract}
\noindent
We consider the hydrodynamic evolution of a quark-gluon gas with non-zero quark masses and net baryon number in its phase diagram. For far-off-equilibrium initial conditions the expansion trajectories appear to violate simple rules based on the second law of thermodynamics that were previously established for ideal or weakly dissipative fluids. For Bjorken flow we present a detailed analysis within kinetic theory that provides a full microscopic understanding of these macroscopic phenomena and establishes their thermodynamic consistency. We point out that, for certain far-off-equilibrium initial conditions, the well-known phenomenon of ``viscous heating'' turns into ``viscous cooling'' where, driven by dissipative effects, the temperature decreases faster than in adiabatic expansion.
\end{abstract}

\maketitle

\section{Introduction}
\label{sec1}

One of the primary goals of relativistic heavy-ion collisions is to study the phase diagram of Quantum Chromodynamics (QCD). In a simplified version which ignores additional dimensions \cite{Stephanov:2006zvm} associated with strangeness and isospin (which are conserved by the strong interactions), this phase diagram reduces to a plane spanned by the temperature $T$ and the baryon chemical potential $\mu_B$. Based on first principles lattice QCD calculations at vanishing $\mu_B$ \cite{Aoki:2006we, Bazavov:2009zn, Borsanyi:2010cj, Bazavov:2011nk} and theoretical modeling at large $\mu_B$ \cite{Stephanov:2006zvm}, it is conjectured that there exists a first-order phase transition line in this plane which terminates at a critical point at finite $(T_\mathrm{cr}, \mu_{B,\mathrm{cr}})$ \cite{Stephanov:1998dy, Stephanov:1999zu}, turning into a continuous crossover at smaller $\mu_B$ values. The search for the QCD critical point has been an area of intense research over the last two decades \cite{Busza:2018rrf, Bzdak:2019pkr}. 

A key feature common to equilibrated systems near a critical point is the enhancement of thermodynamic fluctuations accompanied by diverging correlation length of these fluctuations \cite{hohenberg1977theory}. Accordingly, the QCD critical point search program using heavy-ion collisions is driven by exploration of final-state observables that are sensitive to the correlation length. One such class of observables are higher-order cumulants of net-proton fluctuations, which are predicted to exhibit non-monotonic behavior as the freeze-out region of collision fireballs is systematically varied in the $(T,\mu_B)$ plane from low to high $\mu_B$ \cite{Stephanov:1998dy, Stephanov:1999zu, Hatta:2003wn, Stephanov:2008qz, Stephanov:2011pb, Luo:2017faz}. This strategy has been pursued in the two-stage Beam Energy Scan (BES) program at RHIC, by varying the collision energy $\sqrt{s_{NN}}$ from $200$\,GeV down to $7.7$\,GeV (for BES-I) and $3.0$\,GeV (for BES-II) \cite{Bzdak:2019pkr, STAR:2017sal}.  

However, this simple concept is complicated along several fronts by the realities of relativistic heavy-ion collisions \cite{Nahrgang:2011mg, Nahrgang:2011mv, Mukherjee:2015swa, Herold:2016uvv, Stephanov:2017ghc, Hirano:2018diu, Singh:2018dpk, Nahrgang:2018afz, Yin:2018ejt, An:2021wof}. The main complication arises from the explosive expansion dynamics of the fireballs formed in the collisions which drive the fireball medium out of local thermal equilibrium. At the lower BES energies additional considerations make thermalization even more elusive \cite{Gupta:2022phu}. Equally important is the observation \cite{Steinheimer:2007iy, Du:2021zqz} (reviewed in \cite{Shen:2020gef, An:2021wof}) that different fluid cells inside the expanding medium follow different expansion trajectories through the phase diagram such that the measured final-state particles represent an average over contributions with different chemical compositions at freeze-out. This has led to revived interest in fireball expansion trajectories through the QCD phase diagram.

Such expansion trajectories were first studied over 35 years ago in the context of ideal fluid dynamics where both entropy and baryon number are conserved and the  expansion therefore occurs at fixed entropy per baryon, $S/A{\,=\,}s/n{\,=\,}$const. (where $S$ is the total entropy, $A$ is the nuclear mass (baryon) number, $s=S/V$ is the entropy density, and $n=A/V$ is the net baryon density) \cite{Heinz:1987ca, Lee:1992hn, Cho:1993mv, Hung:1997du}. From these studies it is known that a thermalized gas of massless quarks, antiquarks and gluons expands isentropically along lines of constant $\mu_B/T$, ending at the origin $(T_\infty=0,\mu_{B\infty}=0)$, whereas for a thermalized gas of hadron resonances the adiabatic expansion trajectories end at $(T_\infty=0,\,\mu_{B\infty}=m_N)$ (where $m_N$ is the mass of the nucleon, the lightest baryon number carrying hadron). It was also understood that, at constant temperature $T>0$, an increase in specific entropy $s/n$ leads in local thermal equilibrium to a {\it decrease} of the chemical potential $\mu_B$, irrespective of the mass of the constituents. It came therefore to the present authors as a surprise when in a recent study \cite{Dore:2020jye} of {\it dissipative} expansion trajectories, where viscous heating leads to an increase of entropy with time, for some initial conditions in the quark-gluon plasma phase the chemical potential $\mu_B$ did {\it not} initially decrease, but rather grow with decreasing temperature. Clearly, along such trajectories the {\it equilibrium} value of the specific entropy $s/n$ decreased initially \cite{Dore:2022qyz}. What causes this decrease, and how can this be compatible with the second law of thermodynamics asserting that the total entropy can never decrease (Boltzmann's ``H-theorem")? The authors of Ref.~\cite{Dore:2022qyz} pointed to non-equilibrium entropy corrections as the likely culprit for this phenomenon. In this paper we show that this conjecture was correct and explain in detail all of the possible manifestations of dissipative effects in the heavy-ion fireball expansion trajectories through the QCD phase diagram. 

We point out that the above phenomenon of decreasing {\it equilibrium} entropy is associated with more rapid cooling along the expansion trajectory than expected for adiabatic expansion. One usually associates dissipative effects with the phenomenon of {\it ``viscous heating''} --- a form of internal friction that causes the medium to cool down more slowly in dissipative evolution than in adiabatic expansion. To distinguish the far-off-equilibrium dynamics discovered numerically by Dore {\it et al.} \cite{Dore:2020jye, Dore:2022qyz} from this expected behavior we introduce the concept of {\it ``viscous cooling''} where the dissipative system cools more rapidly than the ideal fluid.

To allow for a largely analytic discussion we follow Ref.~\cite{Dore:2020jye} and assume 1-dimensional Bjorken expansion, but our insights are generic and generalize to arbitrary expansion geometries that can only be studied numerically. Moreover, Dore {\it et al.} employed a Lattice QCD based equation of state that included the effects of criticality \cite{Parotto:2018pwx}. Since a realistic equation of state will not be crucial for our discussion, we simplify the system further and consider a weakly interacting gas of {\it massive} quarks and gluons at finite chemical potential. Such a gas does not have a phase transition or a critical point, but it has non-vanishing transport coefficients arising from the weak interactions among its constituents. We calculate these transport coefficients from kinetic theory self-consistently with our model assumptions.

Dividing the entropy (or any other macroscopic hydrodynamic quantity) into equilibrium and non-equilibrium contributions requires a definition of the temperature $T$ and chemical potential $\mu_B$ associated with the equilibrium part. This matching of equilibrium parameters to the non-equilibrium state is ambiguous. An important aspect of the work presented here is a comparison between macroscopic and microscopic descriptions; for the former we use second-order hydrodynamics, whereas for the latter we implement the relativistic Boltzmann equation with a collision kernel in Relaxation Time Approximation (``RTA Boltzmann"). To ensure conservation of energy-momentum and baryon number by the RTA collision kernel \cite{Romatschke:2011qp, Denicol:2012cn, Florkowski:2013lya, Jaiswal:2013npa, Romatschke:2017acs}\footnote{%
    This works for a momentum independent relaxation time in the RTA collision kernel as used here. For a momentum-dependent relaxation time the RTA collision term must be generalized to remain compatible with these conservation laws under Landau matching \cite{Teaney:2013gca, Rocha:2021zcw, Rocha:2021lze, Dash:2021ibx}.}
we employ the Landau matching conditions  \cite{DeGroot:1980dk} $e = e_\mathrm{eq}(T,\mu_B)$ and $n = n_\mathrm{eq}(T,\mu_B)$, i.e. we assign $T$ and $\mu_B$ such that the associated equilibrium values reproduce the actual energy and net baryon densities of the fluid cell in its non-equilibrium state.

This paper is organized as follows: In Sec.~\ref{sec2} we describe the non-conformal hydrodynamic equations studied in this work. Since we study a \textit{multi-component quantum gas} at finite baryon chemical potential $\mu_B$, its transport coefficients differ from those used in Refs.~\cite{Dore:2020jye} for a single-component Boltzmann gas without conserved charge \cite{Denicol:2014vaa}; these transport coefficients are derived in App.~\ref{appa}. Section~\ref{sec3} briefly summarizes the characteristics of the isentropic expansion trajectories in ideal fluid dynamics for conformal and non-conformal systems. This is followed by a discussion of dissipative expansion trajectories for conformal systems in Sec.~\ref{sec4} and for non-conformal systems in Sec.~\ref{sec5}. In each case two different subsections describe and analyze the hydrodynamic and kinetic theory solutions for these trajectories; for the non-conformal case the two are compared in App.~\ref{appc}. We explain how, for certain types of far-off-equilibrium initial conditions, the system can expand following trajectories along which {\it the equilibrium part of the entropy} initially decreases. For the conformal case, the splitting of the entropy current into equilibrium and non-equilibrium parts is described in App.~\ref{appb}. Our conclusions are presented in Sec. \ref{sec6}.

\section{Dissipative hydrodynamics with Bjorken flow}
\label{sec2}

Bjorken expansion is defined by the flow velocity field $v^x = v^y = 0, v^z = z/t$ \cite{Bjorken:1982qr}. Symmetries dictate that all macroscopic quantities are independent of transverse position $(x,y)$ and space-time rapidity $\eta_s \equiv \tanh^{-1}(z/t)$, and thus depend only on Milne proper time $\tau \equiv \sqrt{t^2 - z^2}$. Together with invariance under longitudinal reflections ($\eta_s \to - \eta_s$), these symmetries force the dissipative part of the conserved baryon current to vanish, and the shear stress tensor reduces to a single independent component $\pi(\tau) \equiv \pi^{\eta_s}_{\eta_s} (\tau)$. The hydrodynamic evolution equations reduce to \cite{Dore:2020jye, Jaiswal:2014isa}
\begin{align}
    \frac{de}{d\tau} &= - \frac{1}{\tau} \left( e + P + \Pi - \pi \right),
\label{e_evol}\\
    \frac{dn}{d\tau} &= - \frac{n}{\tau}, 
\label{n_evol}\\
    \frac{d\Pi}{d\tau} & = - \frac{\Pi}{\tau_R} - \frac{\beta_\Pi}{\tau} - \delta_{\Pi\Pi} \, \frac{\Pi}{\tau} + \lambda_{\Pi\pi} \, \frac{\pi}{\tau},
\label{PI_evol}\\
    \frac{d\pi}{d\tau} &= - \frac{\pi}{\tau_R} + \frac{4}{3} \, \frac{\beta_\pi}{\tau} - \left( \frac{1}{3} \, \tau_{\pi\pi} + \delta_{\pi\pi} \right) \frac{\pi}{\tau} + \frac{2}{3} \, \lambda_{\pi\Pi} \, \frac{\Pi}{\tau}, 
\label{pi_evol}
\end{align}
where $e$ and $n$ are the energy and net quark density, $P$ is the thermal equilibrium pressure, given by the equation of state $P(e,n)=P(T,\mu)$, and $\Pi$ and $\pi$ are the bulk and shear viscous stresses. $\tau_R$ is the microscopic relaxation time in an underlying kinetic description to be discussed below, and $\beta_\pi,\, \beta_\Pi,\, \delta_{\pi\pi},\, \delta_{\Pi\Pi},\, \tau_{\pi\pi},\, \lambda_{\pi\Pi},\, \lambda_{\Pi\pi}$ are associated transport coefficients, derived in Appendix~\ref{appa} using the Chapman-Enskog expansion \cite{Jaiswal:2014isa}.

In Bjorken flow the system volume grows linearly with Milne time $\tau$. Eq.~(\ref{n_evol}) thus reflects conservation of the comoving net quark number per unit transverse area, $n \tau$, owing to the lack of net quark diffusion out of a fluid cell. Equation~(\ref{e_evol}) shows that the same is not true for the energy density: As long as the effective longitudinal pressure $P{+}\Pi{-}\pi$ is positive, the comoving energy density per unit area, $e\tau$, decreases with $\tau$ due to work done by the pressure which converts some of the thermal energy into kinetic motion energy associated with the collective longitudinal expansion.

The evolution equations (\ref{PI_evol},\ref{pi_evol}) for the bulk and shear viscous stresses are derived by iteratively solving (up to second-order in velocity gradients) the Boltzmann equation in relaxation-time approximation, followed by coarse-graining and partial resummation of the gradient terms \cite{Jaiswal:2014isa, Jaiswal:2013npa}. We parametrize the relaxation time $\tau_R$ in terms of the local temperature $T$ as $T\tau_R = 5 C$ where $C$ is a unitless constant.\footnote{%
    This is strictly justified only in the massless limit where the temperature is the only energy scale in the problem but we checked that corrections in powers of $m/T$ are small enough to not affect any of the qualitative features of our new results.}%
    $^,$\footnote{%
    For a conformal system at vanishing chemical potential, the relation $\tau_R \beta_\pi \equiv \eta$ (where $\eta$ is the shear viscosity), along with the parametrization $\tau_R = 5C/T$, implies $C = \eta/s$. However, for $\mu \neq 0$, the same pair of relations implies $C = \eta/s \times (1 + (\mu/T) (n/s) )^{-1}$.}%

As we will be interested in obtaining the hydrodynamic trajectories in the $(T,\mu)$-plane we shall directly solve Eqs.~(\ref{e_evol})--(\ref{pi_evol}) in terms of the Lagrange parameters, $x_a \equiv \{T, \mu\}$, instead of the associated densities $X_a = \{e, n\}$. Making use of the Landau matching conditions (\ref{e_eq and n_eq}) (see App.~\ref{appa}) we express $X_a$ in terms of $x_a$ using the equilibrium distributions
\begin{align}\label{f_eq}
    f^{q}_\mathrm{eq} &= \frac{1}{\exp( \beta E_p -  \alpha) + 1}, \,\,
    f^{\bar{q}}_\mathrm{eq} = \frac{1}{\exp(\beta E_p +  \alpha) + 1}, \nonumber \\
    f^{g}_\mathrm{eq} &= \frac{1}{\exp(\beta E_p) - 1}.
\end{align}
Here $\beta$ is the inverse temperature, $\alpha \equiv \beta \mu$, and $E_p = \sqrt{\bm{p}^2+m^2}$ is the energy of a particle of momentum $\bm{p}$ and mass $m$ in the fluid rest frame. 

For massless quarks and gluons with degeneracy factors $g_q$ and $g_g$, respectively, the desired relations are simply 
\begin{align}
    e_\mathrm{eq}(T,\mu) & =  T^4 \Bigg[ \frac{(4g_g + 7g_q) \pi^2}{120} + \frac{g_q}{4} \left(\frac{\mu}{T}\right)^2  + \frac{g_q}{8\pi^2} \left(\frac{\mu}{T}\right)^4  \Bigg] \nonumber \\
    & = 3 P_{\mathrm{eq}}(T, \mu),
\label{e_eq} \\
    n_\mathrm{eq}(T,\mu) &= T^3 \left[\frac{g_q}{6} \left(\frac{\mu}{T}\right) + \frac{g_q}{6\pi^2} \left(\frac{\mu}{T}\right)^3 \right],
\label{n_eq}
\end{align}
which we use to compute $dx_a = M_{a}^{\,b} \, dX_b$ and convert Eqs.~(\ref{e_evol},\ref{n_evol}) into evolution equations for $T$ and $\mu$. We use $g_q = 2\times N_c\times N_f = 12$ (with $N_c{\,=\,}3$ colors and $N_f{\,=\,}2$ flavors) for quarks and antiquarks and $g_g = 2 \times (N_c^2 - 1) = 16$ for gluons. 

Giving the (anti-)quarks a mass $m\ne0$ changes these relations to 
\begin{align}
    e_\mathrm{eq}(T,\mu) &= \frac{T^4}{2\pi^2} \, \int_0^{\infty} du \, u^2 \, \Bigg\{ \sqrt{u^2 + \frac{m^2}{T^2}} \, g_q \Bigg[\tilde{f}^q_\mathrm{eq} \left(u, \frac{m}{T}, \frac{\mu}{T}\right) \nonumber \\
    & + \tilde{f}^{\bar{q}}_\mathrm{eq} \left(u, \frac{m}{T}, \frac{\mu}{T} \right) \Bigg] + u \, g_g \, \tilde{f}^g_\mathrm{eq}(u) \Bigg\}, 
\label{e_nc} \\
    P_\mathrm{eq}(T,\mu) &= \frac{T^4}{6\pi^2} \, \int_0^{\infty} du \, u^3 \, \biggl\{ \frac{u}{\sqrt{u^2 + \frac{m^2}{T^2}}} \, g_q \Bigg[ \tilde{f}^q_\mathrm{eq} \left(u, \frac{m}{T}, \frac{\mu}{T}\right) \nonumber \\
    & + \tilde{f}^{\bar{q}}_\mathrm{eq} \left(u, \frac{m}{T}, \frac{\mu}{T} \right) \Bigg] + \, g_g \, \tilde{f}^g_\mathrm{eq}(u) \biggr\}, 
\label{P_nc} \\
    n_\mathrm{eq}(T,\mu) &= \frac{T^3}{2\pi^2} \, \int_0^{\infty} du \, u^2 \, g_q \Bigg[ \tilde{f}^q_\mathrm{eq} \left(u, \frac{m}{T}, \frac{\mu}{T}\right) \nonumber \\
    & -  \tilde{f}^{\bar{q}}_\mathrm{eq} \left(u, \frac{m}{T}, \frac{\mu}{T} \right) \Bigg],
\label{n_nc}
\end{align}
where $\tilde{f}^i_\mathrm{eq}(u,z,\alpha) = \bigl[\exp\bigl(\sqrt{u^2{+}z^2} - \alpha_i\bigr) -\theta_i\bigr]^{-1}$ with $\alpha_q = -\alpha_{\bar{q}} = \alpha,\ \alpha_g=0$ as well as $\theta_q=\theta_{\bar{q}}=-1,\ \theta_g=1$.

\section{Expansion trajectories in ideal hydrodynamics}
\label{sec3}

\begin{figure}[t!]
\centering
  \includegraphics[width=0.8\linewidth]{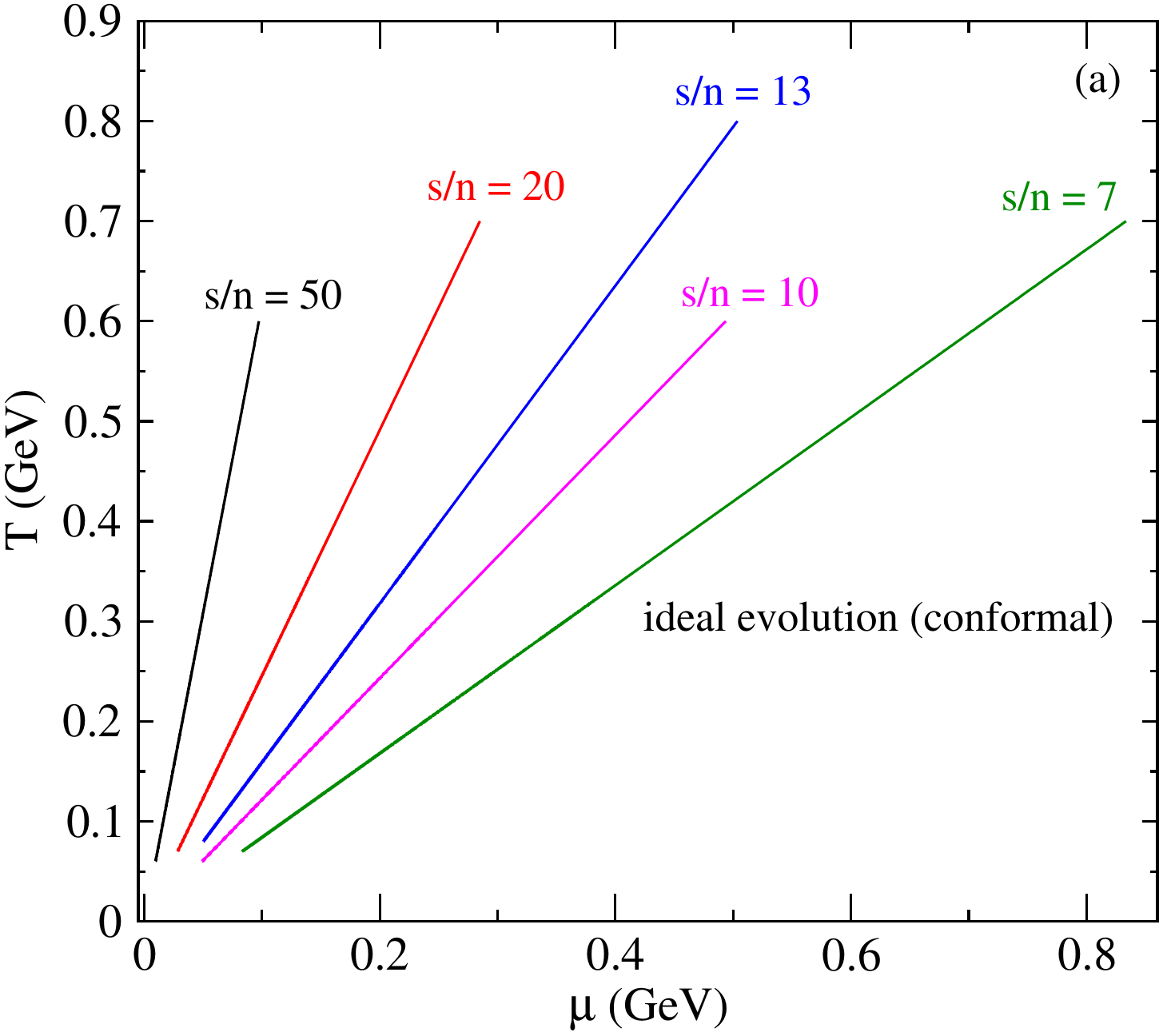}
  \includegraphics[width=0.8\linewidth]{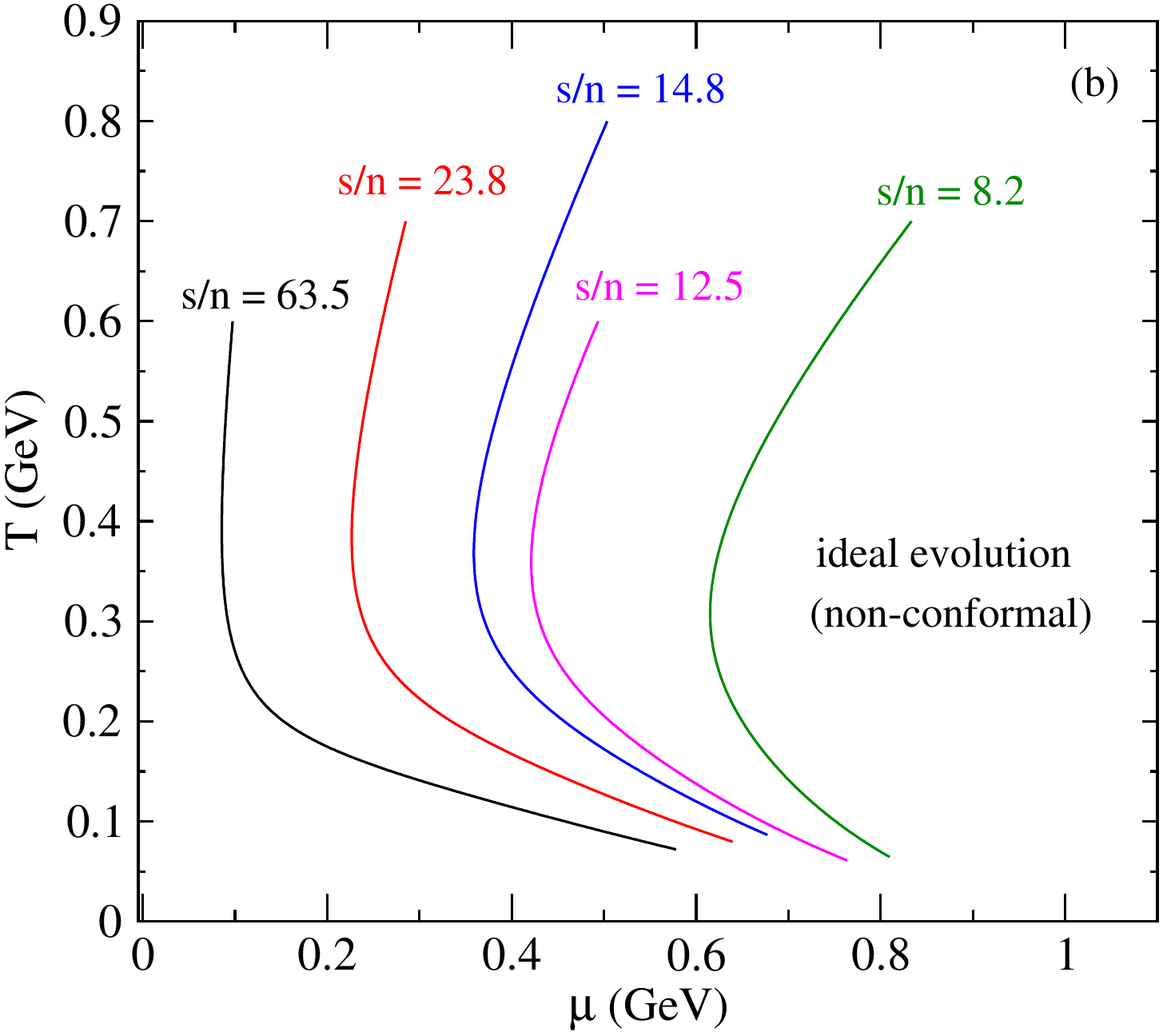}
\vspace{-3mm}
 \caption{Ideal hydrodynamic expansion trajectories in the $(T,\mu)$ plane for a conformal ($m=0$, panel (a)) and a non-conformal ($m_q=m_{\bar q}=1$\,GeV, $m_g=0$, panel (b)) quark-gluon gas with non-zero conserved net baryon charge, for identical sets of initial values $(T_0,\mu_0)$ in (a) and (b). The corresponding conserved entropy/baryon charge values $s/n$ are indicated.
	\label{F1}
	}
\end{figure}

We first solve the hydrodynamic equations in the absence of dissipation, by setting the shear and bulk viscous stresses to zero throughout the evolution: $\pi=\Pi=0$. The ideal expansion trajectories are shown in Fig.~\ref{F1}, for the massless (conformal) case in panel (a) and for a quark-gluon gas with massive ($m=1$ GeV) quarks and antiquarks in panel (b). Throughout this work, initial conditions are imposed at initial Milne time $\tau_0 = 0.1$\,fm/$c$. 
%
%
In panel (a) the initial temperature and chemical potentials $(T_0,\mu_0)$ are chosen to yield the integer $s/n$ ratios indicated.\footnote{%
    Note that $n=\frac{1}{3}n_B$ is the {\it net} quark density such that $s/n= \textstyle{\frac{1}{3}}(S/A)$ where $S$ and $A$ are the total entropy and net baryon number of the system.}
For the non-conformal case, the same initial values $(T_0,\mu_0)$ correspond to the larger $s/n$ ratios listed in panel (b). In ideal fluids both entropy and net quark number of a fluid element in its local rest frame are conserved during evolution, and the decrease of the associated comoving densities arises entirely from volume expansion. This follows directly from the thermodynamic relations $de = T \, ds + \mu \, dn$ and $s = (e+P -\mu n)/T$ which allow to re-cast Eqs.~(\ref{e_evol}), (\ref{n_evol}) as
\begin{align}
    \frac{ds}{d\tau} = - \frac{s}{\tau}, \qquad
    \frac{dn}{d\tau} = - \frac{n}{\tau}, \label{s_n_evol}
\end{align}
with the straightforward solution $s/n = \mathrm{const}$. 
Since for Bjorken flow Milne coordinates serve as a comoving coordinate system, $s/n$ remains constant as a function of Milne time $\tau$ which can be used as a line parameter along each of the trajectories shown in Fig.~\ref{F1}. 

For the conformal case shown in panel (a), with massless constituents yielding the EoS (\ref{e_eq},\ref{n_eq}),
lines of constant $s/n$ correspond to lines of constant $\mu/T$. This follows from conformality: in a system devoid of any dimensionful microscopic scales, $s$ and $n$ (which share the same units) must be related by
\begin{equation}
    s(T,\mu) = g(T,\mu)\,n(T,\mu) 
\label{relate_s_n}
\end{equation}
where $g$ can depend on $T$ and $\mu$ only through the dimensionless ratio $\alpha=\mu/T$. Thus $s/n = g(\alpha) = \mathrm{constant}$ implies constancy of $\alpha$. As a result, all the trajectories in Fig.~\ref{F1}a are straight lines passing through the origin.\footnote{%
    For each trajectory, its mirror image with respect to the temperature axis is itself a solution with the opposite sign of $s/n$, reflecting the oddness of the function $g(\alpha)$.}
    
The analogous expansion trajectories for a non-conformal quark-gluon gas are shown in Fig.~\ref{F1}b. In this case, since the function $g=s/n$ now depends on two dimensionless ratios $\mu/T$ and $m/T$, they are no longer straight lines. To illustrate the effect we have chosen a rather large quark mass of 1\,GeV such that the quark mass effects are clearly visible at all temperatures shown. While at high temperatures $T/m \sim 1$ the expansion trajectories still exhibit approximately linear behavior as observed in the massless case, they drastically change shape at lower temperatures $T/m \sim 0.2-0.3$ where they veer sharply to the right. This is forced by the existence of a non-vanishing net baryon charge which, due to Fermi statistics, requires $\mu\to m=1$\,GeV in the zero temperature limit $T\to 0$.\footnote{%
    This also explains the massless case where $\mu\to m=0$ as $T\to 0$.}
The rightward bend of the trajectories near $T/m \sim 0.2-0.3$ reflects the transition from a high-temperature state where quarks, antiquarks and gluons contribute democratically to the thermodynamic quantities, to a low-temperature state where antiquark and gluon contributions to $n$ are exponentially suppressed relative to those from quarks by combined mass and fugacity effects.

The late-time slopes of the various trajectories can be obtained from thermodynamic arguments: At low temperatures, quarks dominate over antiquarks and gluons, and their rest mass dwarfs their kinetic energy, such that $P \ll e$. As a result, work done by the pressure can be neglected and the energy density falls off $\propto 1/\tau$ while the ratio between energy and entropy density, $e/s$, stays approximately constant. Using this in the thermodynamic relation $Ts = e + P - \mu n$ yields
\begin{align}
    T \approx - \frac{1}{(s/n)} \, \mu + {\cal C}
\end{align}
where ${\cal C} = e/s \approx \, \mathrm{constant}$. Thus, at late-times the isentropic trajectories are straight lines with negative slopes whose magnitude is inversely proportional to $s/n$, as borne out in Fig.~\ref{F1}b.

We note for later use that it follows from Fig.~\ref{F1} that in ideal fluids at fixed temperature $g=s/n$ is a monotonic function of $\mu$, i.e. $\mu$ {\it decreases} when the specific entropy $s/n$ {\it increases}.

\section{Dissipative expansion trajectories for conformal systems with non-zero baryon chemical potential}
\label{sec4}

We now proceed to discuss the expansion trajactories for dissipative systems. We begin in this section with conformal systems without bulk viscous pressure, $\Pi=0$. Non-conformal systems with non-vanishing values for both the shear and bulk viscous stresses will be studied in the following Section~\ref{sec5}. In both sections, we will begin with a discussion of macroscopic hydrodynamic phenomena, followed by an in-depth microscopic analysis using kinetic theory.

\subsection{Conformal Bjorken hydrodynamics}
\label{sec4.1}

%
\begin{figure}[!t]
\begin{center}
  \includegraphics[width=0.8\linewidth]{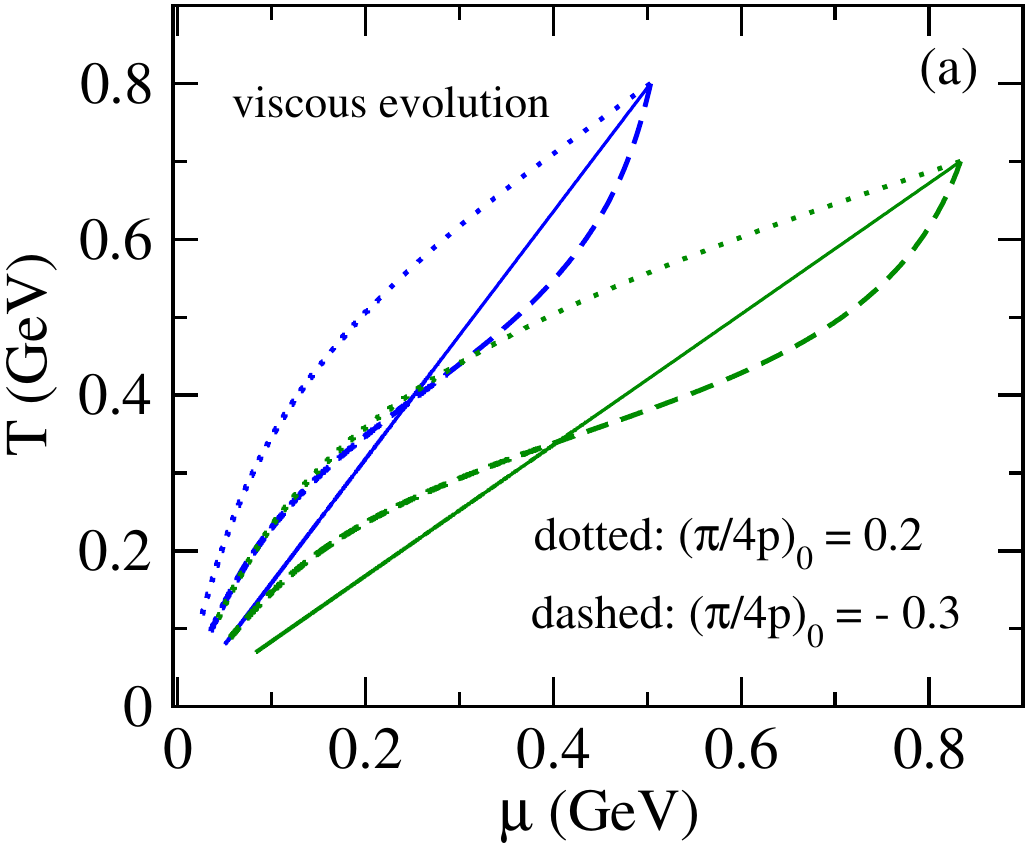}
  \includegraphics[width=0.8\linewidth]{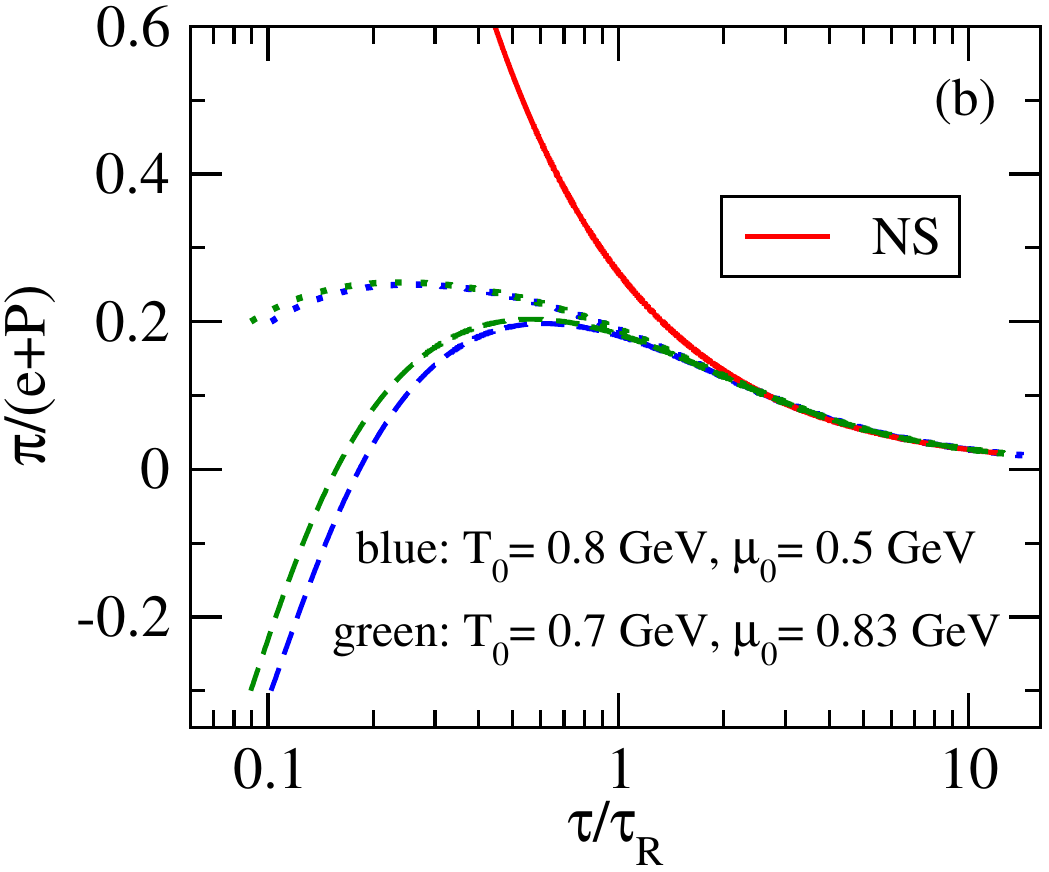}
\end{center}
\vspace{-0.6 cm}
 \caption{%
    (a) Expansion trajectories of a conformal gas in the phase diagram using dissipative hydrodynamics. Solid lines show adiabatic expansion for comparison. 
    (b) Scaled time evolution of the normalised shear stress tensor.
	\label{F2}
	}
\end{figure} 
%

We turn on dissipation by letting the shear stress tensor evolve via Eq.~(\ref{pi_evol}), with $\Pi$ set to zero. For illustration the constant $C$ appearing in $\tau_R = 5C/T$ is set to $C=10/4\pi$, and we consider two sets of initial temperatures and chemical potentials at initial time $\tau_0 = 0.1$\,fm/$c$: $(T_0, \mu_0) = (0.8, 0.5)$\,GeV and $(0.7, 0.83)$\,GeV, shown as blue and green curves in Fig.~\ref{F2}. For both sets we further consider two different values for the initial normalised shear stress: $(\pi/4p)_0 = 0.2$ and $(\pi/4p)_0 = -0.3$. The resulting dissipative expansion trajectories are shown as dotted and dashed lines in Fig.~\ref{F2}a, together with solid lines for ideal expansion starting from the same point $(T_0,\mu_0)$. Fig.~\ref{F2}b shows the evolution of the normalized shear stress $\pi/(e{+}P)$ as a function of the scaled time $\tau/\tau_R$. As the initial temperatures of the green and blue sets of curves are slightly different, their starting values of $\tau/\tau_R = \tau T/(5C)$ differ correspondingly. The solid red line in Fig.~\ref{F2}b shows the Navier-Stokes solution $\pi_{NS} = 4\eta/3\tau$ which can also be written as $\pi_{NS}/(e{+}P)=4 \tau_R/(15 \tau)$. It is a universal function of the scaled time and provides a late-time attractor for the evolution of $\pi/(e{+}P)$ \cite{Heller:2015dha, Romatschke:2017vte, Romatschke:2017acs, Jaiswal:2019cju, Kurkela:2019set}.

Looking at panel (a) we notice immediately that, in spite of the production of additional entropy by dissipation, only the dotted lines (where the initial shear stress $\pi_0$ is positive) exhibit the well-known phenomenon of {\it viscous heating} where the dissipative fluid cools more slowly than the ideal one. The dashed lines, corresponding to negative initial shear stress, instead  deviate from the ideal expansion trajectories initially {\it towards the right} (i.e. towards larger $\mu$ or lower $T$, depending on your point of view), i.e. they  initially exhibit {\it viscous cooling} where the dissipative fluid cools more rapidly than the ideal one. Only at later times do the dashed lines cross over to the left, towards smaller $\mu$.\footnote{%
    Qualitatively similar features were first observed by the authors of \cite{Dore:2020jye} (see also \cite{Dore:2022qyz}). In their work, which also includes bulk viscous effects, some dissipative expansion trajectories even started moving towards the right {\it of the starting point} $(T_0,\mu_0)$, not just right of the ideal isentrope with $(s/n)(\tau)=(s/n)_0$. We will return to this observation further below.}

%
\begin{figure}[htb!]
\begin{center}
  \includegraphics[width=0.85\linewidth]{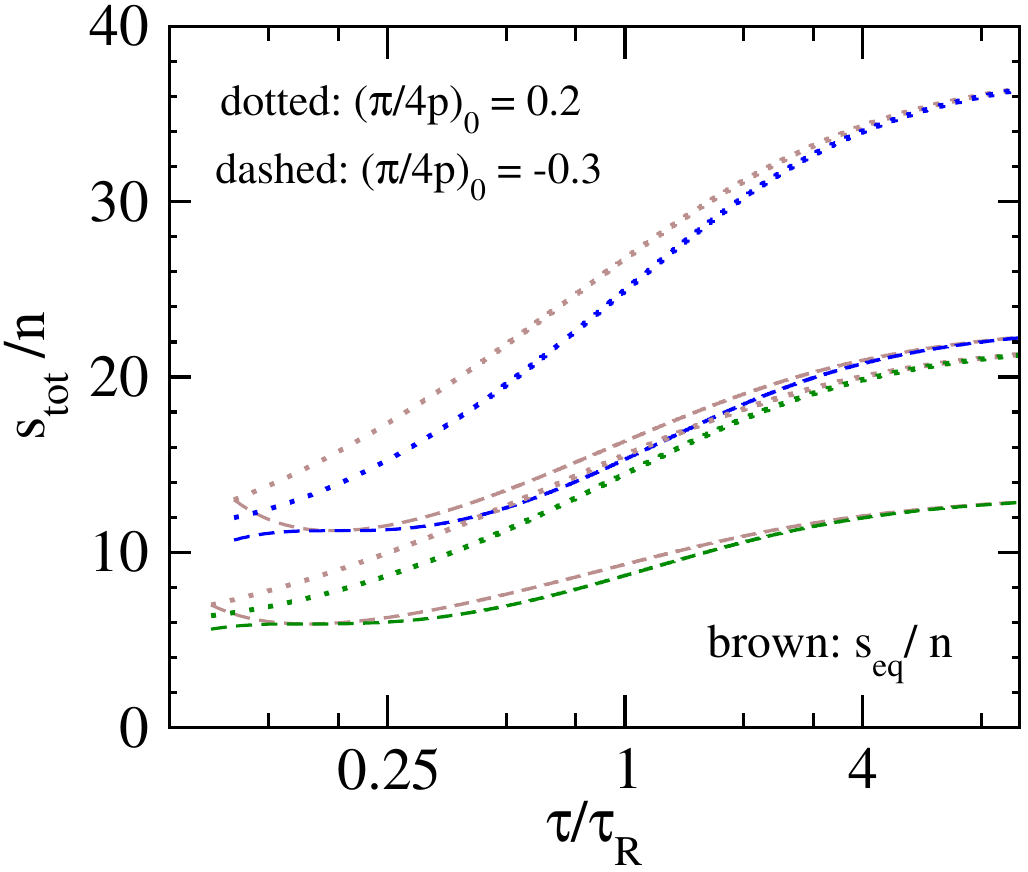}
\end{center}
 \vspace{-5mm}
 \caption{Evolution of the total specific entropy $s_\mathrm{tot}/n$ computed from Eq.~(\ref{s_2}), along trajectories of corresponding line style and color shown in Fig.~\ref{F2}a. The brown curves are obtained using the equilibrium definition. 
	\label{F3}
	\vspace*{-2mm}
	}
\end{figure} 
%

At first sight this feature appears to contradict the argument from the preceding subsection that points with larger $(s/n)(\tau) > (s/n)_0$ should be located in the phase diagram to the left of the isentropic expansion trajectory with $(s/n)_\mathrm{eq}=(s/n)_0$. When (incorrectly) interpreted within the ideal thermodynamic framework used in that subsection, the specific entropy $s/n$ initially seems to {\it decrease} along the dashed trajectories! The flaw in that argument is that it does not recognize the fact that the dissipative stresses describe deviations of the microscopic phase-space distributions from local thermal equilibrium which manifest themselves in non-equilibrium corrections to the entropy density. Initial conditions with non-vanishing shear stress (or, for that matter, any other dissipative flows) have different $(s/n)_0$ than an ideal fluid prepared with the same initial temperature and chemical potential $(T_0,\mu_0)$. An initial condition with a large negative non-equilibrium correction to the entropy density can produce additional total entropy by dissipative heating (thus satisfying the second law of thermodynamics) while, {\it at the same time}, reducing the {\it equilibrium entropy}, by transferring negative entropy from the non-equilibrium to the equilibrium part as the system thermalizes and moves closer to local equilibrium. 

A little manipulation of Eqs.~(\ref{e_evol},\ref{n_evol}) in the presence of dissipation leads (for $\Pi=0$) to
\begin{equation}
\label{s_eq_evol}
    \frac{d(s_\mathrm{eq} \tau)}{d\tau} = \frac{\pi}{ T}. 
\end{equation}
Here $s_\mathrm{eq}$ is the \textit{equilibrium} entropy density, related to $e,\,p,$ and $n$ via the fundamental relation which holds in thermal equilibrium. Clearly, the equilibrium entropy per transverse area of a fluid cell \textit{can decrease} with time whenever the shear stress $\pi$ is negative (i.e. far from its positive first-order Navier-Stokes limit for Bjorken flow). This is precisely the case for the dashed curves in Fig.~\ref{F2}a. Since in these cases it takes some time for $\pi$ to turn positive (see Fig.~\ref{F2}b), the growth with time of the total specific entropy by viscous heating is reflected also in its equilibrium contribution only at later times.

For a conformal gas of quarks and gluons at finite chemical potential, the second-order out-of-equilibrium entropy four-current using kinetic theory is\footnote{%
        This is derived in Appendix~\ref{appb}, along with the coefficients $c_{nn}$ and $c_{n\pi}$.}
\begin{align}
    S^\mu &= s_\mathrm{eq} \, u^\mu - \alpha \, n^\mu - \frac{\beta}{4\beta_\pi} \, u^\mu \, \pi^{\alpha\beta} \, \pi_{\alpha\beta} + c_{nn} \, u^\mu \, n^\alpha \, n_\alpha \nonumber \\
    & \quad  + \, c_{n\pi} \, \pi^{\mu\alpha} \, n_\alpha,
\end{align}
where $\pi^{\mu\nu}$ is the shear stress tensor and $n^\mu$ is the net-quark diffusion current. Due to the vanishing of baryon diffusion in Bjorken flow, the second-order entropy density simplifies to
\begin{equation}
    s_\mathrm{tot} = s_\mathrm{eq} - \frac{3\beta}{8\beta_\pi} \pi^2,
\label{s_2}
\end{equation}
where the second term is a (negative!) non-equilibrium correction. The entropy density is largest in local thermal equilibrium and reduced by dissipative corrections: $s_\mathrm{tot} < s_\mathrm{eq}$. The second law of thermodynamics applies only to the total specific entropy $s_\mathrm{tot}/n$ but not to the equilibrium part $s_\mathrm{eq}/n$ in isolation. This is illustrated in Fig.~\ref{F3} where only $s_\mathrm{tot}/n$ is seen to always grow monotonically with time while for the dashed lines (corresponding to initially negative shear stress) the equilibrium part $s_\mathrm{eq}/n$ initially decreases with time. As the system approaches local thermal equilibrium at late times, dissipative effects become small and the total and equilibrium specific entropies converge and saturate. 

We next turn to solving kinetic theory for Bjorken flow in order to find out whether, and to what extent, the features observed in hydrodynamics also manifest themselves in the underlying microscopic theory from which CE hydrodynamics is obtained.

\subsection{Kinetic theory for a system of massless quarks and gluons}
\label{sec4.2}

Let us consider a system of massless quarks, anti-quarks, and gluons whose evolution is governed by the Boltzmann equation with a relaxation-type collision kernel, sharing a common relaxation time $\tau_R$:
\begin{equation}
\label{RTA_generic}
    p^\mu \partial_\mu f^i(x,p) = - \frac{u \cdot p}{\tau_{R}} \left( f^i - f^i_\mathrm{eq} \right).
\end{equation}
Their equilibrium distributions are given by Eqs.~(\ref{f_eq}), setting $m=0$. Energy-momentum and net-baryon number conservation by the RTA collision kernel are ensured (see, e.g., \cite{Romatschke:2011qp, Romatschke:2017acs, Rocha:2021zcw, Rocha:2021lze, Dash:2021ibx}) if the macroscopic parameters $\beta = 1/T$, $\alpha = \mu/T$, and fluid four-velocity $u^\mu$ are defined by the Landau matching conditions:
\begin{align}
    u_\mu u_\nu T^{\mu\nu} &= \int dP \, (p{\,\cdot\,}u)^2\, \Bigl[ g_q \left(f^q {+} f^{\bar{q}}\right) + g_g f^g \Bigr] \nonumber \\
    & = e_\mathrm{eq}(T,\mu) ,
\\
    u_\mu N^\mu &= \int dP \, (p{\,\cdot\,}u) \, g_q \left(f^q {-} f^{\bar{q}} \right) = n_\mathrm{eq}(T,\mu),
\end{align}
with the right hands sides given by Eqs.~(\ref{e_eq}-\ref{n_eq}). The Lorentz invariant integration measure is defined as $dP \equiv d^3p/[(2\pi)^3 E_p]$, with $E_p = p$ here.

For a system undergoing Bjorken expansion, Eq.~(\ref{RTA_generic}) reduces in Milne coordinates to the ordinary differential equation
\begin{align}
\label{RTABoltzBjorken}
    \frac{df^i}{d\tau} = - \frac{f^i - f^{i}_\mathrm{eq} }{\tau_R},
\end{align}
where the superscript $i$ denotes the species considered. The formal solution of the above equation is \cite{Baym:1984np, Florkowski:2013lya}
\begin{align}
\label{distribution_sol}
    f^{i}(\tau; p_T, w) &= D(\tau,\tau_0) f^{i}_\mathrm{in}(\tau; p_T, w) \nonumber \\
    & + \int_{\tau_0}^{\tau} \frac{d\tau'}{\tau_R(\tau')} \, D(\tau, \tau') \, f^{i}_\mathrm{eq}(\tau';p_T, w),
\end{align}
where $w\equiv p_\eta$ is the longitudinal momentum in Milne coordinates and the damping function $D$ is defined by
\begin{equation}
    D(\tau_2, \tau_1) = \exp\biggl( - \int_{\tau_1}^{\tau_2} \frac{d\tau'}{\tau_R(\tau')} \biggr).
\end{equation}
For the initial momentum distributions we choose a Romatschke-Strickland type parametrization \cite{Romatschke:2003ms}:\footnote{\label{fn9}%
    We note that this parametrization does not allow for a full exploration of all possible values for the bulk viscous pressure $\Pi$ allowed in kinetic theory \cite{Jaiswal:2021uvv}. Very large negative values of $\Pi$ require an additional fugacity factor that allows for oversaturation of at least one of the three particle species considered here. However, for reasons discussed below, we are not primarily interested in this work in large negative $\Pi$ values -- the hydrodynamic expansion trajectories develop their most striking features when the initial value for $\Pi$ is positive.}\label{no_large_bulk}
\begin{align}
    f_\mathrm{in}^{q} &= \frac{1}{\exp\left( \sqrt{p_T^2 + (1 + \xi_0) w^2/\tau_0^2+m^2}/\Lambda_0 - \frac{\nu_0}{\Lambda_0} \right) + 1}, 
\label{RS_q}\\
    f_\mathrm{in}^{\bar{q}} &= \frac{1}{\exp\left( \sqrt{p_T^2 + (1 + \xi_0) w^2/\tau_0^2+m^2}/\Lambda_0 + \frac{\nu_0}{\Lambda_0} \right) + 1}, 
\label{RS_aq}\\
    f_\mathrm{in}^{g} &= \frac{1}{\exp\left( \sqrt{p_T^2 + (1 + \xi_0)w^2/\tau_0^2}/\Lambda_0 \right) - 1},
\label{RS_g}
\end{align}
where in this subsection we set the quark mass to $m=0$. For simplicity we have taken a common anisotropy parameter $\xi_0$ and momentum scale $\Lambda_0$ for all three species, and the same parameter $\nu_0$ for quarks and anti-quarks characterizing a non-zero initial net-quark density.
To solve for temperature and chemical potential we use the Landau matching conditions \cite{Florkowski:2013lya}
\begin{align}
\label{sol_T_c} 
    & e_\mathrm{eq}(T(\tau),\mu(\tau)) = D(\tau, \tau_0) {\cal E}(\tau) 
\\\nonumber 
    &\qquad\quad + \int_{\tau_0}^{\tau} \frac{d\tau'}{\tau_R(\tau')} \, D(\tau, \tau')
      \, {\cal H}_{e}\Bigl(\frac{\tau'}{\tau} \Bigr) \, e_\mathrm{eq}\bigl(T(\tau'), \mu(\tau')\bigr), 
\\
    & n_\mathrm{eq}(T(\tau),\mu(\tau)) = \frac{\tau_0}{\tau}\, n_\mathrm{in}. 
\label{sol_mu_c}
\end{align}
Noting the absence of net-quark diffusion in Bjorken flow we directly used the non-diffusive solution for the net-quark density in the number matching condition. The ${\cal H}_e $ function is defined as
\begin{align}
    {\cal H}_e(x) \equiv \frac{1}{2} \left(x^2 +  \frac{\tanh^{-1} \left( \sqrt{1 - \frac{1}{x^2}} \right) }{\sqrt{1 - \frac{1}{x^2}}} \right).
\end{align}
The time-dependent function ${\cal E}$ arises from the momentum integral of the initial distributions:
\begin{align}
    {\cal E}(\tau) = {\cal H}_e\left( \frac{\tau_0}{\tau \sqrt{1+\xi_0}} \right) \, e_\mathrm{eq}(\Lambda_0, \nu_0).
\end{align}
The initial net-quark density is given by $n_\mathrm{in} = n_\mathrm{eq}(\Lambda_0, \nu_0)/\sqrt{1+\xi_0}$. 

The integral equations (\ref{sol_T_c},\ref{sol_mu_c}) are solved iteratively for the temperature and chemical potential by numerical quadrature. After finding the solutions $(T(\tau), \mu(\tau))$ we use them to compute the shear stress tensor component $\pi \equiv \pi^{\eta}_{\eta}$:
\begin{align}
    & \pi(\tau) = D(\tau, \tau_0) \, {\cal C}(\tau) \nonumber \\
    & + \int_{\tau_0}^{\tau} \frac{d\tau'}{\tau_R(\tau')} \, D(\tau, \tau') \, {\cal H}_{\pi} \Bigl( \frac{\tau'}{\tau} \Bigr)\, e_\mathrm{eq}\bigl(T(\tau'), \mu(\tau')\bigr),
\end{align}
where we defined
\begin{align}
    {\cal H}_\pi(x) &\equiv \frac{1}{6 \left(  1 - x^2\right)} \Bigg[ x^2 \, \left(1 + 2\,x^2 \right) \nonumber \\
    & + \left( 1 - 4x^2 \right) \frac{\tanh^{-1} \left( \sqrt{1 - \frac{1}{x^2}} \right) }{\sqrt{1 - \frac{1}{x^2}}} \Bigg]
\end{align}
as well as
\begin{align}
    {\cal C}(\tau) =  {\cal H}_{\pi}\left( \frac{\tau_0}{\tau \sqrt{1+\xi_0}} \right) e_\mathrm{eq}(\Lambda_0, \nu_0).
\end{align}

We are particularly interested in computing the evolution of the out-of-equilibrium entropy density in kinetic theory. For this, we start from the following definition of the entropy current of a gas of quarks, anti-quarks, and gluons:
\begin{align}
\label{entropy}
    S^\mu(x) = - \sum_{i=1}^{3} g_i \int dP \, p^\mu \, \phi_{i}[f^i]. 
\end{align}
Here $i = 1, 2, 3$ labels quarks, antiquarks, and gluons, respectively. The function $\phi_i[f^i]$ is defined by
\begin{align}\label{phi_i_f}
    \phi_{i}[f^i] = f^i \ln\left(f^i\right) - \frac{1 + \theta_i \, f^i}{\theta_i} \ln\left( 1 + \theta_i \, f^i \right). 
\end{align}
It distinguishes between statistics of the constituents via the parameter $\theta_i$: $\theta_1 = \theta_2 = -1$ (Fermi-Dirac) and $\theta_3 = 1$ (Bose-Einstein). For Bjorken flow, the entropy density $s_{\mathrm{tot}} \equiv u_\mu S^\mu$ is given by
\begin{align}
\label{entropy_Bjorken_c}
    s_{\mathrm{tot}}(\tau) = - \sum_{i=1}^3 g_i \int \frac{dp_T \, dw \, p_T}{4\pi^2 \tau}  \, \phi_{i}[f^i(\tau;p_T,w)]
\end{align}
where $f^i(\tau;p_T,w)$ is the solution given in Eq. (\ref{distribution_sol}).

\begin{figure}[htb!]
\centering
  \includegraphics[width=0.8\linewidth]{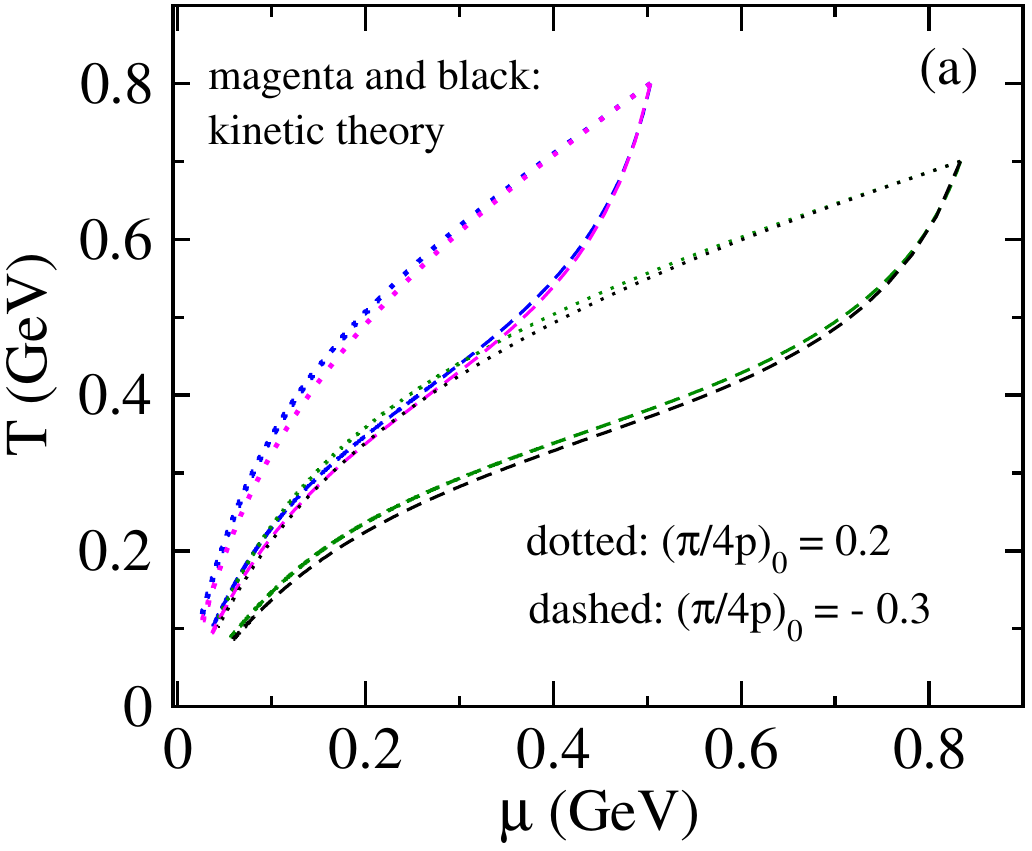}
  \includegraphics[width=0.8\linewidth]{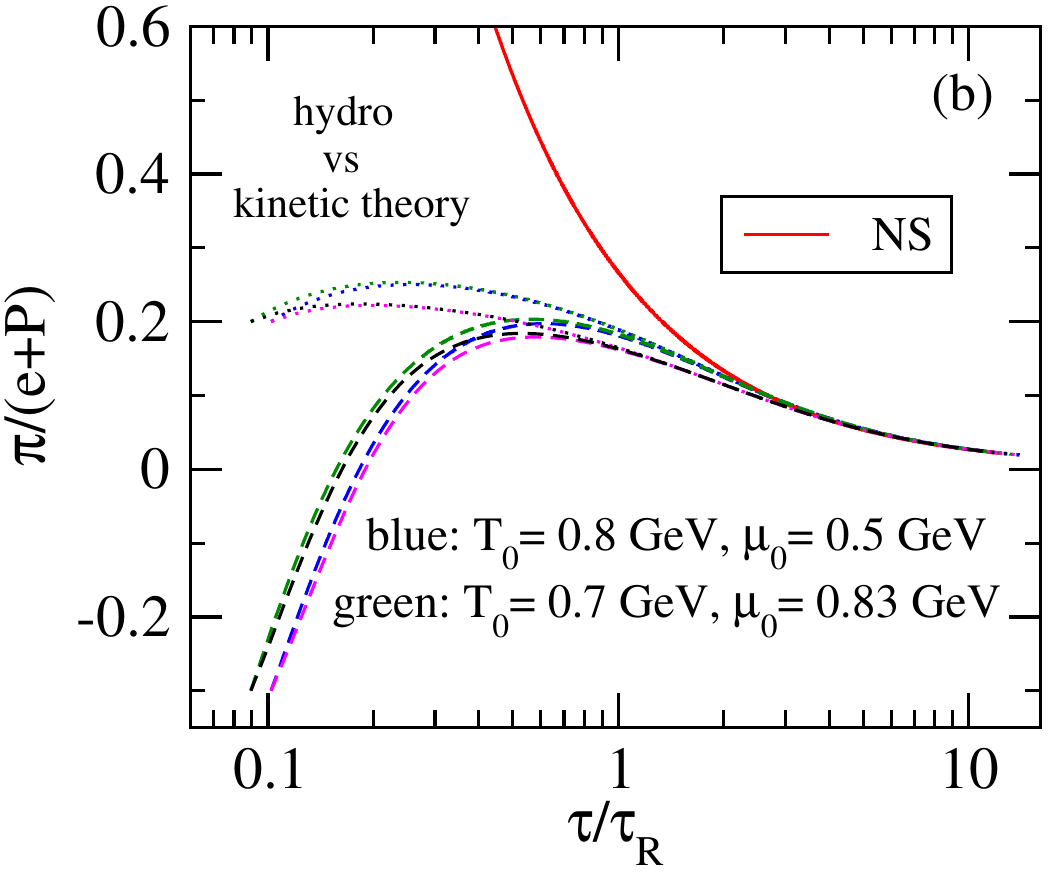}
 \caption{%
    Comparison of the expansion trajectories (panel (a)) and the evolution of the inverse Reynolds number (panel (b)) between exact kinetic theory (magenta and black curves) and their second-order conformal hydrodynamic approximation (blue and green curves).
	\label{F4}
	}
\end{figure} 

In Fig. \ref{F4} we compare, for identical initial conditions, the dissipative expansion trajectories through the phase diagram (panel (a)) and the evolution of the shear inverse Reynolds number $\pi/(e{+}P)$ (panel (b)) obtained from the exact solution (\ref{sol_T_c},\ref{sol_mu_c}) of conformal kinetic theory (magenta and black curves) with those computed with second-order conformal hydrodynamics (blue and green curves, same as those shown in Fig.~\ref{F2}). This comparison tests the accuracy of the hydrodynamic approximation to the underlying microscopic dynamics described by the RTA Boltzmann equation. In particular, the phenomenon of {\it viscous cooling} exhibited by the dashed-line trajectories in panel (a) is seen to be a robust feature of the exact solution to the underlying microscopic kinetic theory and not an artefact of the macroscopic hydrodynamic approximation. The corresponding initial parameters $(\Lambda_0, \nu_0, \xi_0)$ in the initial distribution functions (\ref{RS_q})--(\ref{RS_g}) are listed in Table \ref{table:IC_c}.
%
\begin{table}[ht!]
 \begin{center}
 \resizebox{0.7 \columnwidth}{!}{
  \begin{tabular}{|c|c|c|c|}
   \hline
   Color &  $\Lambda_0$ (GeV)  & $\nu_0$ (GeV)  &  $\xi_0$  \\
   \hline
    Magenta dotted & $5.671$ & $4.142$ & $9.258$ \\
   \hline
    Magenta dashed & $2.651$ & $1.997$ & $- 0.885$ \\
   \hline
    Black dotted & $4.856$ & $6.892$ & $9.258$ \\
   \hline
    Black dashed & $2.257$ & $3.326$ & $- 0.885$ \\
    \hline
  \end{tabular}}
  \caption{Initial parameters $(\Lambda_0,\nu_0,\xi_0)$ used to generate the kinetic theory trajectories in Fig.~\ref{F4}. Due to conformal invariance, $\bar\pi_0$ depends only on $\xi_0$. Hence, the $\xi_0$ values for the dotted ($\bar\pi_0 = 0.2$) and dashed ($\bar\pi_0 = -0.3$) pairs of curves are each identical.}
  \label{table:IC_c}
 \end{center}
 \vspace*{-.6cm}
\end{table}
%

In panel (a) the magenta and black curves for kinetic theory lie slightly to the right of the corresponding blue and green hydrodynamic trajectories, suggesting that their specific entropies are somewhat smaller in kinetic theory than in the hydrodynamic approximation. This is further substantiated in Fig.~\ref{F5} below. Consistent with this observation, the normalised shear stress shown in panel (b) is somewhat smaller in kinetic theory than for hydrodynamics which is seen to over-predict the deviation from equilibrium, especially if the initial shear stress is large and positive (dotted curves). Up to these differences, hydrodynamics shows excellent agreement with kinetic theory for conformal Bjorken evolution.

\begin{figure}[h!]
\centering
  \includegraphics[width=0.8\linewidth]{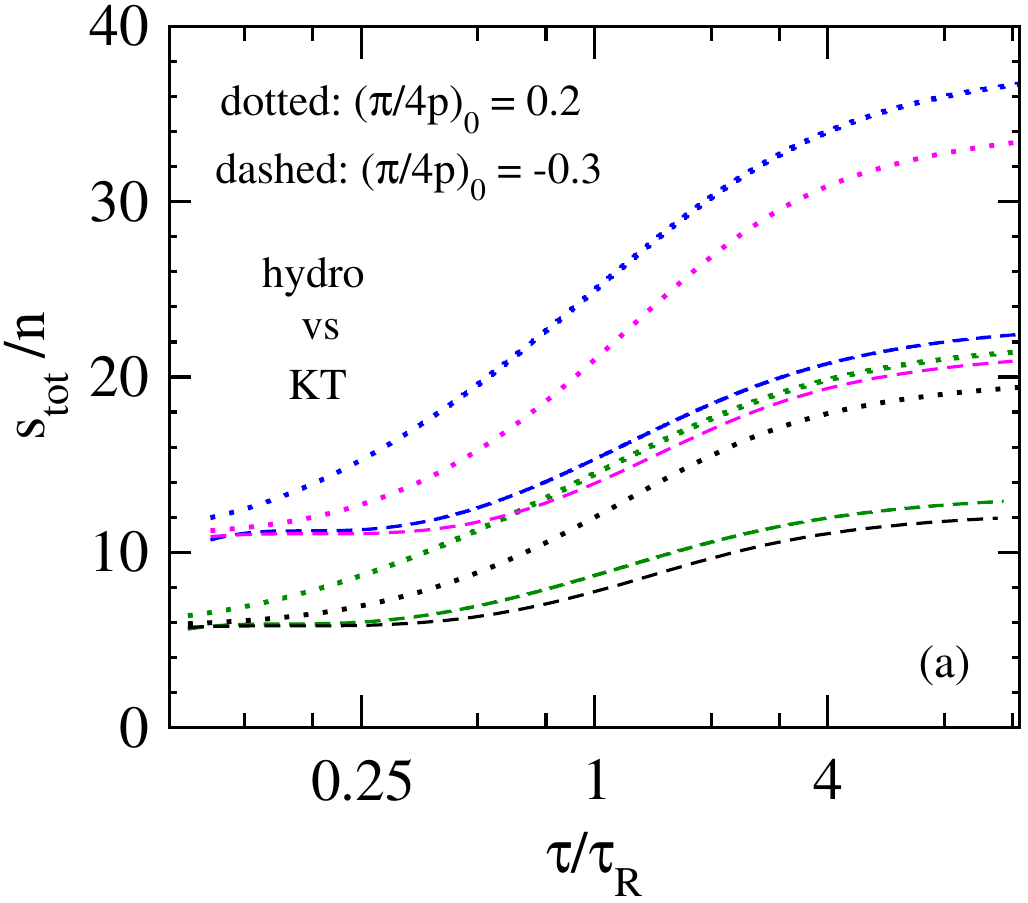}
  \includegraphics[width=0.8\linewidth]{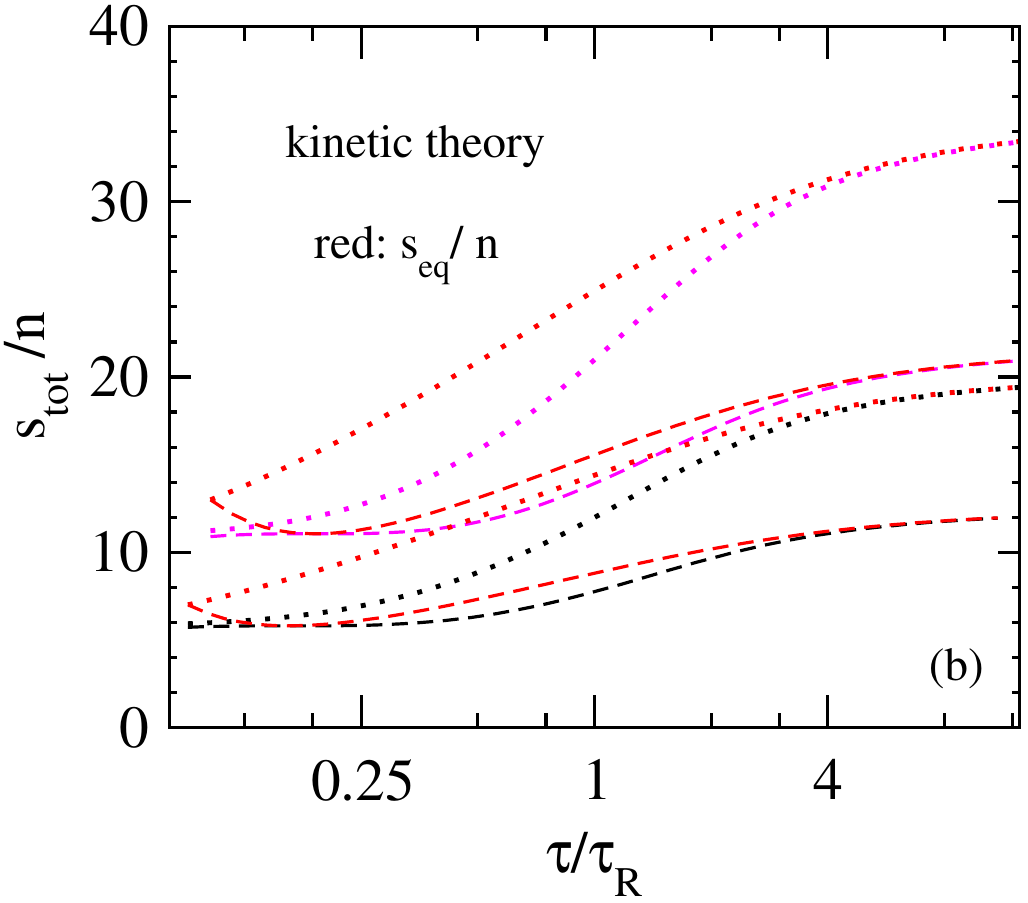}
 \vspace{-2mm}
 \caption{%
    (a) Comparison of the out-of-equilibrium specific entropy evolution in conformal RTA Boltzmann theory and its hydrodynamic approximation. Line styles and colors are identical to Fig.~\ref{F4}. (b) Comparison of the kinetic evolution of $s_\mathrm{tot}/n$ with its equilibrium part $s_\mathrm{eq}/n$ (red lines) along the magenta and black trajectories shown in Fig. \ref{F4}a. 
	\label{F5}
	}
\end{figure} 

As noted earlier in a different context \cite{Chattopadhyay:2018apf}, differences between hydrodynamics and kinetic theory become more readily apparent in the evolution of the out-of-equilibrium specific entropy. Its time evolution is shown in Fig.~\ref{F5}a. Similar to Fig.~\ref{F4}, the black and magenta curves in Fig. \ref{F5}a denote kinetic theory trajectories, whereas the green and blue curves correspond to hydrodynamics. The hydrodynamic solutions are identical to those shown in Fig.~\ref{F3}. All kinetic theory solutions lie below the hydrodynamic ones, consistent with the observations made in Figs.~\ref{F4}a and \ref{F4}b. For example, for $\bar\pi_0 = 0.2$ (dotted lines) the initial value of the total specific entropy, $s_\mathrm{tot}/n$, computed in kinetic theory is about $7\%$ percent less than the one obtained within hydrodynamics. For the dotted black and magenta curves, kinetic theory yields $(s_{\mathrm{tot}}/n)_0 = s_\mathrm{in}(\Lambda_0,\nu_0)/n_\mathrm{in}(\Lambda_0, \nu_0) \approx 5.92$ and $11.23$, respectively, whereas the dotted green and blue hydrodynamic curves correspond to $(s_{\mathrm{tot}}/n)_0 \approx 6.39$ and $11.98$. This initial difference between dotted kinetic theory and dotted hydrodynamic results persists throughout the evolution, resulting in an error of approximately 10 percent at late times. 

For the other initial condition, $\bar\pi_0 = -0.3$ (dashed lines), the kinetic theory and hydrodynamic trajectories start from approximately the same initial value for $s_{\mathrm{tot}}/n$. However, the hydrodynamic trajectories are characterised by larger entropy production and therefore differ at late times from the microscopic theory prediction by $\sim 7\%$. It is fascinating to note the sensitivity of entropy production to the initial conditions of the fluid: in spite of starting from nearly equal initial values of $s_{\mathrm{tot}}/n$, the dotted and dashed kinetic theory solutions (corresponding to $\bar\pi_0 = 0.2$ and $\bar\pi_0 = - 0.3$, respectively) differ appreciably in the late time values of $s_{\mathrm{tot}}/n$. For the dashed black and magenta trajectories in Fig.~\ref{F4}a, the late-time $s/n$ is approximately twice its initial value, whereas for the dotted trajectories the viscous heating factor is more than 3 times. The reason for this is the following: the total amount of specific entropy $\Delta (s/n)$ that can be generated during the fluid's equilibration is
\begin{align}
\label{entropy_generate}
    \Delta \left( \frac{s}{n} \right) = - \left( \frac{\delta s_{\mathrm{neq}}}{n} \right)_0 + \frac{1}{(n \tau)_0} \int_{ \tau_0}^{\infty} \frac{d\tau}{T(\tau)} \, \pi(\tau),
\end{align}
where $\delta s_{\mathrm{neq}} = s_{\mathrm{tot}} - s_{\mathrm{eq}}$. Although both dotted and dashed kinetic theory trajectories have nearly equal $(\delta s_{\mathrm{neq}}/n)_0$ initially, the dotted lines are characterised by positive shear throughout, leading to a larger value of the integral in Eq.~(\ref{entropy_generate}) than for the dashed ones where the integrand has both negative and positive contributions.   

In Fig.~\ref{F5}b we compare the evolution of the \textit{equilibrium} specific entropy ($s_\mathrm{eq}/n$, red curves) with the \textit{out-of-equilibrium} $s_\mathrm{tot}/n$ (magenta and black), both computed using kinetic theory. The magenta and black curves are the same as those in panel (a). The observed features are qualitatively similar to Fig.~\ref{F3} obtained within hydrodynamics. For initialisation $\bar\pi_0=0.2$ (dotted lines), $s_\mathrm{eq}/n$ increases monotonically ---  consistent with the dotted magenta and green phase trajectories in Fig.~\ref{F4}a lying to the left of the isentropic straight line trajectories shown in Fig.~\ref{F2}. In contrast, the dashed red curves for $\bar\pi_0 = - 0.3$ show $s_\mathrm{eq}/n$ evolving non-monotonically: the equilibrium part of the specific entropy initially decreases before starting to increase after $\tau/\tau_R \approx 0.3$. This reflects a transfer of large negative specific entropy from the non-equilibrium to the equilibrium sector at early times. During this stage the local momentum distribution forms a prolate ellipsoid whose deformation is continuously decreasing via free-streaming which red-shifts the longitudinal momenta. Once the momentum distribution reaches local isotropy, the sign of the shear stress $\pi$ flips and, according to Eq.~(\ref{s_eq_evol}), so does the direction of entropy flow between the non-equilibrium and equilibrium sectors. In Figure~\ref{F4}a the initial decrease of $s_\mathrm{eq}/n$ manifests itself by causing the dashed magenta and black curves to first move to the right of the straight-line ideal expansion trajectory before veering left at late times, eventually crossing the ideal line (see Fig.~\ref{F2}) due to the overall production of entropy by viscous heating and the approach to local equilibrium at late times. 

This transfer of entropy from the non-equilibrium to the equilibrium sector is a novel effect which to our knowledge has not been previously described.
The second law of thermodynamics demands (i) $s/n \leq s_\mathrm{eq}/n$ and (ii) $d(s/n)/d\tau \geq 0$, where $s=s_\mathrm{tot}$ stands for the {\it total} specific entropy (i.e. the sum of its equilibrium contribution $\bigl(s_\mathrm{eq}/n\bigr)(T,\mu)$ and its non-equilibrium correction which depends on the dissipative flows). As time evolves, entropy can be transferred from the (negative) non-equilibrium correction to the equilibrium part. Therefore $d(s_\mathrm{eq}/n)/d\tau$ does {\it not} need to be positive definite.

Comparison of the dashed and dotted curves in Fig.~\ref{F4}b shows that the rate (sign and magnitude) at which entropy flows from the non-equilibrium to the equilibrium sector depends on the values of the dissipative fluxes. Within the hydrodynamic framework this is expressed by Eq.~(\ref{s_eq_evol}). However, while the dashed lines in Fig.~\ref{F4}a move to the right of the ideal expansion trajectories, they do not move towards larger chemical potential as has been observed for some expansion trajectories shown in Refs.~\cite{Dore:2020jye, Dore:2022qyz}. In the following subsection we will therefore discuss the criteria that must be satisfied for obtaining dissipative expansion trajectories that start out by moving towards larger $\mu$. We will see that this very hard to achieve in conformal systems but much easier in non-conformal systems. This will set the stage for discussion of non-conformal expansion trajectories in Section~\ref{sec5}.

\subsection{Criteria for generating trajectories with increasing chemical potential}
\label{sec4.3}
    
Let us explore the necessary conditions required to generate Bjorken expansion trajectories where the chemical potential increases with Milne time. The discussion will be general, including both shear and bulk viscous stresses. We start with the thermodynamic relations among differentials of the Lagrange parameters $x_a = \{T,\mu\}$ and the densities $X_a = \{e,n \}$:
\begin{equation}
    \begin{pmatrix}
      dT \\
      d\mu
    \end{pmatrix} = \frac{1}{\mathcal{A}}
\begin{pmatrix}
  a_4 & -a_2 \\
    - a_3 & a_1  
\end{pmatrix} 
\begin{pmatrix}
  de\\
  dn
\end{pmatrix},
\end{equation}
with the determinant $\mathcal{A} \equiv (a_1 a_4 - a_3 a_2)$ and the positive definite thermodynamic response functions
\begin{align}
\nonumber
    a_1 & = \left( \frac{\partial e}{\partial T} \right)_{\mu},\
    a_2 = \left( \frac{\partial e}{\partial \mu} \right)_{T},
\end{align} 
    %
\begin{align}
\label{a1toa4}
    a_3 & = \left( \frac{\partial n}{\partial T} \right)_{\mu},\
    a_4 = \left( \frac{\partial n}{\partial \mu} \right)_{T}.
\end{align}
The determinant $\mathcal{A}$ is also positive since a thermodynamic transformation $(de{\,>\,}0,\, dn{\,=\,}0)$ must lead to $dT{\,>\,}0$.\footnote{%
    Similarly, a transformation where $de=0$ and  $dn>0$ must yield $d\mu > 0$. We spot-checked both expectations numerically.}
Thus, the criterion for an increase in chemical potential, $d\mu/d\tau \geq 0$, is
\begin{align}
        - a_3 \, \frac{de}{d\tau} + a_1 \, \frac{dn}{d\tau} \geq 0\  
        \implies\  a_3 \frac{e+P-\phi}{\tau} - a_1 \frac{n}{\tau} \geq 0,
\end{align}
where we used the equations of motion (\ref{e_evol}, \ref{n_evol}) and introduced the shorthand $\phi \equiv \pi - \Pi$. Accordingly, the amount of normalised dissipative stress, $\bar\phi \equiv \phi/(e{+}P)$, required for generating $d\mu>0$ trajectories is
\begin{align}
\label{bound}
        \bar\phi \leq 1 - \frac{a_1}{a_3} \, \frac{n}{(e{+}P)}\equiv \psi.
\end{align}
This obviously prefers negative shear stress values, $\pi<0$, and positive bulk viscous pressure, $\Pi>0$. Both are ``unnatural'' for Bjorken expansion where the Navier-Stokes values for the dissipative fluxes have the opposite signs.

%
\begin{figure}[h]
\centering
  \includegraphics[width=0.8\linewidth]{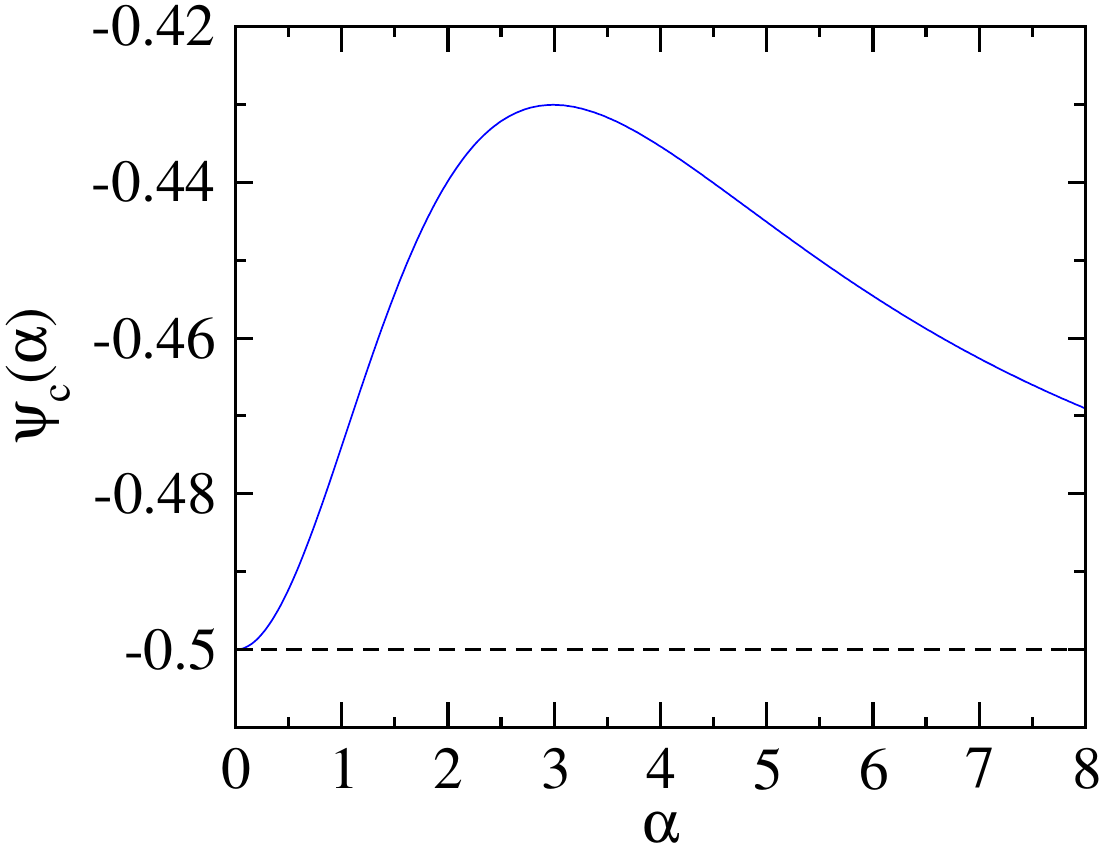}
 \vspace{-3mm}
 \caption{%
    The solid blue line shows the upper bound on the normalised shear stress that ensures $d\mu/d\tau > 0$ in conformal Bjorken dynamics. The dashed black line, $\psi_c = - 0.5$, represents the lower bound on $\bar\pi$ allowed by kinetic theory.
\label{F6}
	}
\end{figure}%
%

Let us first explore the implications of this criterion for the simpler case of a conformal system $(\Pi{\,=\,}0)$ of massless quarks, anti-quarks, and gluons at non-zero net baryon density.\footnote{%
    Note that the upper limit of $\bar\phi$ depends on the equation of state of the system. Accordingly, the bound changes somewhat when replacing quantum statistics (which we use here) by classical Boltzmann statistics, or changing the degeneracy factors for (anti-)quarks or gluons.} 
In this case the response functions $a_1$ and $a_3$ simplify to
\begin{align}
    a_1 = 3\, s_\mathrm{eq}, \quad
    a_3 = \frac{g_q}{3} \, T \, \mu,
\end{align}
such that the condition (\ref{bound}) yields
\begin{align}
\label{psi}
    \bar\pi \leq 1 - 3\, \Bigl(\frac{1}{2} + \frac{\alpha^2}{2\pi^2}\Bigr) \Big/ 
    \Bigl(1 + \alpha \frac{n}{s_\mathrm{eq}}\Bigr) 
    \equiv \psi_c(\alpha).
\end{align}
A plot of $\psi_c(\alpha)$ is shown in Fig.~\ref{F6}. For the trajectory to move towards the right in the phase diagram, $d\mu>0$, $\bar\pi$ must lie below the blue curve. In kinetic theory there is a lower bound on the normalised shear, $\bar\pi \geq - 0.5$ (arising from the condition of non-negative effective transverse pressure, $P_T = P + \pi/2\geq0$), indicated by the black dashed line. In conformal kinetic theory the condition $d\mu>0$ thus restricts $\bar\pi$ to a very thin sliver of parameter space bounded by the solid blue and dashed black lines which covers only a small fraction of the overall parameter space $\bar\pi \in [-0.5, 0.25]$. Moreover, even if the system is initialized within this sliver at early times, it quickly moves out of it as the normalised shear stress rapidly approaches the free-streaming attractor $\bar\pi \approx 0.25$. Thus, in \textit{conformal} kinetic theory it is nearly impossible to find trajectories characterised by substantial increase in chemical potential.

\begin{figure}[h!]
\centering
\includegraphics[width=0.75\linewidth]{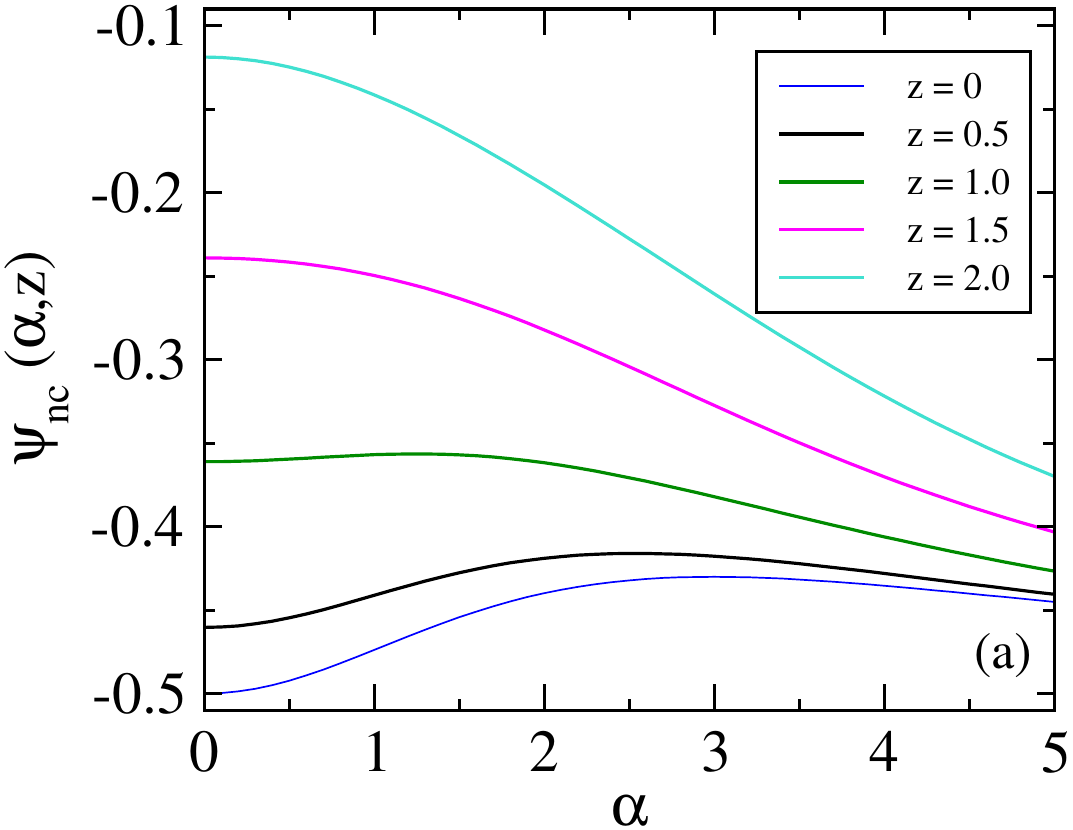}
\vspace{0.3 cm}

\includegraphics[width=0.75\linewidth]{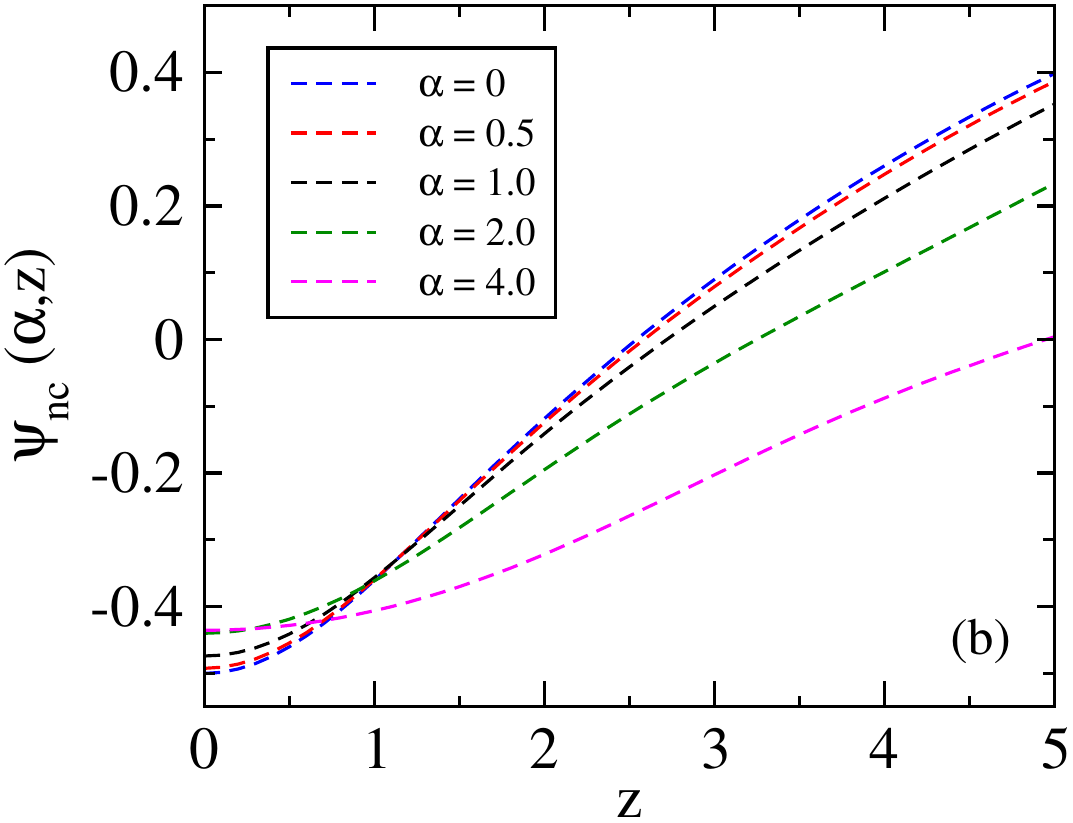}
 \vspace{-3mm}
\caption{%
    The upper bound $\psi_\mathrm{nc} (\alpha,z)$ in non-conformal systems, as (a) a function of $\alpha=\mu/T$ for fixed values of $z=m/T$ and (b) a function of $z$ for fixed values of $\alpha$. For $\bar\phi < \psi_\mathrm{nc}$ expansion trajectories are characterised by $d\mu/d\tau>0$.
	\label{F7}
	}
\end{figure} 

However, the introduction of masses for the quarks changes the equation of state of the system, and accordingly all thermodynamic quantities determining the upper bound of $\bar\phi$ on the r.h.s. of Eq.~(\ref{bound}) change, too. As $\bar\phi$ is dimensionless, the bound can depend only on the dimensionless variables $\alpha=\mu/T$ and $z \equiv m/T$. A plot of $\psi_\mathrm{nc}(\alpha,z)$ (defined by the r.h.s. of the inequality (\ref{bound})) is shown in Fig.~\ref{F7}, where in panel (a) $\alpha$ is varied while keeping $z$ constant, and vice-versa in panel (b). Clearly, with increasing quark mass the allowed domain of $\bar\phi$ that allows for trajectories with $d\mu/d\tau>0$ significantly increases. For instance, for typical initial values $\alpha_0 \sim 1$ and $m/T_0 \sim 1$, one needs $\bar\phi_0 \lesssim - 0.36$ which is much less restrictive than the requirement  $\bar\phi_0 \lesssim - 0.49$ for the same $\alpha_0$ in the conformal case. 

\section{Dissipative expansion trajectories for non-conformal systems with a conserved charge} 
\label{sec5}
\subsection{Non-conformal Bjorken hydrodynamics} 
\label{sec5.1}

\begin{figure}[h!]
\begin{center}
\includegraphics[width=0.8\linewidth]{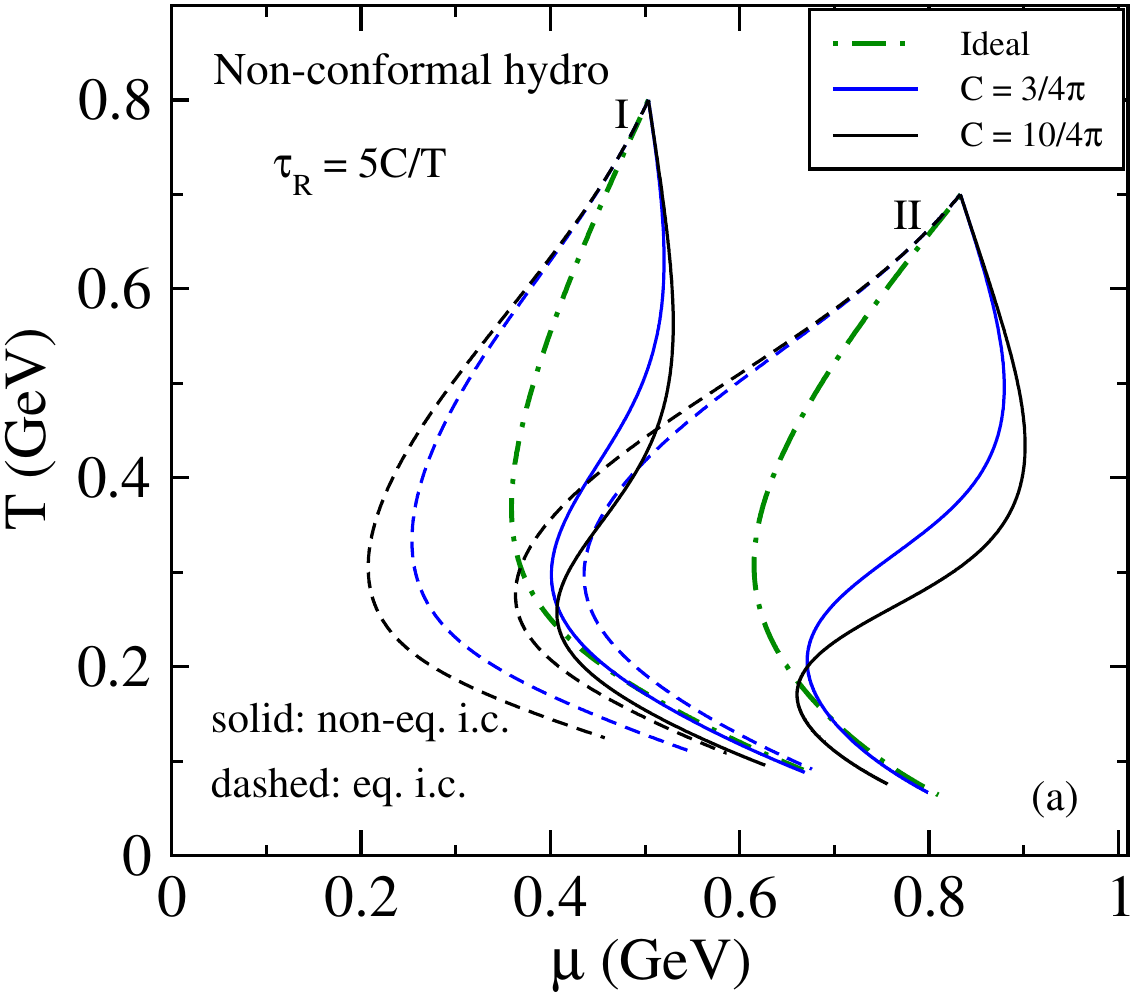}
\includegraphics[width=0.85\linewidth]{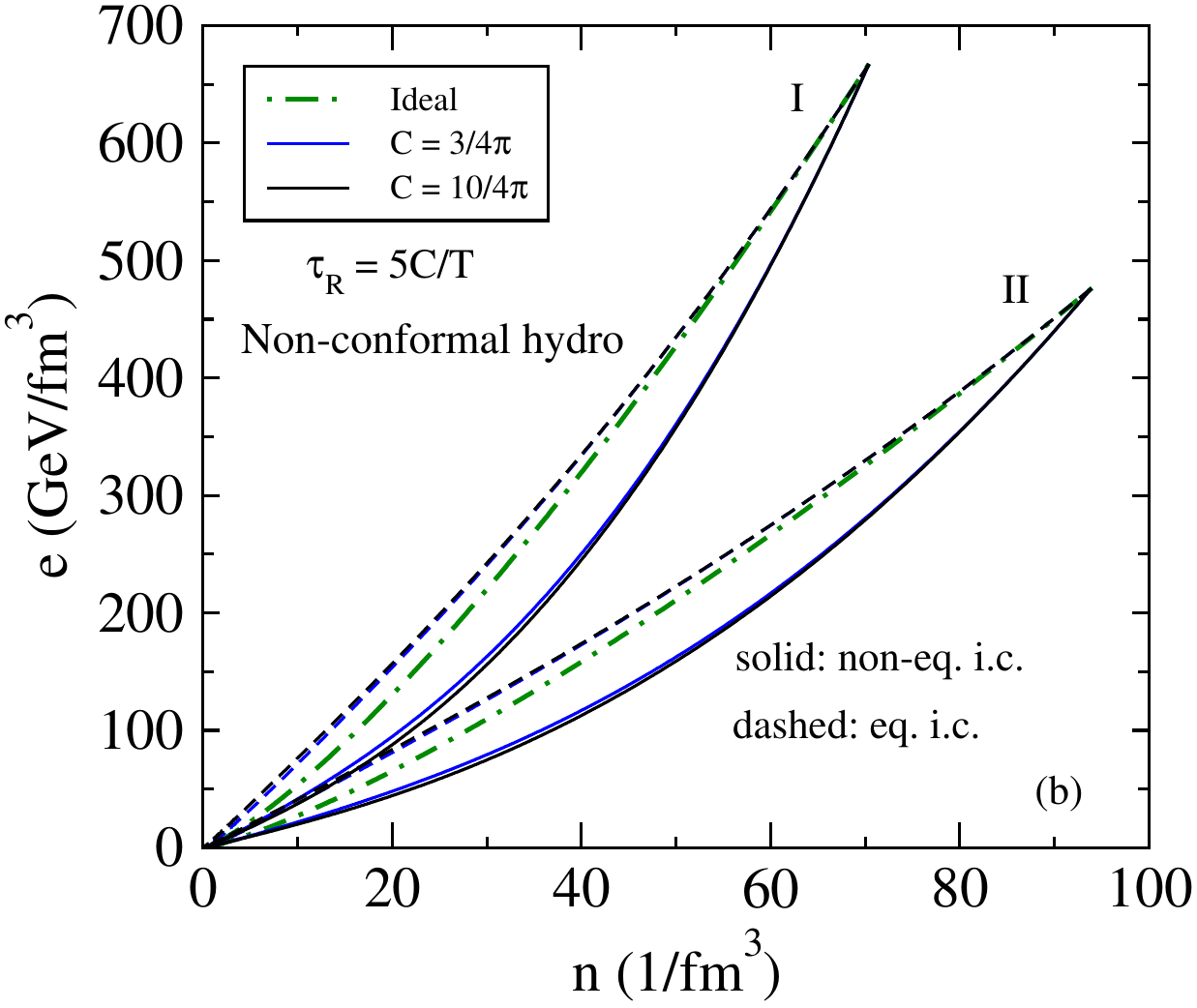}
\end{center}
\vspace{-4mm}
\caption{%
   Expansion trajectories from second-order non-conformal hydrodynamics for a massive quark-gluon gas with non-zero net quark number, in the $T{-}\mu$ plane (a) and $e{-}n$ plane (b), respectively. The green dash-dotted lines show the corresponding ideal expansion trajectories from Fig.~\ref{F1}b (for $s/n = 14.8$ and 8.2, respectively) for comparison. The other lines are explained in the text.
	\label{F8}
	\vspace*{-4mm}
	}
\end{figure} 

We now relax the restriction to conformal symmetry and restore the bulk viscous pressure $\Pi$ in the second-order viscous hydrodynamic equations  (\ref{e_evol}-\ref{pi_evol}) and solve them with initial conditions imposed at $\tau_0 = 0.1$\,fm/$c$ using the transport coefficients obtained in Appendix \ref{appa} for a massive quark-gluon gas at $\mu \neq 0$. The resulting expansion trajectories in the temperature-chemical potential plane (panel a) and the energy-(quark)density plane (panel b) are shown in  Fig.~\ref{F8}. We consider two sets of initial conditions taken from Fig.~\ref{F1}, with $(T_0,\mu_0)\approx(0.8,0.5)$\,GeV (set I) and $\approx(0.7,0.83)$\,GeV (set II), respectively.\footnote{%
        These correspond to initial energy and net-quark densities of $(e_0,n_0)\approx (667\, \mathrm{GeV/fm}^3, 70/\mathrm{fm}^{3})$ (set I) and $\approx(476\,\mathrm{GeV/fm}^3, 94/\mathrm{fm}^{3})$ (set II).}
The corresponding ideal expansion trajectories from Fig.~\ref{F1}b are shown as green dash-dotted lines for orientation. In both panels the blue and black curves show dissipative expansion trajectories for two choices of the coupling strength, parametrized by the relaxation time constant $C$ as given in the legend, while solid and dashed lines distinguish between different initial conditions for the dissipative flows: dashed lines use local equilibrium initial conditions $\bar\pi_0 = \bar\Pi_0 = 0$ whereas solid lines correspond to $\bar\pi_0 \approx - 0.437$, $\bar\Pi_0 \approx 0.012$ for set~I and $\bar\pi_0 \approx - 0.439$, $\bar\Pi_0 \approx 0.010$ for set~II, such that for both sets $\bar\phi_0 \equiv \bar\pi_0 - \bar\Pi_0 = - 0.45$.\footnote{%
    The specific choice of these somewhat arbitrary-looking values is motivated by the comparison with kinetic theory in Sec.~\ref{sec5.2} below. As explained in footnote~\ref{fn9}, the \textit{three-parameter} Romatschke-Strickland distributions (\ref{RS_q}-\ref{RS_g}) limit the range of initial dissipative flows that can be generated; in particular, this parametrization takes away the freedom of choosing $\bar\pi_0$ and $\bar\Pi_0$ independently once their difference $\bar\phi_0$ is specified.}

Similar to the conformal case shown in Fig.~\ref{F2}, the dashed and solid lines in Fig.~\ref{F8} form two qualitatively different classes of dissipative expansion trajectories. The differences between them are much more pronounced in the $T{-}\mu$ planel (panel (a)) than in the $e{-}n$ plane (panel (b)): the dashed lines, corresponding to equilibrium initial conditions, first graze along the ideal trajectories before {\it viscous heating} drives them left towards larger equilibrium specific entropies. Since larger $C$ values (longer relaxation times) describe more weakly coupled systems with larger shear and bulk viscosities, they lead to stronger viscous entropy production. The solid lines, on the other hand, which correspond to large negative initial values for the (normalized) shear and bulk stress combination $\bar\phi \equiv \bar\pi - \bar\Pi$, exhibit {\it viscous cooling}: in panel (a) they move initially {\it right}, i.e. towards larger chemical potentials, as anticipated in Sec.~\ref{sec4.3}, and cool more rapidly than the ideal fluid. Again, this phenomenon lasts longer for more viscous systems described by larger $C$ values.\footnote{%
    In the $e{-}n$ plane (panel (b)) the differences arising from different coupling strengths $C$ are almost unrecognizable: the blue and black dashed trajectories are indistinguishable, and also their solid analogues (corresponding to off-equilibrium initial conditions) are very close to each other.
} 
The decrease in equilibrium specific entropy along these curves is due to their initialisation with large negative $\phi$, as dictated by the non-conformal version of Eq.~(\ref{s_eq_evol}),
\begin{align}
\label{seq_evol_nc}
    \frac{d(s_\mathrm{eq} \tau)}{d\tau} = \frac{\phi}{T} = \frac{\pi-\Pi}{T}\,.
\end{align}
As described in Secs.~\ref{sec4.1} and \ref{sec4.2}, it reflects the transfer of negative viscous entropy between the non-equilibrium and equilibrium contributions to the total specific entropy.

\begin{figure}[t!]
\vspace*{3mm}
\centering
  \includegraphics[width=0.8\linewidth]{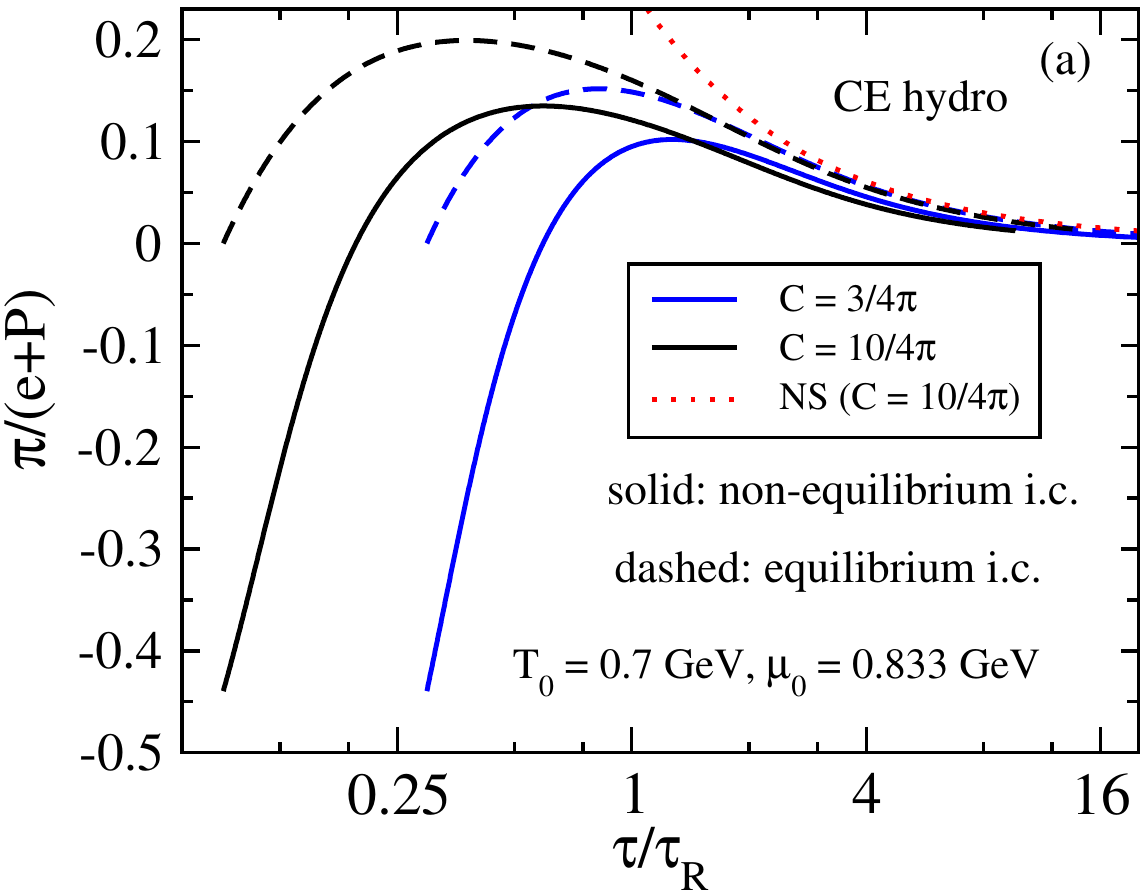}
  \includegraphics[width=0.8\linewidth]{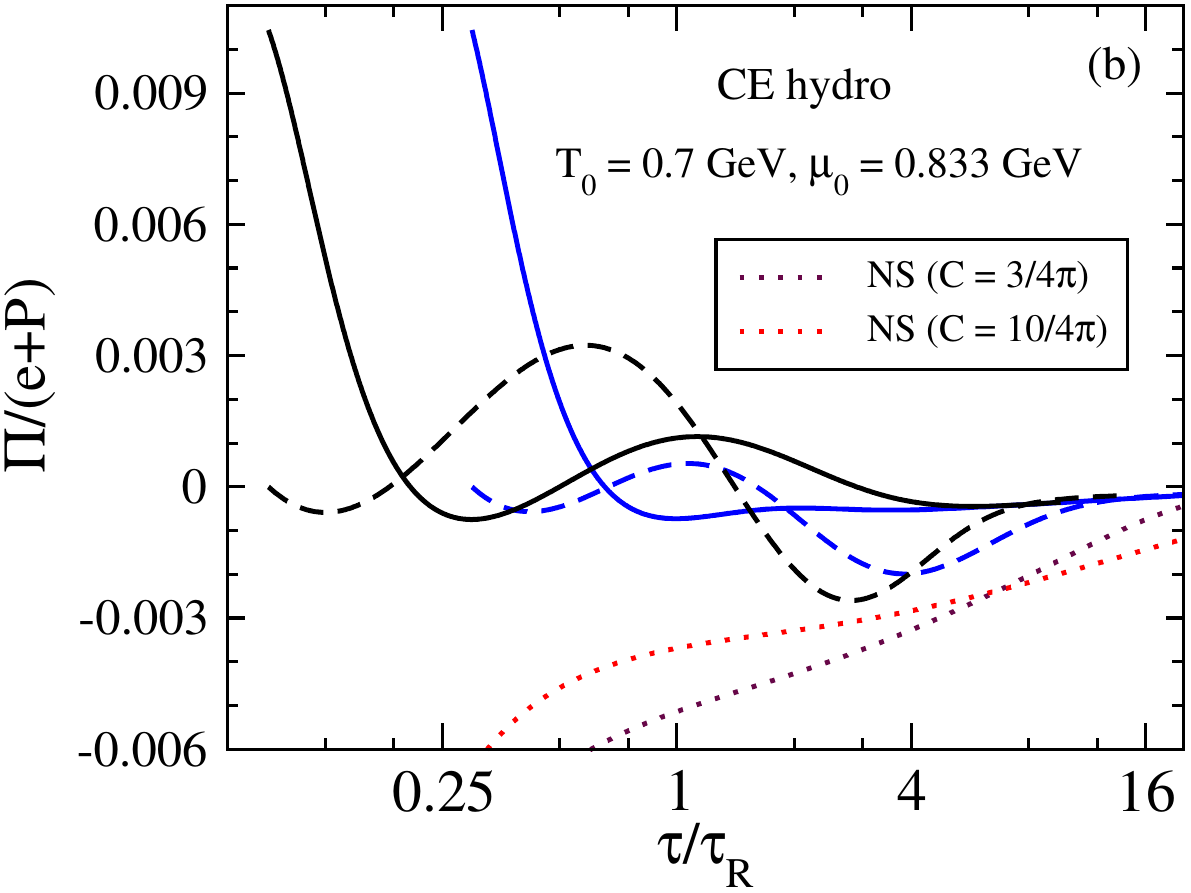}
 \vspace{-4mm}
 \caption{Scaled time evolution of (a) shear and (b) bulk inverse Reynolds numbers obtained from second-order hydrodynamics, corresponding to the expansion trajectories (of same color and line-style) in set II of Fig. \ref{F8}. The dotted curves show the Navier-Stokes limit. (Note that, for fixed $\tau_0=0.1$\,fm/$c$,  different $C$ values result in different scaled initial times $\tau_0/\tau_R.$)
	\label{F9}
	\vspace*{-4mm}
	}
\end{figure} 

As the system evolves one expects the shear and bulk viscous pressures to approach their Navier-Stokes limits (which is positive for shear stress and negative for the bulk viscous pressure). This is illustrated in Fig.~\ref{F9}.\footnote{%
    To avoid clutter we show the evolution of $\bar\pi$ and $\bar\Pi$ only for set II. The results for set I are qualitatively similar.}
When this happens, $\phi$ turns positive and $s_\mathrm{eq}/n$ increases again, forcing the solid curves in Fig.~\ref{F8} to turn left, towards regions of smaller chemical potential. As the overall result of dissipation is an increase of total entropy, the solid lines eventually have to cross the ideal expansion trajectories and end up to their left where $s_\mathrm{eq}/n>(s/n)_0$. 

As the system approaches local thermal equilibrium $s_\mathrm{eq}/n$ saturates, for different initial conditions generally at different values, and the expansion trajectories settle on isentropic trajectories corresponding to these asymptotic $s/n$ values. Therefore, once $T/m$ drops below about 0.3, all trajectories again turn to the right, due to the nonzero mass of the carriers of the conserved baryon charge as explained in Sec.~\ref{sec3}. A deeper understanding of this on a microscopic level will be obtained in the next subsection where the same phenomena are seen to arise in non-conformal kinetic theory.

We close this subsection with some additional discussion of Fig.~\ref{F9}. Remember that the dashed trajectories start with equilibrium initial conditions whereas the solid lines correspond to large negative initial shear combined with a much smaller initial bulk viscous pressure. Panel (a) shows that for both types of initial conditions $\bar\pi$ initially grows rapidly to significantly large positive values, on a rapid time scale $\tau \lesssim \tau_R$. i.e. before microscopic collisions become effective. This is caused by the rapid longitudinal expansion which red-shifts the longitudinal momenta and reduces the longitudinal pressure of the system \cite{Jaiswal:2021uvv, Jaiswal:2019cju, Kurkela:2019set}.\footnote{%
    This argument holds strictly only as long as $\bar\Pi$ is small which applies here.}
At much later times $\tau\gg \tau_R$ the shear stress joins its first-order Navier-Stokes limit (here shown for $C = 10/4\pi$) $\bar\pi_\mathrm{NS} = 4/3 \times [\beta_\pi/ (e+P)]/(\tau/\tau_R)$ where the dimensionless quantity $\beta_\pi/(e{+}P)$ is a function of $\mu/T$ and $m/T$ that approaches 1/5 in the massless limit.

Panel (b) of Fig.~\ref{F9} shows that the bulk viscous pressure evolves in a rather complex fashion which, for the case of equilibrium initial conditions, even features oscillations driven by the bulk-shear coupling terms in Eqs.~(\ref{PI_evol},\ref{pi_evol}) (qualitatively consistent with hydrodynamic results obtained in \cite{Jaiswal:2014isa, Jaiswal:2022udf} for a single-species massive Boltzmann gas at vanishing chemical potential).\footnote{%
    Due to the smallness of $\bar\Pi$ its back-reaction on the evolution of $\bar\pi$ is negligible.}
Also, it approaches its first-order Navier-Stokes limit $\bar\Pi_\mathrm{NS} = - [\beta_\Pi/(e{+}P)]/(\tau/\tau_R)$
(shown as dotted lines for both choices of $C$ in Fig.~\ref{F9}b) only at much later times than the shear stress (even outside the time range shown here). We checked that this is, too, is caused by bulk-shear coupling \cite{Denicol:2014vaa} with a shear stress that exceeds the magnitude of $\bar\Pi$ by more than an order of magnitude. 

\subsection{Kinetic theory for a non-conformal quark-gluon gas with net baryon charge}
\label{sec5.2}
\subsubsection{Formal solution}
\label{sec5.2.1}

The RTA Boltzmann equation (\ref{RTABoltzBjorken}) for a system of gluons and massive quarks with Boltzmann statistics undergoing Bjorken flow has been solved in Refs.~\cite{Florkowski:2014sfa, Florkowski:2014txa}. For reasons of uniformity of notation we briefly review this solution, generalizing it along the way to include quantum statistics in the distribution functions. The solution will be expressed in terms of macroscopic quantities, i.e. the temperature, chemical potential, shear, and bulk viscous pressures. The initial conditions for the shear and bulk viscous pressures are generated from Eqs.~(\ref{RS_q}-\ref{RS_g}). With these initial distributions, the exact evolutions of the temperature and chemical potential are obtained by the Landau matching conditions (\ref{sol_T_c},\ref{sol_mu_c}), generalized to non-zero quark mass:   
\begin{align}
    & e_\mathrm{eq}(T(\tau),\mu(\tau)) = D(\tau, \tau_0) \frac{\Lambda_0^4}{4\pi^2} \tilde{H}_e\left[ \frac{\tau_0}{\tau\sqrt{1+\xi_0}} , \frac{m}{\Lambda_0}, \frac{\nu_0}{\Lambda_0}\right] \nonumber \\
    &+ \frac{1}{4\pi^2} \int_{\tau_0}^{\tau} \frac{d\tau'}{\tau_R(\tau')} \, D(\tau, \tau') \, T(\tau')^4 \, \tilde{H}_e\left[ \frac{\tau'}{\tau} , \frac{m}{T(\tau')}, \frac{\mu(\tau')}{T(\tau')}\right],
\label{sol_T_nc} \\
    & n_\mathrm{eq}(T(\tau),\mu(\tau)) = \frac{\tau_0}{\tau} \, n_\mathrm{in}.
\label{sol_mu_nc}
\end{align}
Here
\begin{align}
    \tilde{H}_e(y,z,\alpha) &\equiv \int_0^{\infty}\!\!\! du \, u^3 \Bigg[ g_g \, \tilde{f}^g_\mathrm{eq}(u) \, H_e(y, 0) \nonumber \\
    & + \, g_q \left( \tilde{f}^q_\mathrm{eq}(u,z,\alpha) + \tilde{f}^{\bar{q}}_\mathrm{eq}(u,z,\alpha) \right) \, H_e(y,z/u) \Bigg],
\end{align}
with $\tilde{f}_\mathrm{eq}$ from Eq.~(\ref{e_nc}-\ref{n_nc}) above and
\begin{align}
    H_e(y,z) \equiv y \left( \sqrt{y^2 + z^2} + \frac{1+z^2}{\sqrt{y^2-1}} \, \tanh^{-1}\sqrt{\frac{y^2-1}{y^2+z^2}} \right).
\end{align}
Eqs.~(\ref{sol_T_nc}-\ref{sol_mu_nc}) are solved by numerical iteration to obtain solutions for $T(\tau)$ and $\mu(\tau)$. Once these are known the distribution function $f(\tau;p_T,w)$ itself can be determined at any $\tau$ using Eq.~(\ref{distribution_sol}), and all macroscopic hydrodynamic quantities can be obtained by taking appropriate moments of $f$. For example, the out-of-equilibrium entropy density is given by Eq.~(\ref{entropy_Bjorken_c}). The shear and bulk viscous stresses are obtained from the effective longitudinal and transverse pressures, $P_L$ and $P_T$, using $\pi \equiv 2 (P_T - P_L)/3$ and $\Pi \equiv (P_L + 2 P_T - 3P)/3$, where
\begin{align}
    & P_{L,T} = D(\tau, \tau_0) \frac{\Lambda_0^4}{4\pi^2} \tilde{H}_{L,T}\left[ \frac{\tau_0}{\tau\sqrt{1+\xi_0}} , \frac{m}{\Lambda_0}, \frac{\nu_0}{\Lambda_0}\right] \nonumber \\
    & + \frac{1}{4\pi^2} \int_{\tau_0}^{\tau} \frac{d\tau'}{\tau_R(\tau')} \, D(\tau, \tau') \, T^4(\tau') \, \tilde{H}_{L,T}\left[ \frac{\tau'}{\tau} , \frac{m}{T(\tau')}, \frac{\mu(\tau')}{T(\tau')}\right],
\end{align}
with
\begin{align}
\nonumber
\!\!\!\!
& \tilde{H}_{L,T}(y,z,\alpha) \equiv \int_0^{\infty} \!\!\! du \, u^3 \Bigg[ g_g \tilde{f}^g_\mathrm{eq}(u) H_{L,T}(y, 0) \nonumber \\
& + \, g_q \left( \tilde{f}^q_\mathrm{eq}(u,z,\alpha) + \tilde{f}^{\bar{q}}_\mathrm{eq}(u,z,\alpha) \right) H_{L,T}\left(y,\frac{z}{u}\right)  \Bigg], 
\end{align}
where
\begin{align}
    H_{L}(y,z) &= \frac{y^3}{(y^2-1)^{3/2}} \Bigg( \sqrt{(y^2-1)(y^2+z^2)} \nonumber \\
    & - (z^2 + 1) \, \tanh^{-1}\sqrt{\frac{y^2-1}{y^2+z^2}} \Bigg), \\
    H_{T}(y,z) &= \frac{y}{2 (y^2-1)^{3/2}} \Bigg( - \sqrt{(y^2-1)(y^2+z^2)} \nonumber \\
    & + (z^2 + 2y^2 - 1) \, \tanh^{-1}\sqrt{\frac{y^2-1}{y^2+z^2}} \Bigg).
\end{align}

\subsubsection{Numerical results}
\label{sec5.2.2}

\begin{figure}[h!]
\centering
  \includegraphics[width=0.8\linewidth]{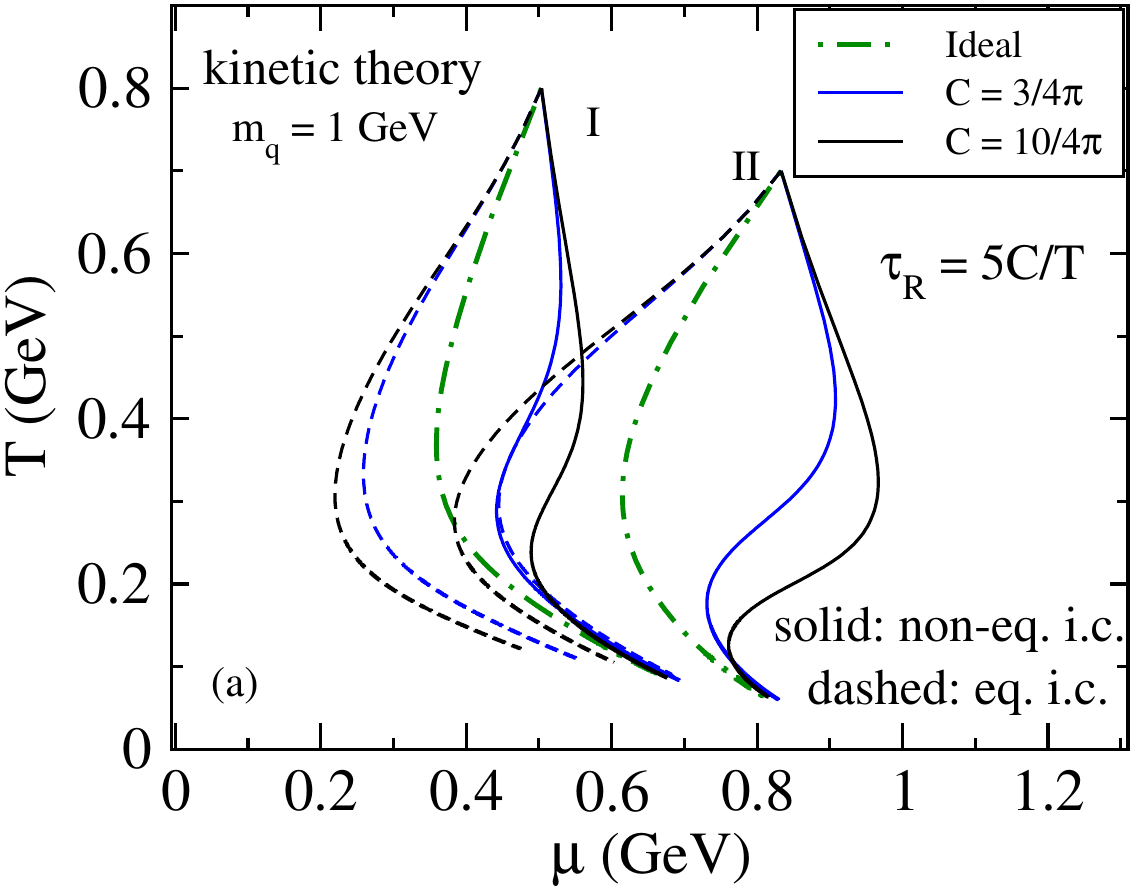}
  \includegraphics[width=0.8\linewidth]{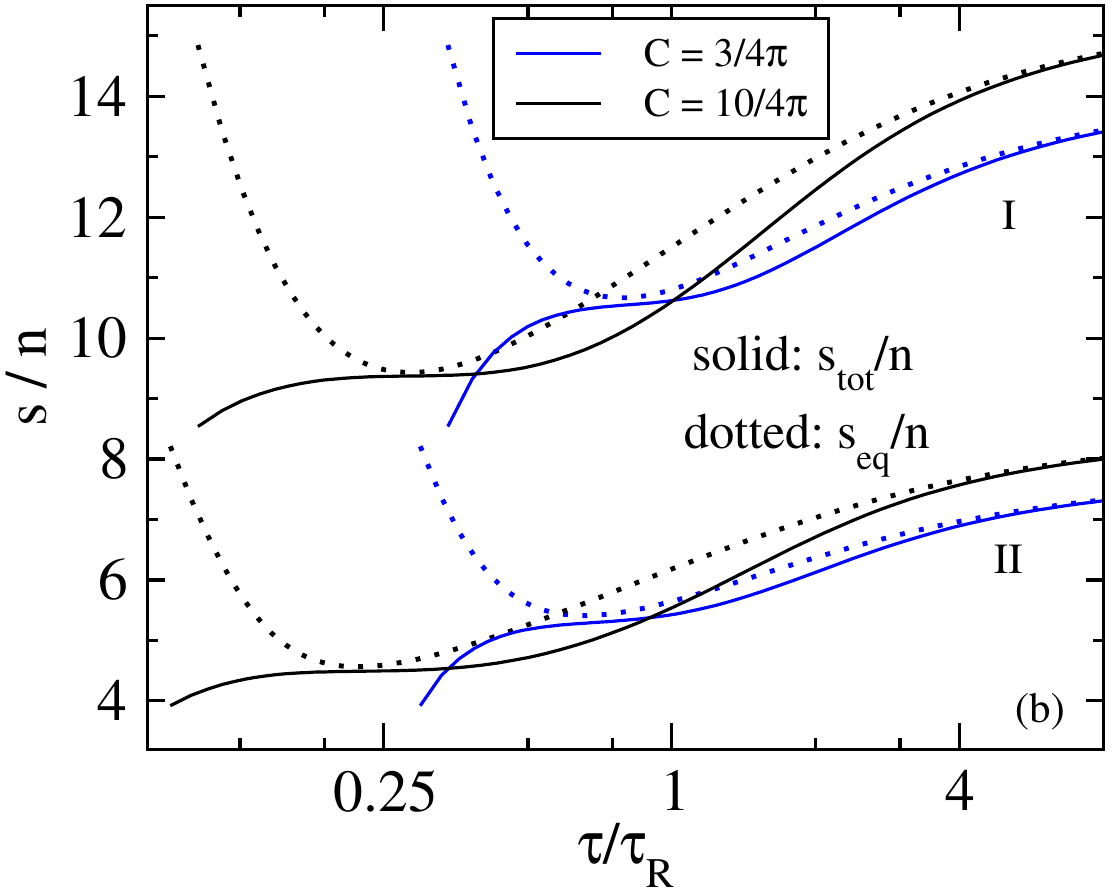}
 \vspace{-3mm}
 \caption{%
    (a) Expansion trajectories from non-conformal kinetic theory with quark mass $m=1$\,GeV. Initial conditions are the same and line colors and styles have the same meaning as in Fig.~\ref{F8}a.
    (b) Solid lines: evolution of the full (non-equilibrium + equilibrium) specific entropy $s_\mathrm{tot}/n$ obtained from kinetic theory for the identically colored solid lines in panel (a). Dotted lines show the corresponding equilibrium contributions $s_\mathrm{eq}/n$ for comparison. 
	\label{F10}
	\vspace*{-4mm}
	}
\end{figure}

\begin{figure}[thb!]
\centering
  \includegraphics[width=0.8\linewidth]{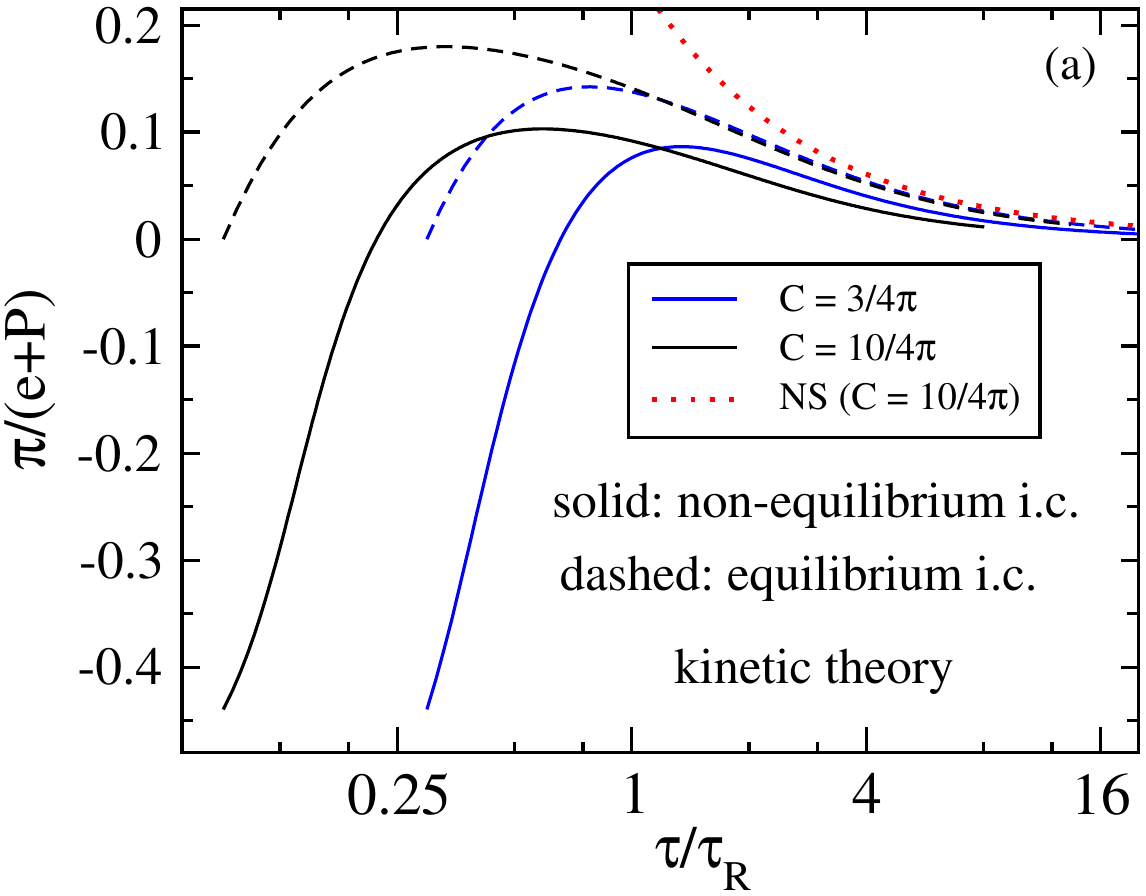}
  \includegraphics[width=0.8\linewidth]{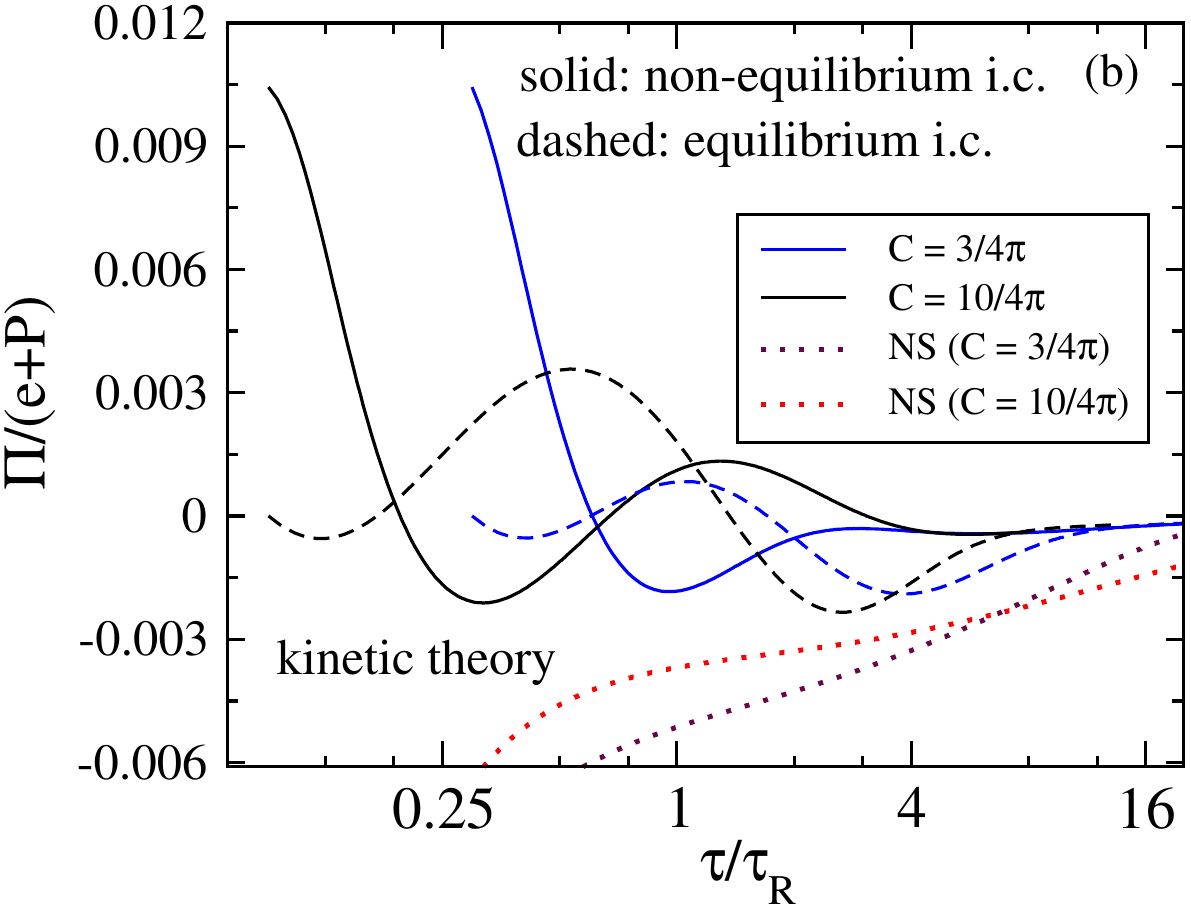}
 \vspace{-2mm}
 \caption{Same as Fig.~\ref{F9} but for the solutions obtained from kinetic theory.
	\label{F11}
	}
\end{figure}

Figure~\ref{F10}a shows the expansion trajectories obtained from non-conformal kinetic theory, for the same initial conditions as for the hydrodynamic solutions presented in the preceding subsection, and the associated scaled-time evolution of the shear and bulk viscous stresses is shown in Fig.~\ref{F11}. Line styles and colors have the same meaning as for the corresponding hydrodynamic solutions shown in Figs.~\ref{F8}, \ref{F9}.
%
%
The initial Romatschke-Strickland parameters for the non-equilibrium initial conditions used to generate the solid lines are listed Table \ref{table:IC_nc}.

\begin{table}[b!]
 \begin{center}
 \resizebox{0.7 \columnwidth}{!}{
  \begin{tabular}{|c|c|c|c|c|c|}
   \hline
   Set &  $\Lambda_0$ (GeV)  & $\nu_0$ (GeV)  &  $\xi_0$  & $\bar{\phi}_0$ & $\psi_{nc,0}$ \\
   \hline
    I & $0.342$ & $0.597$ & $- 0.981$ & $- 0.45$ & $- 0.301$ \\
   \hline
    II & $0.209$ & $1.001$ & $- 0.986$ & $- 0.45$ & $-0.269$ \\
   \hline
  \end{tabular}}
  \caption{Initial parameters $(\Lambda_0,\nu_0,\xi_0)$ used to generate the blue and black solid curves in the sets I and II of Fig.~\ref{F10}. As before $\bar\phi_0$ denotes the corresponding initial value of $\bar\phi = \bar\pi-\bar\Pi$. $\psi_{nc,0}$ denotes the maximum value of $\bar\phi_0$ that leads to trajectories for which initially $d\mu/d\tau>0$.}
  \label{table:IC_nc}
 \end{center}
 \vspace*{-6mm}
\end{table}
%
The qualitative and quantitative similarities between the kinetic theory results shown in Figs.~\ref{F10}a, \ref{F11} and the hydrodynamic results shown earlier in Figs.~\ref{F8}, \ref{F9} are obvious. Some of the quantitative differences are discussed in Appendix~\ref{appc} but they do not impact our qualitative observations. Additional insights into the mechanisms at work in Fig.~\ref{F10}a can be gleaned from Fig.~\ref{F10}b.\footnote{%
    Lacking an expression for the non-equilibrium entropy density in non-conformal second-order dissipative hydrodynamics, there was no analogous plot provided  for the hydrodynamic solutions presented in Sec.~\ref{sec5.2.1}.}
In kinetic theory Eq.~(\ref{entropy_Bjorken_c}) provides an unambiguous definition of the total entropy density, and it can be easily split into an equilibrium part and a non-equilibrium correction. For the expansion trajectories corresponding to the set of far-from-equilibrium initial conditions listed in Table~\ref{table:IC_nc} and shown as solid lines in Fig.~\ref{F10}a, panel (b) plots the scaled time evolution of the full specific entropy using solid lines, together with the corresponding equilibrium contributions $s_\mathrm{eq}/n$ using dotted lines. Note that, up to overall larger specific entropies at smaller $\mu/T$ values, the qualitative characteristics of the curves shown in Fig.~\ref{F10}b are the same for initial condition sets I and II, and we will therefore discuss them together. 

These expansion trajectories are characterised by large initial dissipative fluxes, resulting in large negative non-equilibrium corrections to the initial specific entropy (i.e. the solid lines start at much lower $s/n$ values than the dotted lines). The large negative deformation parameters $\xi_0$, close to their absolute lower limits $\xi_\mathrm{min}=-1$ (see Table~\ref{table:IC_nc}), imply that initially the longitudinal local momentum distribution is much wider than the transverse one and almost flat. Accordingly, the total $(s/n)_0$ is much smaller than its initial equilibrium contribution $(s_\mathrm{eq}/n)_0$. Note that, for both sets I and II, the starting values of $s/n$ (both equilibrium and non-equilibrium parts) are the same for the blue and black lines; they start, however, at different $\tau/\tau_R$ due to different values for $C$. At first sight, the steep initial drop of all the dotted $s_\mathrm{eq}/n$ lines seems to indicate a rapid approach towards local thermal equilibrium, on a time scale $\tau\lesssim \tau_R$, but once $s_\mathrm{eq}/n$ reaches the solid line showing the total $s/n$, it doesn't stay there but ``bounces off'', settling on the solid line for good only much later at $\tau\gg\tau_R$.

The clue to understanding this phenomenon is the observation that the initial ``fake'' equilibrium state (when $s_\mathrm{tot}/n$ and $s_\mathrm{eq}/n$ coincide for the first time) happens at $\tau < \tau_R$, i.e. before microscopic collisions have had much of an effect. It is rather a consequence of approximate free-streaming at very early times in Bjorken flow \cite{Jaiswal:2021uvv, Jaiswal:2019cju, Kurkela:2019set,  Chattopadhyay:2021ive} which red-shifts the longitudinal momenta, causing the momentum distribution to eventually cross from its initially strongly prolate ($P_L>P_T$) to an oblate ($P_L<P_T$) ellipsoidal shape.\footnote{%
     Note that a non-interacting (i.e. free-streaming) system initialised with Romatschke-Strickland distributions characterised by two energy scales $(\Lambda_0, \nu_0)$ and a negative $\xi_0$ passes through such a ``fake thermal equilibrium" state. This happens at $\tau_\mathrm{fte} = \tau_0/\sqrt{1+\xi_0}$ when the free-streaming distribution function takes a thermal form: $f^i(\tau_\mathrm{fte}; p_T, p_z) = f_\mathrm{RS}^i\bigl(\tau; p_T, (\tau_\mathrm{fte}/\tau_0)p_z\bigr) = f_\mathrm{eq}^i(\Lambda_0, \nu_0; p)$.}
Near the crossing point, when the momentum distribution is spherical, the full specific entropy $s_\mathrm{tot}/n$ approximately\footnote{%
    i.e. up to deviations caused by a non-exponential energy dependence and a non-equilibrium normalization value.}
coincides with its equilibrium part $s_\mathrm{eq}/n$ but since collisions require more time they cannot keep the system in this ``fake'' equilibrium state --- instead, the momentum distribution becomes anisotropic again by turning oblate. Only much later are collisions able to locally isotropize the momentum distribution and keep it there. At that stage the dotted lines for $s_\mathrm{eq}$ merge with the solid lines for $s_\mathrm{tot}$, and both measures for $s/n$ saturate at their common asymptotic, global thermal equilibrium value.

The time when $s_\mathrm{eq}/n$ approaches its minimum roughly marks the point where {\it viscous cooling} turns into {\it viscous heating} and the initially rightward moving expansion trajectories in Fig.~\ref{F10}a turn again towards the left of the phase diagram.

%
%
The solid lines show that the total specific entropy grows in two spurts: The first of these (the {\it viscous cooling} stage) occurs early, before the system reaches its ``fake'' equilibrium state, and is driven by longitudinal momentum-stratification via free-streaming. The second spurt is caused by local thermalization via microscopic interactions which lead to {\it viscous heating}.\footnote{%
    We note that the first spurt is larger for the more strongly coupled (less viscous) system with $C=3/4\pi$ but the second spurt and overall entropy production is higher for the less strongly coupled (more viscous) system with $C=10/4\pi$.}
From the work of Israel and Stewart \cite{Israel:1979wp} it is known that, in viscous hydrodynamics, the rate of entropy production is quadratic in the bulk and shear stresses, $\propto \pi^2/2\eta+\Pi^2/\zeta$. The evolution of $\pi$ and $\Pi$ for the initial condition set II is shown in Fig.~\ref{F11}. Comparison of the solid lines for set II in Fig.\ref{F10}b with those in Fig.~\ref{F11} confirms that the plateau for $s/n$ indeed lies close to the time when $\pi$ passes through zero, although slightly shifted by the contribution from the bulk viscous entropy production rate.

\subsection{(Absence of) early-time universality}
\label{sec5.3}

In Refs.~\cite{Jaiswal:2021uvv, Chattopadhyay:2021ive} it was shown that in non-conformal RTA kinetic theory with Bjorken flow there is an early-time, far-from-equilibrium attractor for the longitudinal pressure $P_L=P+\Pi-\pi$ (driven by the approximate free-streaming dynamics at early times) while no such attractor exists for the shear and bulk viscous stresses $\pi$ and $\Pi$ separately. That the early-time evolution of $\pi$ and $\Pi$ is not ruled by a universal attractor at early times is seen in Fig.~\ref{F11} which shows strong sensitivity to the initial conditions, with convergence to the first-order Navier-Stokes attractor only at very late times $\tau/\tau_R\gg 1$ (i.e. small Knudsen number).\footnote{%
        This brief analysis does not rule out the existence of an early-time attractor in the full phase-space of variables, $(T, \mu, \pi, \Pi)$, as often considered in the theory of non-autonomous dynamical systems. Attractors using such generalized definitions have been pursued for conformal and non-conformal fluids at vanishing chemical potential \cite{Behtash:2017wqg, Kamata:2022jrc}.
} In contrast,
%
\begin{figure}[htb!]
\centering
   \includegraphics[width=0.8\linewidth]{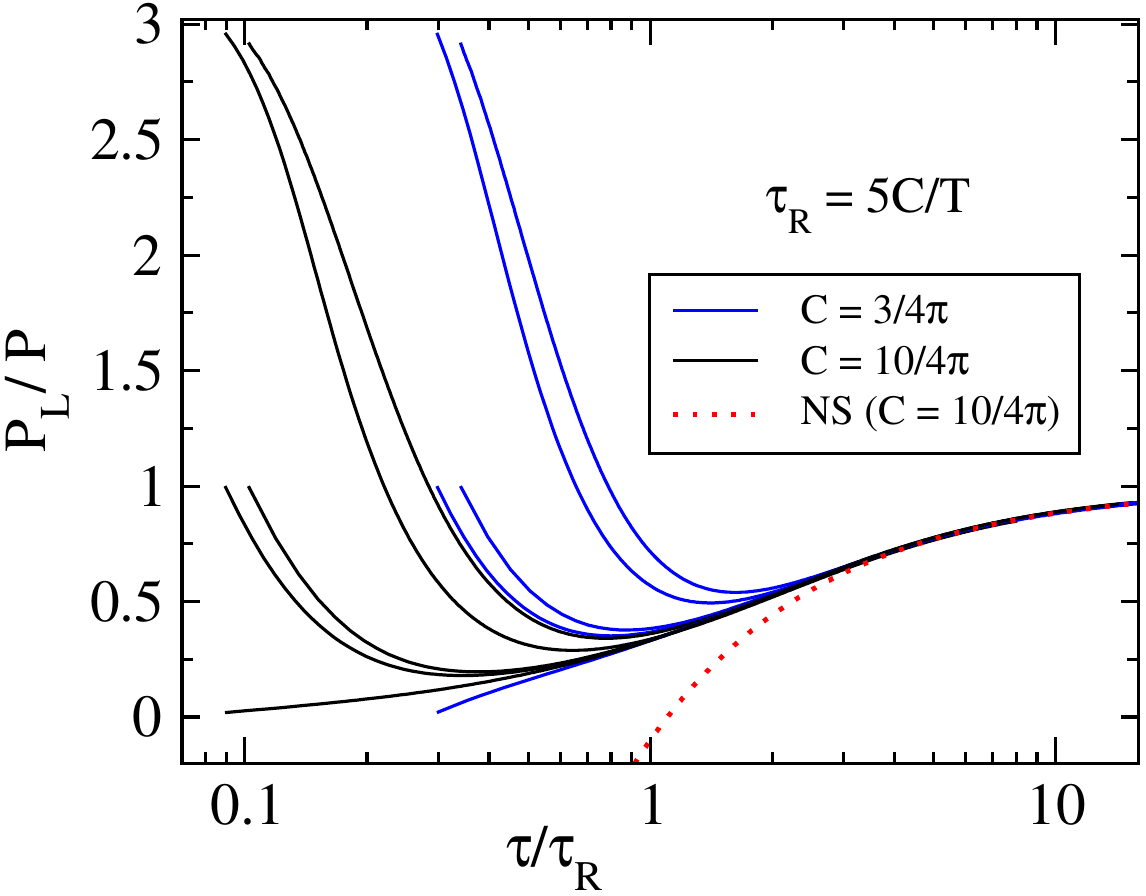}
 \vspace{-3mm}
 \caption{Evolution of the scaled effective longitudinal pressure $P_L/P$ as a function of the inverse Knudsen number $\tau/\tau_R$. The red-dotted line is the Navier-Stokes solution obtained with initial $(T_0, \mu_0)$ corresponding to set II of Fig.~\ref{F8}. 
	\label{F12}
	}
\end{figure} 
%
an early-time, far-from-equilibrium attractor for $P_L/P$, to which all initial conditions converge on very short time scales $\tau \lesssim \tau_R$, is seen in Fig.~\ref{F12}. Figure~\ref{F14} in Appendix~\ref{appc} demonstrates that this attractor is not reproduced by the hydrodynamic approximation studied in Sec.~\ref{sec5.1}. All this is consistent with earlier work \cite{Jaiswal:2021uvv, Chattopadhyay:2021ive} and extends it to systems with non-zero net baryon number.

\section{Conclusions}
\label{sec6}

In this paper we explored the evolution of phase trajectories of a system of quarks and gluons undergoing Bjorken expansion, comparing second-order Chapman-Enskog hydrodynamics and kinetic theory. Using this bare bones model we were able to reproduce the findings in Ref.~\cite{Dore:2020jye} which, under certain circumstances, suggest at first sight a loss of entropy, violating the second law of thermodynamics. This phenomenon is accompanied by a novel feature of far-off-equilibrium expansion dynamics which we call {\it viscous cooling}: the dissipative fluid cools more rapidly than an ideal one, contrary to the more common internal friction effect in dissipative fluids known as {\it viscous heating}. We found that viscous cooling arises only for far-off-equilibrium initial conditions where both the shear and bulk viscous stresses have the opposite signs of their Navier-Stokes values, to which they relax over time. Its most dramatic manifestation, where the cooling system initially moves towards {\it larger} $\mu_B$ whereas an ideal fluid would have evolved towards {\it smaller} $\mu_B$, requires non-zero quark masses, large negative shear stress and (relatively) large positive initial bulk viscous pressures.\footnote{%
    We stress that the natural scales for the (smaller) bulk and (larger) shear viscous pressures differ by about an order of magnitude in our system, see Figs.~\ref{F9} and \ref{F11}.}

As the origin of this phenomenon we identified a transfer of negative entropy from the (large) dissipative correction in the initial state to the equilibrium sector. For the massless (conformal) case we succeeded in deriving a macroscopic expression for the dissipative entropy correction, and hence we were able to check this mechanism in both the macroscopic hydrodynamic and microsopic kinetic theory approaches. For the massive (non-conformal) system our detailed quantitative analysis relied entirely on the unambiguous microscopic definition (\ref{entropy_Bjorken_c}) from kinetic theory. On a qualitative level, though, the sign (i.e. the direction) of this entropy transfer can in both cases be obtained from the macroscopic equation (\ref{seq_evol_nc}). We confirmed that the {\it total} entropy (i.e. the sum of the equilibrium contribution and the non-equilibrium correction) always increases with time, as required by the H-theorem. 

It is worth contemplating how the phenomenon of viscous (or more generally: dissipative) cooling generalizes to systems undergoing arbitrary 3-dimensional expansion, without Bjorken symmetry. A particularly interesting problem arising in this context is to identify a general criterion for far-off-equilibrium initial conditions (for both the flow velocity profile and the dissipative fluxes) that lead to dissipative cooling. For systems that ultimately move towards local thermal equilibrium we expect dissipative cooling to be a transient phenomenon that turns to dissipative heating once the dissipative fluxes approach their Navier-Stokes values and become weak. But there may exist (possibly externally driven) flow profiles where this does not happen. It will be very interesting to establish conditions for dissipative cooling that can be realized in laboratory experiments and make the phenomenon directly observable. These questions are the subject of ongoing research.\\ 

\noindent\textit{\textbf{Acknowledgements:}}
The authors thank Lipei Du, Amaresh Jaiswal, and Sunil Jaiswal for fruitful discussions. This work was supported by the U.S. Department of Energy, Office of Science, Office for Nuclear Physics under Awards No. \rm{DE-SC0004286} (U.H.) and DE-FG02-03ER41260 (T.S. and C.C.).


\appendix

\section{Second-order hydrodynamics for a massive quark-gluon gas}
\label{appa}

The energy-momentum tensor of a gas of quarks, anti-quarks, and gluons is
\begin{align}
    T^{\mu\nu} = \sum_{i=1}^3 \int dP_i \, p^\mu_i \, p^\nu_i \, f^i = e \, u^\mu u^\nu - \left(P + \Pi\right) \Delta^{\mu\nu} + \pi^{\mu\nu},
\end{align}
where $i$ differentiates between the species which have masses $m_1 = m_2 = m$ (quarks and antiquarks) and $m_3 = 0$ (gluons), and $dP_i \equiv g_i d^3p/[(2\pi)^3 E_{p,i}]$ with $E_{p,i} = \sqrt{p^2 + m_i^2}$. In the constitutive relation, $e$ is the energy density in the fluid rest frame, $P$ and $\Pi$ are the equilibrium and bulk viscous pressures, respectively, and $\pi^{\mu\nu}$ is the shear stress tensor. The projection tensors $u^\mu u^\nu$ and $\Delta^{\mu\nu} = g^{\mu\nu} - u^\mu u^\nu$ select the components of $T^{\mu\nu}$ along the temporal and spatial directions in the local fluid rest frame, defined by $u^\mu_\mathrm{LRF}=(1,0,0,0)$. The conserved net-quark charge current is given by
\begin{align}
    N^\mu = n \, u^\mu + n^\mu = N^{\mu}_1 - N^{\mu}_2 
\end{align}
where $n$ is the LRF net quark density and $n^\mu$ the quark number diffusion current, with
\begin{align}
    N^{\mu}_i = \int dP_i \, p^\mu_i \, f^i. 
\end{align}
In the Landau matching scheme the energy and net quark densities are written in terms of the local equilibrium distributions as 
\begin{align}\label{e_eq and n_eq}
    e &= u_\alpha \, u_\beta \, \sum_{i=1}^3 \int \, dP_i \, p^\alpha_i \, p^\beta_i \, f_\mathrm{eq}^i, \\
    n &= u_\alpha \, \sum_{i=1}^2 \, (-1)^{i-1}  \, \int \, dP_i \, p^\alpha_i \, f_\mathrm{eq}^i, 
\end{align}
which defines the values for the inverse temperature and chemical potential in the latter:
\begin{align}\label{f_eq_appendix}
    f_\mathrm{eq}^i = \frac{1}{\exp\left(\beta u{\,\cdot\,}p_i - \alpha_i \right) - \theta_i}.
\end{align}
Here $\beta=1/T$ is the inverse temperature and $\alpha_i=\mu_i/T$ the normalized chemical potential of species $i$ ($\alpha_1 = -\alpha_2 = \alpha$, $\alpha_3 = 0$), and $\theta_1 = \theta_2 = -1$, $\theta_3 = 1$ distinguishes between Fermi-Dirac and Bose-Einstein statistics. The dynamics of the system is governed by conservation of energy-momentum and charge current: $\partial_\mu T^{\mu\nu} = 0$ and $\partial_\mu N^\mu = 0$. Projection of the former along the fluid velocity, $u_\mu \partial_\nu T^{\mu\nu} = 0$, and the latter yield the comoving evolution of energy and number densities, whereas projection of the former orthogonal to the flow, $\Delta^\gamma_\nu \partial_\mu T^{\mu\nu} = 0$, determines the acceleration of a fluid element: 
\begin{align}
    \dot{e} &= - \left(e + P\right) \theta - \Pi \, \theta + \pi^{\mu\nu} \,\sigma_{\mu\nu}, \label{dot_e}\\
    \dot{n} &= - n \theta - \partial_\mu n^\mu. \label{dot_n} \\
     \left( e{+}P \right) \dot{u}^\gamma &= \nabla^\gamma P_\mathrm{eq} - \Pi \, \dot{u}^\gamma + \nabla^\gamma \Pi - \Delta^{\gamma}_{\nu} \partial_\mu \pi^{\mu\nu} \label{dot_u}.
\end{align}
Here we defined the comoving time derivative $\dot{A} = u^\mu \partial_\mu A$, the spatial derivative in the fluid rest-frame $\nabla^\mu = \Delta^{\mu\nu} \partial_\nu$, the fluid expansion rate $\theta = \nabla_\mu u^\mu$, 
and the velocity shear tensor $\sigma^{\mu\nu} = \Delta^{\mu\nu}_{\alpha\beta} \, \nabla^\alpha u^\beta$, where
\begin{align}
    \Delta^{\mu\nu}_{\alpha\beta} = \left( \Delta^\mu_\alpha \, \Delta^\nu_\beta + \Delta^\mu_\beta \, \Delta^\nu_\alpha \right)/2 - \frac{1}{3} \, \Delta^{\mu\nu} \, \Delta_{\alpha\beta}
\end{align}
projects any rank-2 tensor on its traceless part and locally spatial components.

In Eqs.~(\ref{dot_e}-\ref{dot_u}), the first terms on r.h.s. stem from ideal hydrodynamic constitutive relations, while the others arise due to dissipation. We shall eventually express comoving derivatives of energy density and net-quark density in terms of derivatives of inverse temperature $\beta$ and normalised chemical potential $\alpha$. For this, we use the definitions given by Eqs. (\ref{e_eq and n_eq}) to write
\begin{align}\label{invert_dot_e_n}
    \dot{e} &= - \dot{\beta} \, I_{3,0}^{(0)} - \dot{\alpha} \sum_{i=1}^2 \, (-1)^i \, I_{2,0}^{(0),i}, \nonumber \\
    \dot{n} &=  \dot{\beta} \sum_{i=1}^2 \, (-1)^i \, I_{2,0}^{(0),i} + \dot{\alpha} \sum_{i=1}^2 I_{1,0}^{(0),i},
\end{align}
where we used the notation $I_{n,q}^{(r)} = \sum_{i=1}^3 I_{n,q}^{(r), i}$, with\footnote{%
    The superscript index $r$ is somewhat redundant in the notation as the moment depends only on the difference $n{-}r$, and one could have simply defined, $J_{n,q}^i = I_{n,q}^{(0),i}$. However, we continue using it for ease of comparison of our transport coefficients with those obtained previously for a single component massive Boltzmann gas at vanishing chemical potential \cite{Jaiswal:2014isa}. 
    } 
\begin{align}
    I_{n,q}^{(r),i} \equiv \frac{1}{(2q+1)!!} \int dP_i \, \left( u\cdot p_i \right)^{n - r -2q} \, \left( \Delta_{\alpha\beta} \, p^\alpha_i \, p^\beta_i \right)^q \, {\cal F}^i_\mathrm{eq}.
\end{align}
The function ${\cal F}^i_\mathrm{eq}$ stems from taking a derivative of $f_\mathrm{eq}^i$ and is defined as ${\cal F}^i_\mathrm{eq} = f_\mathrm{eq}^i \tilde{f}_\mathrm{eq}^i$ where $\tilde{f}_\mathrm{eq}^i = 1 + a_i f_\mathrm{eq}^i$.\footnote{\label{calFeq}%
        For later use we note that the function ${\cal F}^i_\mathrm{eq}$ has the following properties: $ {\cal F}^i_\mathrm{eq} = - 1/(u\cdot p_i) \times \partial f^i_{\mathrm{eq}}/\partial \beta$, ${\cal F}^q_\mathrm{eq} = \partial f^q_\mathrm{eq}/\partial \alpha$, and ${\cal F}^{\bar{q}}_\mathrm{eq} = - \partial f^{\bar{q}}_\mathrm{eq}/\partial \alpha$.}

Using Eqs.~(\ref{invert_dot_e_n}) along with the time evolution equations for $e$ and $n$ in Eqs.~(\ref{dot_e}-\ref{dot_n}) we obtain
\begin{align}\label{dot_beta_dot_alpha}
    \dot{\beta} &= - {\cal G}_\beta \, \theta - {\cal H}_\beta \, \left( \Pi \, \theta - \pi^{\mu\nu} \, \sigma_{\mu\nu} \right), \nonumber \\
    \dot{\alpha} &= - {\cal G}_\alpha \, \theta - {\cal H}_\alpha \, \left( \Pi \, \theta - \pi^{\mu\nu} \, \sigma_{\mu\nu} \right).
\end{align}
The functions ${\cal G}$ and ${\cal H}$ are defined as
\begin{align}
    {\cal G}_\beta &= \frac{1}{{\cal D}} \Bigl[ \left( e + P_\mathrm{eq}\right) \, \sum_{i=1}^2 I_{1,0}^{(0),i}   + n \, \sum_{i=1}^2 (-1)^i I_{2,0}^{(0), i} \Bigr],
\label{G_beta} \\
    {\cal G}_\alpha  &= - \frac{1}{{\cal D}} \Bigl[  \left( e + P_\mathrm{eq}\right) \sum_{i=1}^2 (-1)^i I_{2,0}^{(0), i}   +  n \, I_{3,0}^{(0)} \Bigr],
\label{G_alpha} \\
    {\cal H}_\beta &= \frac{\sum_{i=1}^2 \, I_{1,0}^{(0), i}}{{\cal D}}, \,\,\,\,\, {\cal H}_\alpha = - \frac{\sum_{i=1}^2 (-1)^i \, I_{2,0}^{(0), i}}{{\cal D}}, 
\label{H_beta_alpha} \\
    {\cal D} &\equiv \Bigl( \sum_{i=1}^2 (-1)^i \, I_{2,0}^{(0), i} \Bigr)^2 - I_{3,0}^{(0)} \, \sum_{i=1}^2 I_{1,0}^{(0), i}. 
\label{D}
\end{align}
In the conformal case, the coefficients ${\cal G}_\beta$ and ${\cal G}_\alpha$ simplify to ${\cal G}_\beta = -\beta/3$, ${\cal G}_\alpha = 0$, consistent with \cite{Jaiswal:2015mxa}.\footnote{%
        We use the conformal relations, $e(\beta,\alpha)$ and $n(\beta,\alpha)$, given in Eqs.~(\ref{e_eq}-\ref{n_eq}) to simplify some of the terms appearing in ${\cal G}_\beta$ and ${\cal G}_\alpha$: $\sum_{i=1}^2 (-1)^i I_{2,0}^{(0),i} = - \partial e/\partial \alpha = -3 n/\beta$ and $I_{3,0}^{0} = - \partial e/\partial \beta = 4e/\beta$.}
To compute the bulk viscous pressure and shear stress tensor of the system we start from their respective definitions
\begin{align}
    \Pi &= - \frac{\Delta_{\alpha\beta}}{3} \, \sum_{i=1}^{3} \, \int dP_i \, p^\alpha_i \, p^\beta_i \, \delta f^i, \nonumber \\
    \pi^{\mu\nu} &= \Delta^{\mu\nu}_{\alpha_\beta} \, \sum_{i=1}^{3} \, \int \, dP_i \, p^{\alpha}_i \, p^{\beta}_i \, \delta f^i. 
\label{bulk_shear}
\end{align}
These off-equilibrium corrections are obtained by solving the RTA Boltzmann equation 
\begin{align}
    p^\mu_i \, \partial_\mu \, f^i = - \frac{u\cdot p_i}{\tau_R} \, \left( f^i - f^i_\mathrm{eq} \right)
\end{align}
iteratively in powers of $\delta f^i=f^i-f_\mathrm{eq}^i$.
The leading order correction is
\begin{align}
    \delta f^i_{(1)} = - \frac{\tau_R}{u \cdot p_i}  p^\mu \partial_\mu f_\mathrm{eq}^i = - \tau_R \left( \dot{f}_\mathrm{eq}^i + \frac{1}{u\cdot p_i} p^\mu \nabla_\mu f_\mathrm{eq}^i \right). \label{deltaf_1}  
\end{align}
With Eq.~(\ref{f_eq_appendix}) this can be written as
\begin{align}
    \delta f^i_{(1)} & =  \tau_R \Big[ (u\cdot p_i) \dot{\beta}  - \dot{\alpha_i} 
      + \frac{ \beta \, \theta}{3 (u\cdot p_i)} \Delta_{\mu\nu} \, p^\mu_i \, p^\nu_i \Big] {\cal F}_\mathrm{eq}^i \nonumber \\
      & + \frac{\tau_R \, \beta}{(u\cdot p_i)} p^\mu_i \, p^\nu_i \, \sigma_{\mu\nu} \, {\cal F}_\mathrm{eq}^i 
\nonumber \\
    & + \tau_R \Big[ \beta \left( \dot{u} \cdot p_i \right)  + p^\mu_i \nabla_\mu \beta    -  \frac{p^\mu_i}{u\cdot p_i} \nabla_\mu \alpha_i \, \Big] {\cal F}_\mathrm{eq}^i.
\end{align}
The terms in the first square bracket give rise to the bulk viscous pressure, the second term involving $\sigma^{\mu\nu}$ yields the shear stress tensor, and the terms in the second line result in the first-order quark diffusion current. Since it is not needed for Bjorken flow we shall neglect the quark diffusion current in the following. Also, we convert time derivatives of $\beta$ and $\alpha$ into velocity gradients using Eqs.~(\ref{dot_beta_dot_alpha}) to obtain
\begin{align}
    \delta f^i_{(1)} & =   \tau_R \, \theta \Big[ - {\cal G}_\beta \, (u \cdot p_i)  + {\cal G}^i_\alpha  
    + \frac{\beta \Delta_{\mu\nu}}{3 (u\cdot p_i)} \,  p^\mu_i \, p^\nu_i \Big] {\cal F}_\mathrm{eq}^i \nonumber \\
    & + \frac{\tau_R \, \beta}{(u\cdot p_i)} p^\mu_i \, p^\nu_i \, \sigma_{\mu\nu} \, {\cal F}_\mathrm{eq}^i
\nonumber \\
       & + \tau_R \, \left( \Pi \, \theta - \pi^{\mu\nu} \,\sigma_{\mu\nu} \right) \, \left( {\cal H}_\alpha^i - {\cal H}_\beta \, (u \cdot p_i) \right);
\label{delta fi_2}
\end{align}
here we defined ${\cal G}_\alpha^1 = - {\cal G}_\alpha^2 = {\cal G}_\alpha, {\cal G}_\alpha^3 = 0$ and ${\cal H}_\alpha^1 = - {\cal H}_\alpha^2 = {\cal H}_\alpha, {\cal H}_{\alpha}^3 = 0$, with ${\cal G}_\alpha$ and ${\cal H}_\alpha$ given by Eqs.~(\ref{G_alpha}-\ref{H_beta_alpha}). Note that the terms in the first line of the r.h.s. of Eq.~(\ref{delta fi_2}) are of first order in velocity gradients whereas those in the second line are of second order. Using Eq.~(\ref{delta fi_2}) in the definitions (\ref{bulk_shear}) of the bulk viscous pressure and shear stress tensor and retaining terms only up to first order in gradients, we obtain the Navier-Stokes expressions for $\Pi$ and $\pi^{\mu\nu}$: 
\begin{align}
\label{NS_massive}
    \Pi = - \, \tau_R \, \beta_\Pi \, \theta , \qquad
    \pi^{\mu\nu} = 2 \tau_R \, \beta_\pi \, \sigma^{\mu\nu}.
\end{align}
The coefficients $\beta_\Pi$ and $\beta_\pi$ are given by\footnote{%
        We have checked that in the conformal limit $\beta_\pi \to 4P/5$ as obtained in  \cite{Jaiswal:2015mxa} for a \textit{conformal} quark-gluon gas at finite $\mu_B$. Also, in this limit $e = 3P$ and $c_s^2 = 1/3$ such that $\beta_\Pi\to 0$ as expected.}
\begin{align} 
\label{betaPI}
    \beta_\Pi &= \frac{5}{3} \, \beta\,   I_{4,2}^{(1)} \, + \sum_{i=1}^3 \, {\cal G}_\alpha^i \, I_{2,1}^{(0),i} - {\cal G}_\beta \, I_{3,1}^{(0)} \nonumber \\
    & = \frac{5}{3} \beta_\pi - \left( e + P \right) c_s^2, \nonumber \\
    \beta_\pi &= \beta \, I_{4,2}^{(1)}.
\end{align}
We simplified the coefficient $\beta_\Pi$ using the definition of squared speed of sound,\footnote{%
    To obtain the following relation, we write $dP/de = \partial P/\partial \beta \times d\beta/de + \partial P/\partial \alpha \times d\alpha/de$, and then use the condition of fixed specific entropy, i.e, $d(s/n)_\mathrm{eq} = 0 \implies n \, de = (e+P) \, dn$, to express the derivatives $d\beta$ and $d\alpha$ in terms of $de$.}
\begin{align}
\label{cs2}
     c_s^2 \equiv \left( \frac{\partial P}{\partial e} \right)_{s/n} & = \frac{1}{(e+P) {\cal D}} \Bigg[ \frac{\partial P}{\partial \beta} \, \, \left(  (e+P) \, \frac{\partial n}{\partial \alpha} + n \, \frac{\partial n}{\partial \beta} \right) \nonumber \\
    & + \frac{\partial P}{\partial \alpha} \left( (e+P) \, \frac{\partial e}{\partial \alpha} + n \, \frac{\partial e}{\partial \beta} \right) \Bigg],
\end{align}
where ${\cal D}$ is given by Eq.~(\ref{D}).
Moreover, using the properties of ${\cal F}^i_\mathrm{eq}$ one finds the following relations:
\begin{align}\label{I_21 to I_31}
    I_{2,1}^{(0),1} = - \frac{\partial P_1}{\partial \alpha}, \qquad 
    I_{2,1}^{(0),2} = \frac{\partial P_2}{\partial \alpha}, \qquad 
    I_{3,1}^{(0)} = \frac{\partial P}{\partial \beta}. 
\end{align}
Here $P_1$ and $P_2$ are the partial pressures of quarks and anti-quarks, respectively. This yields\footnote{%
    To write ${\cal G}_\beta$ and ${\cal G}_\alpha$ in terms of the quantities appearing in Eq.~(\ref{cs2}) we used Eqs.~(\ref{dot_beta_dot_alpha}) to identify $I_{3,0}^{(0)} = - \partial e/\partial \beta$, $ \sum_{i=1}^{2} (-1)^i \, I_{2,0}^{(0), i} = -\partial e/\partial \alpha = \partial n/\partial \beta$, and $\sum_{i=1}^2 I_{1,0}^{(0),i} = \partial n/\partial \alpha$.}
\begin{align}
    \sum_{i=1}^3 {\cal G}_\alpha^i \, I_{2,1}^{(0),i} - {\cal G}_\beta \, I_{3,1}^{(0)} = - {\cal G}_\alpha \, \frac{\partial P}{\partial \alpha} - {\cal G}_\beta \, \frac{\partial P}{\partial \beta} = - (e{+}P) \, c_s^2.
\end{align}

To derive second-order relaxation-type evolution equations for the bulk and shear stresses one takes comoving time-derivative of Eqs.~(\ref{bulk_shear}) and then uses the RTA Boltzmann equation; one finds
\begin{align}
    &\dot{\Pi} = - \frac{\Pi}{\tau_R} 
\\\nonumber
     & + \frac{\Delta_{\alpha\beta}}{3} \sum_{i=1}^{3} \int dP_i \, p^\alpha_i \, p^\beta_i \bigg[ - \frac{\delta f^i_{(1)}}{\tau_R} + \frac{1}{u \cdot p_i} \, p^\gamma_i \nabla_\gamma \delta f^i \bigg],
\label{bulk_CE_1} 
\\
    &\dot{\pi}^{\langle \mu \nu \rangle} = - \frac{\pi^{\langle \mu \nu \rangle}}{\tau_R} 
\\\nonumber
    & - \Delta^{\mu\nu}_{\alpha\beta} \sum_{i=1}^{3} \int dP_i \, p^\alpha_i \, p^\beta_i \bigg[ - \frac{\delta f^i_{(1)}}{\tau_R} + \frac{1}{u \cdot p_i} \, p^\gamma_i \nabla_\gamma \delta f^i \bigg].
\label{shear_CE_1}
\end{align}
The momentum integration over $\delta f^i_{(1)}$ is performed after using Eq.~(\ref{delta fi_2}); as we are interested in computing the dissipative evolution to second order in gradients we also include the second-order terms in Eq.~(\ref{delta fi_2}):
\begin{align}
    \dot{\Pi} &= - \frac{\Pi}{\tau_R} \, - \beta_\Pi \, \theta + {\cal K}_0 \left( \Pi \, \theta - \pi^{\mu\nu} \, \sigma_{\mu\nu}  \right)
    + \frac{1}{3}  \Delta_{\alpha\beta} \, {\cal K}^{\alpha\beta}, 
\\  
    \dot{\pi}^{\langle \mu \nu \rangle} &= - \frac{\pi^{ \mu \nu}}{\tau_R} + 2 \beta_\pi \, \sigma^{\mu\nu} \, - \Delta^{\mu\nu}_{\alpha\beta} \, {\cal K}^{\alpha\beta},
\end{align}
with ${\cal K}_0 \equiv \Big( {\cal H}_\beta \, I_{3,1}^{(0)} - \sum_{i=1}^{3} \, {\cal H}_\alpha^i \, I_{2,1}^{(0),i} \Big)$.\footnote{%
    Using ${\cal H}_\beta = (1/{\cal D}) \times \partial n/\partial \alpha$, ${\cal H}_\alpha = (1/{\cal D}) \times \partial e/\partial \alpha$, along with the definitions (\ref{I_21 to I_31}), the coefficient ${\cal K}_0$ simplifies to ${\cal K}_0 = (\partial P/\partial e)_n$.} 
We introduced a rank two tensor ${\cal K}^{\alpha\beta}$ with contributions from a rank four tensor, $\rho^{\alpha\beta\gamma\delta}_2$, and from gradients of a rank three tensor, $\rho^{\alpha\beta\gamma}_1$:
\begin{align}
\label{K_alphabeta}
    {\cal K}^{\alpha\beta} \equiv \left( \nabla_\gamma \, \rho^{\alpha\beta\gamma}_1 + \, \sigma_{\gamma\delta} \, \rho^{\alpha\beta\gamma\delta}_2
     + \frac{\theta}{3} \, \Delta_{\gamma\delta} \, \rho^{\alpha\beta\gamma\delta}_2 \right).
\end{align}
These tensors are defined in analogy to $I^{\mu_1 \cdots \mu_n}_q$, by $\rho^{\mu_1 \dots \mu_n}_q \equiv \sum_{i=1}^3 \rho^{\mu_1 \cdots \mu_n, i}_q$. The $\rho$-tensors for each species stem from deviations of the corresponding distribution functions from equilibrium:
\begin{align}
    \rho^{\mu_1 \cdots \mu_n, i}_q = \int \frac{dP_i}{(u\cdot p_i)^q} \, p^{\mu_1}_i \cdots p^{\mu_n}_i \, \delta f^i. 
\end{align}
In the rest of this Appendix we compute those $\rho^{\mu_1\cdots \mu_n}_q$ that are required to obtain ${\cal K}^{\alpha\beta}$. We start by noting from Eq.~(\ref{K_alphabeta}) that ${\cal K}^{\alpha\beta}$ is at least of second order in gradients. This implies that we can simply use the first-order terms of $\delta f^i_{(1)}$ in Eq.~(\ref{delta fi_2}). However, it is customary to re-write $\delta f^i_{(1)}$ such that it contains only dissipative fluxes, $\Pi$ and $\pi^{\mu\nu}$, and not gradients of velocity (like $\theta$ and $\sigma^{\mu\nu}$). This is implemented by substituting the Navier-Stokes equations (\ref{NS_massive}) into the expression (\ref{delta fi_2}) for $\delta f^i_{(1)}$ \cite{Jaiswal:2013npa}. We denote the resulting correction as $\delta f^i_{(1),r}$ where the subscript $r$ denotes re-summation:
\begin{align}
\label{firstorderdeltaf}
    \delta f^i_{(1),r} & =  \frac{\Pi}{\beta_\Pi} \Big[ {\cal G}_\beta \, (u \cdot p_i)  - {\cal G}^i_\alpha  
      - \frac{\beta}{3 (u\cdot p_i)} \, \Delta_{\mu\nu} \, p^\mu_i \, p^\nu_i \Big] {\cal F}^i_\mathrm{eq}  \nonumber \\
      & + \frac{ \beta}{2 \, \beta_\pi \, (u\cdot p_i)} p^\mu_i \, p^\nu_i \, \pi_{\mu\nu} \, {\cal F}^i_\mathrm{eq}.
\end{align}
With this we express the $\rho$-tensors as momentum integrals over ${\cal F}^i$:
\begin{align}
    \rho^{\mu_1 \cdots \mu_n}_q &= \frac{\Pi}{\beta_\Pi} \, {\cal G}_\beta \, I^{\mu_1 \cdots \mu_n}_{q-1} -  \frac{\Pi}{\beta_\Pi} \sum_{i=1}^2 \, {\cal G}^i_\alpha \, I^{\mu_1 \cdots \mu_n,i}_q \nonumber \\
    & -  \frac{\Pi \beta}{3 \beta_\Pi} \Delta_{\gamma\delta} \, I^{\mu_1 \cdots \mu_n \, \gamma \, \delta}_{q+1}
     + \frac{\beta}{2 \beta_\pi} \, \pi_{\gamma\delta} \, I^{\mu_1 \cdots \mu_n \, \gamma \, \delta }_{q+1}.
\end{align}
To obtain the evolution of the bulk viscous pressure we evaluate the projection of the tensor ${\cal K}^{\alpha\beta}$ along $\Delta_{\alpha\beta}$. For the sake of clarity we show the contributions from the term involving the gradient of $\rho^{\alpha\beta\gamma}_1$ and those involving $\rho^{\alpha\beta\gamma\delta}_2$ separately:
\begin{align}
    & \frac{1}{3} \, \Delta_{\alpha \beta} \, \nabla_\gamma \, \rho^{\alpha\beta \gamma}_1 = {\cal K}_1 \, \Pi \, \theta + \frac{2}{3}  \, \pi^{\mu\nu} \, \sigma_{\mu\nu}, 
\nonumber \\
    &  \frac{\Delta_{\alpha\beta}}{3} \, \left( \sigma_{\gamma\delta} \, \rho^{\alpha\beta\gamma\delta}_2 + \frac{\theta}{3} \, \Delta_{\gamma\delta} \, \rho^{\alpha\beta\gamma\delta}_2  \right) = {\cal K}_2 \, \Pi \, \theta +                       {\cal K}_3 \, \pi^{\mu\nu} \, \sigma_{\mu\nu}.
\end{align}
The coefficients occurring in these expressions are
\begin{align}
    {\cal K}_1 &= \frac{5}{3\beta_\Pi} \left[ {\cal G}_\beta \, I_{3,1}^{(0)} - \sum_{i=1}^3 \, {\cal G}^i_\alpha \, I_{3,1}^{(1),i} - \frac{5}{3} \, \beta_\pi  \right] = - \frac{5}{3}, 
\nonumber \\
    {\cal K}_2 &= \frac{5}{3\beta_\Pi} \, \left[ {\cal G}_\beta \, I_{4,2}^{(1)} - \sum_{i=1}^3 \, {\cal G}_\alpha^i \, I_{4,2}^{(2),i} - \frac{7 \beta}{3} \, I_{6,3}^{(3)} \right], 
\nonumber \\
    {\cal K}_3 &= \frac{7 \beta}{3 \beta_\pi} \, I_{6,3}^{(3)}.
\end{align}
Similarly, for the shear stress tensor evolution we compute the projection of ${\cal K}^{\alpha\beta}$ along $\Delta^{\mu\nu}_{\alpha\beta}$. This gives rise to the contributions 
\begin{align}
      & \Delta^{\mu\nu}_{\alpha\beta} \, \nabla_\gamma \, \rho^{\alpha\beta\gamma}_1 = {\cal K}_4 \, \Pi \, \sigma^{\mu\nu} - 2 \, \pi^{\langle \mu}_\gamma \, \omega^{\nu\rangle \gamma} + 2 \, \pi^{\langle \mu}_\gamma \, \sigma^{\nu\rangle \gamma} \nonumber \\
     & + \frac{5}{3} \, \pi^{\mu\nu} \,\theta, 
\nonumber \\
      & \Delta^{\mu\nu}_{\alpha\beta} \, \left( \sigma_{\gamma\delta} \, \rho^{\alpha\beta\gamma\delta}_2 + \frac{\theta}{3} \, \Delta_{\gamma\delta} \, \rho^{\alpha\beta\gamma\delta}_2  \right) = {\cal K}_5 \, \Pi \, \sigma^{\mu\nu} \nonumber \\
     & + \frac{4\beta}{\beta_\pi} \, I_{6,3}^{(3)} \, \pi^{\langle \mu}_\gamma \, \sigma^{\nu\rangle \gamma} + {\cal K}_6 \, \pi^{\mu\nu} \, \theta,  
 \end{align}
with
\begin{align}
    {\cal K}_4 &= \frac{2}{\beta_\Pi} \, \left[ {\cal G}_\beta \, I_{3,1}^{(0)} - \sum_{i=1}^3 {\cal G}_\alpha^i \, I_{3,1}^{(1),i} - \frac{5}{3} \, \beta_\pi  \right]  = -2,
\nonumber \\
    {\cal K}_5 &= \frac{2}{\beta_\Pi} \, \left[ {\cal G}_\beta \, I_{4,2}^{(1)} - \sum_{i=1}^3 {\cal G}_\alpha^i \, I_{4,2}^{(2),i} - \frac{7 \, \beta}{3} \, I_{6,3}^{(3)} \right] = \frac{6}{5} \, {\cal K}_2,
\nonumber \\
    {\cal K}_6 &= \frac{7 \beta}{3\beta_\pi} \, I_{6,3}^{(3)} = {\cal K}_3.
\end{align}
Putting everything together we finally arrive at the following second-order non-conformal evolution equations for the bulk and shear viscous stresses for a quark-gluon gas:
\begin{align}
    \dot{\Pi} &= - \frac{\Pi}{\tau_R} - \beta_\Pi \, \theta - \delta_{\Pi\Pi} \, \Pi \, \theta + \lambda_{\Pi\pi} \, \pi^{\mu\nu} \, \sigma_{\mu\nu}, 
\label{bulk_eq_appendix}\\
    \dot{\pi}^{\langle \mu \nu \rangle} &= - \frac{\pi^{\mu\nu}}{\tau_R} + 2 \, \beta_\pi \, \sigma^{\mu\nu} + 2 \, \pi^{\langle \mu}_\gamma \, \omega^{\nu\rangle \gamma} - \tau_{\pi\pi} \, \pi^{\langle \mu}_\gamma \, \sigma^{\nu\rangle \gamma} \nonumber \\
    & - \delta_{\pi\pi} \, \pi^{\mu\nu} \, \theta + \lambda_{\pi\Pi} \, \Pi \, \sigma^{\mu\nu}. \label{shear_eq_appendix}
\end{align}
The transport coefficients are given by
\begin{align}
    \delta_{\Pi\Pi} &= - \left( {\cal K}_0 + {\cal K}_1 + {\cal K}_2 \right) = - {\cal K}_0 - \frac{5}{9} \chi,
\\
    \lambda_{\Pi\pi} &= \frac{2}{3}  - {\cal K}_0 + {\cal K}_3 = \frac{2}{3}  - {\cal K}_0 + \frac{7 \, \beta }{3 \, \beta_\pi} \, I_{6,3}^{(3)}, 
\\
    \tau_{\pi\pi} &= 2 + \frac{4\,\beta }{\beta_\pi} \, I_{6,3}^{(3)}, 
\\
    \delta_{\pi\pi} &= \frac{5}{3} + \frac{7\beta}{3\beta_\pi} \, I_{6,3}^{(3)}, 
\\
    \lambda_{\pi\Pi} &= - \left( {\cal K}_4 + {\cal K}_5 \right) =   - \frac{2}{3} \, \chi, \quad \text{with}
\\
    \chi  & \equiv \frac{3}{\beta_\Pi} \, \Big[  {\cal G}_\beta \left( I_{3,1}^{(0)} + I_{4,2}^{(1)} \right) - \sum_{i=1}^3 {\cal G}_\alpha^i \, \left( I_{3,1}^{(1),i} + I_{4,2}^{(2),i} \right) 
\nonumber \\
    & \qquad\quad - \frac{\beta}{3} \left( 5 \, I_{5,2}^{(2)} + 7 I_{6,3}^{(3)} \right)  \Bigr] 
\label{chi_coeff}.
\end{align}
At vanishing chemical potential and ignoring quantum statistical effects, the expressions for $\tau_{\pi\pi}$ and $\delta_{\pi\pi}$ are identical to those found in \cite{Jaiswal:2014isa} for a single-species massive Boltzmann gas. Moreover, in this limit the coefficient ${\cal K}_0 = (\partial P/\partial e)_n$ becomes equal to the squared speed of sound $c_s^2$, and the coefficient $\chi$ reduces to\footnote{%
    For this, we substitute ${\cal G}_\beta = - (e+P)/I_{3,0}^{(0)} = - \beta c_s^2$ and $ {\cal G}_\alpha = 0$ in Eq.~(\ref{chi_coeff}).}
\begin{align}
    \chi = \frac{\beta}{\beta_\Pi} \left[ - 3 \, c_s^2 \left( I_{3,1}^{(0)} + I_{4,2}^{(1)} \right) - 5 \, I_{5,2}^{(2)} - 7 \, I_{6,3}^{(3)} \right].
\end{align}
Using Eqs.~(8,9) in Ref.~\cite{Jaiswal:2014isa}, this expression for $\chi$ is found to match with their Eq.~(40). Accordingly, the transport coefficients obtained in this section fully reduce to those obtained in \cite{Jaiswal:2014isa} for a single-component Boltzmann gas at $\mu = 0$ in the appropriate limit.

Also, in the conformal limit ($m = 0$), the coefficient $I_{6,3}^{(3)} \to - \beta_\pi/(7\beta)$. Accordingly, $\tau_{\pi\pi} \to 10/7$, $\delta_{\pi\pi} \to 4/3$, such that Eq.~(\ref{shear_eq_appendix}) reduces to the shear stress evolution equation derived in \cite{Jaiswal:2015mxa} for a conformal quark-gluon gas at finite quark density. 

\vspace*{-2mm}
\section{Second-order entropy current from conformal kinetic theory}
\label{appb}
\vspace*{-3mm}

We start with the general definitions (\ref{entropy},\ref{phi_i_f}) for the entropy current in kinetic theory, assuming small deviations of the distribution functions from their equilibrium forms: $f^i = f^{i}_\mathrm{eq} + \delta f^i$. One can then write
\begin{equation}
    S^\mu = s_\mathrm{eq} u^\mu + \delta S^\mu
\end{equation}
where $s_\mathrm{eq}$ is the equilibrium entropy density of the system and the deviation $\delta S^\mu$ can be Taylor expanded about equilibrium as\footnote{%
        Different from App.~\ref{appa} we here shall not include include any degeneracy factors in the integration measure but simply write $dP \equiv d^3p/[(2\pi^3) \, p]$. Also, as all particles are massless, we use a common four-momentum $p^\mu$.
}
\begin{align}
\nonumber
\label{delta entropy current}
    \delta S^\mu = {-} \sum_{i = 1}^{3} g_i \!\!\int\!\! dP \, p^\mu \Big( \phi'_{i}[f^i_\mathrm{eq}] \, \delta f^i   
    %
     + \frac{\phi''_{i}[f^i_\mathrm{eq}]}{2}  (\delta f^i)^2 + \cdots \Big).
\end{align}
Here $\phi'_i[f^i_\mathrm{eq}]$ and $\phi''_{i}[f^i_\mathrm{eq}]$ denote the first and second derivative of $\phi_i[f^i]$ w.r.t. $f^i$, evaluated at $f^i = f^i_\mathrm{eq}$. 

The leading-order term in $\delta S^\mu$ (i.e. the one linear in $\delta f$) can be simplified using the Landau matching conditions. Taking the first derivative of Eq.~(\ref{phi_i_f}), 
\begin{align}
 \phi'_i[f^i_\mathrm{eq}] = \ln\left( \frac{f^i_\mathrm{eq}}{1 + \theta_i f^i_\mathrm{eq}} \right),
\end{align}
and inserting the equilibrium distributions for quarks, anti-quarks, and gluons yields $\phi'_i[f^i_\mathrm{eq}] = - \beta \left(u\cdot p\right) + \alpha_i$, where $\alpha_q=-\alpha_{\bar{q}}=\alpha$, $\alpha_g=0$.
%
With this the first term in $\delta S^\mu$
evaluates to
\begin{align}
    \delta S^\mu_{1} &= \beta u_\nu \int dP \, p^\mu \, p^\nu \, \left[ g_q \left( \delta f^q + \delta f^{\bar{q}} \right) + g_g \, \delta f^g  \right] 
\nonumber \\
    & \quad - \alpha \int dP \, p^\mu \, g_q \left( \delta f^q - \delta f^{\bar{q}}  \right) 
\\
    & = \beta \, u_\nu \delta T^{\mu \nu} - \alpha \, n^\mu = - \alpha \, n^\mu;
\end{align}
here $n^\mu$ is the net-quark diffusion current, and we used the matching condition $u_\nu \delta T^{\mu\nu} = 0$. This result could have been anticipated as to linear-order in dissipation the only available four-vector is $n^\mu$ \cite{Muronga:2003ta}. Moreover, this shows that the non-equilibrium correction to the entropy density $u_\mu S^\mu$ is necessarily of second order in the dissipative fluxes as $u_\mu n^\mu = 0$. This is also expected, as for a system with given values of energy and conserved charge densities the thermal equilibrium state represents a maximum of the entropy density where its first derivatives with respect to the dissipative flows vanish.\\[-2ex]

Let us now compute the corrections of ${\cal O}\bigl((\delta f)^2\bigr)$ in the entropy four-current:
\begin{align}
\nonumber
     S^\mu = s_\mathrm{eq} \, u^\mu -\alpha \, n^\mu - \sum_{i = 1}^{3} g_i \int dP \, p^\mu \, \frac{\phi''_{i}[f^i_\mathrm{eq}]}{2} \, (\delta f^i)^2.
\end{align}
With $\tilde{f}^i_\mathrm{eq} = 1 + \theta_i f^i_\mathrm{eq}$, $\phi''_i[f^i_\mathrm{eq}] = 1/(f^i_\mathrm{eq} \tilde{f}^i_\mathrm{eq})\equiv 1/{\cal F}_\mathrm{eq}^i$, and the ${\cal}(\delta f)^2$ correction to the entropy current is
\begin{align}
\label{deltaS2_appendix}
    \delta S^\mu_{2} = - \sum_{i}^{3} \frac{g_i}{2} \int dP \, p^\mu \frac{(\delta f^i)^2}{{\cal F}_\mathrm{eq}^i}.
\end{align}
From Ref.~\cite{Jaiswal:2015mxa} we know that to first order in the Chapman-Enskog expansion
\begin{align}
\label{deltaf_species_CE}
    \frac{\delta f^q}{{\cal F}_\mathrm{eq}^q} &= {\cal A}_p^{\mu\nu} \pi_{\mu\nu} + {\cal B}_p \, p^\mu \, n_\mu, \\
    \frac{\delta f^{\bar{q}}}{{\cal F}_\mathrm{eq}^{\bar{q}}} &= {\cal A}_p^{\mu\nu} \pi_{\mu\nu} + {\cal \bar{B}}_p \, p^\mu \, n_\mu, \\
    \frac{\delta f^g}{{\cal F}_\mathrm{eq}^q} &= {\cal A}_p^{\mu\nu} \pi_{\mu\nu}, 
\end{align}
with
\begin{align} 
\label{deltaf_A_B}
    {\cal A}_p^{\mu\nu} &\equiv \frac{\beta}{2 (u\cdot p) \beta_\pi} p^\mu p^\nu, \quad
\nonumber\\
    {\cal B}_p &\equiv \frac{1}{\beta_n} \left( \frac{n}{e{+}P} - \frac{1}{u\cdot p}\right),
\nonumber \\
    {\cal \bar{B}}_p &\equiv \frac{1}{\beta_n} \left( \frac{n}{e{+}P} + \frac{1}{u\cdot p}\right);
\end{align}
here $\beta_\pi \equiv 4P/5$ and $\beta_n \equiv \frac{1}{3} (\partial n/\partial \alpha)_\beta - n^2 T/(4P) $. Using these definitions one obtains
\begin{widetext}
\begin{align}
    \sum_{i=1}^{3} g_i \frac{(\delta f^i)^2}{{\cal F}_\mathrm{eq}^i} &= \frac{\beta^2}{4\beta_\pi^2}\pi_{\alpha\beta} \, \pi_{\gamma \delta} \, p^\alpha p^\beta p^\gamma p^\delta \, \frac{1}{(u\cdot p)^2} \, \left[ g_q \left( {\cal F}_{\mathrm{eq}}^q{+}{\cal F}_\mathrm{eq}^{\bar{q}} \right) + g_g \, {\cal F}_\mathrm{eq}^g \right] 
\nonumber \\
    & + \frac{1}{\beta_n^2} n_\alpha \, n_\beta \, p^\alpha \, p^\beta \,  \, g_q \left[ \left( \frac{n^2}{\left( e{+}P \right)^2} + \frac{1}{(u \cdot p)^2} \right) \left( {\cal F}_\mathrm{eq}^q{+}{\cal F}_\mathrm{eq}^{\bar{q}} \right) - \frac{2n}{(e{+}P)} \frac{1}{u \cdot p}  \left( {\cal F}_\mathrm{eq}^q{-}{\cal F}_\mathrm{eq}^{\bar{q}} \right) \right] 
\nonumber \\
    & + \frac{\beta}{\beta_\pi \beta_n}  \pi_{\alpha\beta} \, n_{\gamma} \, p^\alpha p^\beta p^\gamma \frac{1}{u\cdot p} \, g_q
    \left[ \frac{n}{e{+}P} \left( {\cal F}_\mathrm{eq}^q{+}{\cal F}_\mathrm{eq}^{\bar{q}} \right) - \frac{1}{u\cdot p} \left( {\cal F}_\mathrm{eq}^q{-}{\cal F}_\mathrm{eq}^{\bar{q}} \right) \right].
\label{deltaf_squared_appendix} 
\end{align}
\end{widetext}
To obtain $\delta S^\mu_2$ we will substitute the above expression in Eq. (\ref{deltaS2_appendix}). The term quadratic in the shear stress yields
%
\begin{align}
    \delta S^{\mu}_{2, \pi\pi} &=   - \frac{\beta^2}{8 \, \beta_\pi^2} \,  \pi_{\alpha\beta} \, \pi_{\gamma\delta} \, \int \frac{dP}{(u\cdot p)^2} \, p^\mu p^\alpha p^\beta p^\gamma p^\delta \, 
\nonumber \\
    &\qquad\qquad\qquad\qquad
    \times\left[ g_q \left( {\cal F}_\mathrm{eq}^q{+}{\cal F}_\mathrm{eq}^{\bar{q}} \right) + g_g \, {\cal F}_\mathrm{eq}^g \right] 
\nonumber\\
    & = - \frac{\beta^2}{4 \, \beta_\pi^2} u^\mu \, \pi^{\alpha\beta} \, \pi_{\alpha\beta} \, \left[ g_q \left( J_{3,2}^q{+}J_{3,2}^{\bar{q}} \right) + g_g J_{3,2}^g  \right] 
\label{pipi},
\end{align}
where we introduced the thermodynamic integrals
\begin{align}
    J_{n,l}^i \equiv \frac{1}{\left(2l + 1\right)!!} \, \int dP \, \left( u \cdot p \right)^{n} \, {\cal F}_\mathrm{eq}^i.
\end{align}
Noting that
\begin{align}
\label{F_deriv_beta}
    {\cal F}_\mathrm{eq}^i = f_\mathrm{eq}^i \, \tilde{f}_\mathrm{eq}^i = - \frac{1}{u\cdot p} \, \frac{\partial f_\mathrm{eq}^i}{\partial \beta}
\end{align}
(with the derivative to be taken at fixed $\alpha$) we simplify
\begin{align}
    g_q \left( J^q_{3,2} + J^{\bar{q}}_{3,2} \right) + g_g J^g_{3,2} &=  - \frac{1}{15} \sum_i g_i \int dP \, \left( u\cdot p \right)^2 \frac{\partial f_\mathrm{eq}^i}{\partial \beta} \nonumber \\
    & = - \frac{1}{15} \frac{\partial e_\mathrm{eq}}{\partial \beta} = \frac{\beta_\pi}{\beta} 
\end{align}
such that
\begin{align}
    & \delta S^\mu_{2,\pi\pi} = - \frac{\beta}{4\beta_\pi} \, u^\mu \, \pi^{\alpha\beta} \, \pi_{\alpha\beta}. 
\label{deltaS2_pipi_appendix}
\end{align}
The term in Eq.~(\ref{deltaf_squared_appendix}) quadratic in the quark diffusion current is manipulated similarly:
\begin{align}
    &\delta S^\mu_{2,nn} = - \frac{g_q}{2\beta_n^2} \, n_\alpha n_\beta \int dP \, p^\mu p^\alpha p^\beta \bigg[ - \frac{2n\,({\cal F}_\mathrm{eq}^q{-}{\cal F}_\mathrm{eq}^{\bar{q}})}{(e{+}P) (u \cdot p)} \,  \nonumber \\
    & \qquad\qquad\qquad
     + \bigg( \frac{n^2}{\left( e{+}P \right)^2} + \frac{1}{(u \cdot p)^2} \bigg) \left( {\cal F}_\mathrm{eq}^q{+}{\cal F}_\mathrm{eq}^{\bar{q}} \right)
     \bigg]
\nonumber \\
    & =  \frac{g_q}{2\beta_n^2}\, u^\mu \, n^\alpha n_\alpha \bigg[  \frac{n^2}{(e{+}P)^2} \, \left( J_{3,1}^q{+}J_{3,1}^{\bar{q}} \right) + \left( J_{1,1}^q{+}J_{1,1}^{\bar{q}} \right) \nonumber \\
    & \qquad\qquad\qquad\qquad - \frac{2n}{e{+}P} \left( J_{2,1}^q{-}J_{2,1}^{\bar{q}} \right) \bigg].
\end{align}
With Eq.~(\ref{F_deriv_beta}) this simplifies further to
\begin{align}
    g_q \left( J_{3,1}^q + J_{3,1}^{\bar{q}} \right)
    & = - \frac{1}{3} \frac{\partial}{\partial \beta} \int dP \,  \left( u\cdot p \right)^2 g_q \left( f^q_\mathrm{eq} + f^{\bar{q}}_\mathrm{eq} \right) \nonumber \\
    & = - \frac{1}{3} \frac{\partial \left(e^q_\mathrm{eq} + e^{\bar{q}}_\mathrm{eq}\right)}{\partial \beta}.
\end{align}
To compute $g_q (J_{1,1}^q + J_{1,1}^{\bar{q}})$ we use ${\cal F}_\mathrm{eq}^q = (\partial f_\mathrm{eq}^q/\partial \alpha)$, ${\cal F}_\mathrm{eq}^{\bar{q}} = - (\partial f_\mathrm{eq}^{\bar{q}}/\partial \alpha)$
and obtain
\begin{align}
    g_q \left(J_{1,1}^q {+} J_{1,1}^{\bar{q}}\right) & = \frac{1}{3} \, \frac{\partial }{\partial \alpha} \int dP \, (u{\,\cdot\,}p) \, g_q \left( f^q_\mathrm{eq} {-} f^{\bar{q}}_\mathrm{eq} \right) 
    = \frac{1}{3} \frac{\partial n}{\partial \alpha}.
\nonumber
\end{align}
Finally we calculate
\begin{align}
    g_q \left(J_{2,1}^q - J_{2,1}^{\bar{q}}\right) &= \frac{1}{3} \, \frac{\partial}{\partial \alpha} \int dP \, \left( u\cdot p \right)^2 \, g_q \left( f_\mathrm{eq}^q + f_\mathrm{eq}^{\bar{q}} \right) \nonumber \\
    & = \frac{1}{3} \frac{\partial \left(e_\mathrm{eq}^q + e_\mathrm{eq}^{\bar{q}} \right)}{\partial \alpha}.
\end{align}
All this allows to express $\delta S^\mu_{2,nn}$ in compact form:
\begin{align}
\label{deltaS2_nn_appendix}
    \delta S^{\mu}_{2,nn} &= \frac{1}{6\beta_n^2} \, u^\mu \, n^\alpha \, n_\alpha \bigg[- \frac{n^2}{(e+P)^2} \frac{\partial \left(e^q_\mathrm{eq} + e^{\bar{q}}_\mathrm{eq}\right)}{\partial \beta} + \frac{\partial n}{\partial \alpha} \nonumber \\
    & - \frac{2n}{(e+P)} \frac{\partial \left(e_\mathrm{eq}^q + e_\mathrm{eq}^{\bar{q}} \right)}{\partial \alpha}   \bigg].
\end{align}
Finally, we evaluate the cross term between quark 
diffusion and shear stress in Eq.~(\ref{deltaf_squared_appendix}):
%
%
\begin{figure*}
\centering
\hspace*{-3mm}\includegraphics[width=0.8\linewidth]{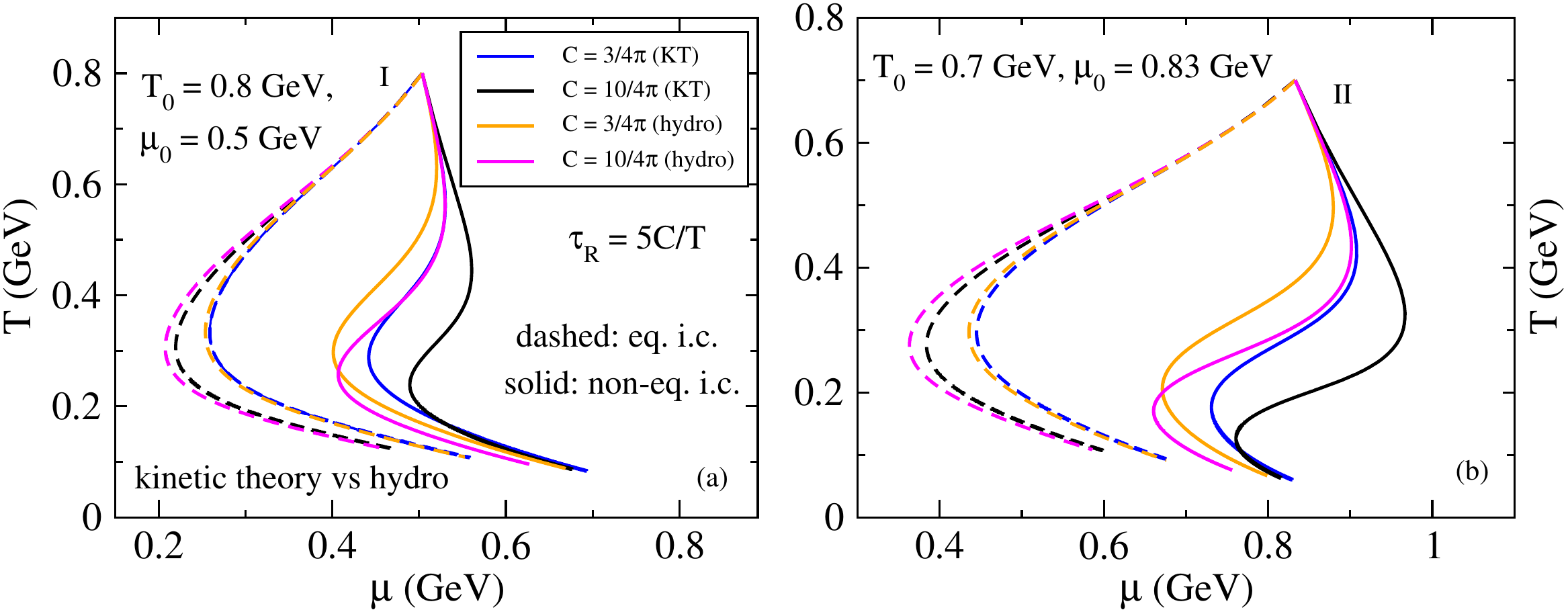}
\includegraphics[width=0.8\linewidth]{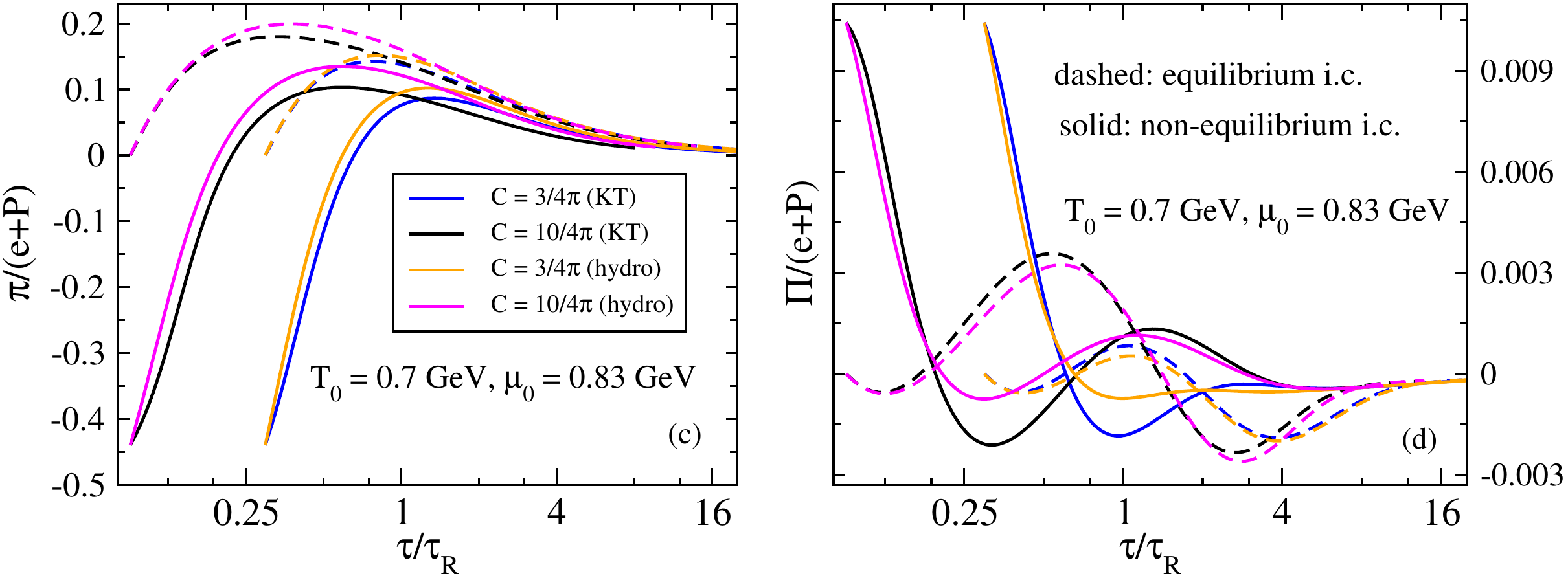}
\vspace{-2mm}
\caption{%
    Upper panels compare phase trajectories from kinetic theory (blue and black) and second-order hydrodynamics (orange and magenta). Panels (a) and (b) correspond, respectively, to initial conditionsets I and II. The lower panels compare the evolution of the shear (panel c) and bulk (panel d) inverse Reynolds numbers for set II of initial conditions.
	\label{F13}
	}
\end{figure*} 
%
%
\begin{align}
    &\delta S^\mu_{2,n\pi} = - \frac{1}{2} \frac{\beta}{\beta_\pi \beta_n} \, \pi_{\alpha\beta} \, n_\gamma \int \frac{dP}{u\cdot p} \, p^\mu p^\alpha p^\beta p^\gamma 
\nonumber\\
    & \times \Bigl[ \frac{n}{e{+}P} \left( {\cal F}_\mathrm{eq}^q{+}{\cal F}_\mathrm{eq}^{\bar{q}} \right) - \frac{1}{u\cdot p} \left( {\cal F}^q{-}{\cal F}_\mathrm{eq}^{\bar{q}} \right) \Bigr] 
\nonumber \\
    & = - \frac{\beta}{\beta_\pi \beta_n} \pi^{\mu\alpha} \, n_\alpha \, g_q \Bigl[ \frac{n}{e{+}P} \left( J_{3,2}^q{+}J_{3,2}^{\bar{q}} \right) - \left( J_{2,2}^q{-}J_{2,2}^{\bar{q}} \right) \Bigr].
\nonumber
\end{align}
Using
\begin{align}
    g_q \left( J_{3,2}^q + J_{3,2}^{\bar{q}} \right) &= - \frac{1}{15} \, \frac{\partial \left(e_\mathrm{eq}^q + e_\mathrm{eq}^{\bar{q}} \right)}{\partial \beta},
\nonumber \\
    g_q \left( J_{2,2}^q - J_{2,2}^{\bar{q}} \right) &= \frac{1}{15} \frac{\partial}{\partial \alpha} \int \, dP \, \left( u \cdot p \right)^2 \, g_q \left( f_\mathrm{eq}^q + f_\mathrm{eq}^{\bar{q}} \right) \nonumber \\
    & = \frac{1}{15} \, \frac{\partial \left(e_\mathrm{eq}^q + e_\mathrm{eq}^{\bar{q}} \right)}{\partial \alpha}
\end{align}
we write
\begin{align}
\label{deltaS2_npi_appendix}
    \delta S^{\mu}_{2,n\pi} &=  \frac{\beta}{15 \, \beta_\pi \beta_n} \, \pi^{\mu\alpha} \, n_\alpha \, \bigg[ \frac{n}{e+P} \frac{\partial \left(e_\mathrm{eq}^q + e_\mathrm{eq}^{\bar{q}} \right)}{\partial \beta} 
\nonumber \\
    &\qquad\qquad\qquad\qquad\qquad
    +  \frac{\partial \left(e_\mathrm{eq}^q + e_\mathrm{eq}^{\bar{q}} \right)}{\partial \alpha} \bigg].
\end{align}
Collecting all contributions we finally obtain for the second-order entropy current of a conformal gas of quarks and gluons at finite chemical potential
\begin{align}\label{entropy_second_order}
   S^\mu &= s_\mathrm{eq} \, u^\mu - \alpha \, n^\mu - \frac{\beta}{4\beta_\pi} \, u^\mu \, \pi^{\alpha\beta} \, \pi_{\alpha\beta} \nonumber \\
   & + c_{nn} \, u^\mu \, n^\alpha \, n_\alpha + c_{n\pi} \, \pi^{\mu\alpha} \, n_\alpha,
\end{align}
with coefficients $c_{nn}$ and $c_{n\pi}$ given by
\begin{align}
    c_{nn} &= \frac{1}{6\beta_n^2} \bigg[ - \left(\frac{n}{e+P}\right)^2 \frac{\partial \left(e^q_\mathrm{eq} + e^{\bar{q}}_\mathrm{eq}\right)}{\partial \beta} + \frac{\partial n}{\partial \alpha} \nonumber \\
    & \qquad\qquad - \frac{2n}{e+P} \, \frac{\partial \left(e_\mathrm{eq}^q + e_\mathrm{eq}^{\bar{q}} \right)}{\partial \alpha} \biggr], 
\\\nonumber
   c_{n\pi} &=  \frac{\beta}{15 \beta_\pi \beta_n} \biggl[  \frac{n}{e+P} \, \frac{\partial \left(e_\mathrm{eq}^q + e_\mathrm{eq}^{\bar{q}} \right)}{\partial \beta} + \frac{\partial \left(e_\mathrm{eq}^q + e_\mathrm{eq}^{\bar{q}} \right)}{\partial \alpha} \biggr].
\end{align}
The last term in Eq.~(\ref{entropy_second_order}) implies that the entropy flux in the fluid rest frame is not along the direction of net-quark diffusion unless the latter  points along an eigen-direction of the shear stress tensor.\footnote{%
   Note that at vanishing $\mu_B$ the second-order entropy density, $s = s_{\mathrm{eq}} - \beta/(4\beta_\pi) \pi^{\alpha\beta} \pi_{\alpha\beta}$, with $\beta_\pi = 4P/5$, is identical in form to that obtained for a conformal Boltzmann gas \cite{Chattopadhyay:2014lya}.
}

\section{Non-conformal kinetic theory versus hydrodynamics} 
\label{appc}

In this Appendix we compare some of the results obtained in Sec.~\ref{sec5} using non-conformal kinetic theory (Sec.~\ref{sec5.2}) and second-order hydrodynamics (Sec.~\ref{sec5.1}). The insights gained from this comparison are not new but confirm the findings reported in Refs.~\cite{Jaiswal:2021uvv, Chattopadhyay:2021ive} and extend them to systems with non-zero net baryon charge.

Figure~\ref{F13} compares the phase trajectories (upper panels) and the evolution of the dissipative flows (lower panels) from second-order viscous hydrodynamics (``hydro'', taken from Fig.~\ref{F8}) and kinetic theory (``KT'', taken from Fig.~\ref{F10}). While the hydrodynamic and kinetic expansion trajectories agree well for equilibrium initial conditions (where dissipative effects manifest themselves by {\it viscous heating}), and this agreement gets better with increasing interaction strength, they differ substantially for the far-off-equilibrium initial conditions for which the expansion initially leads to {\it viscous cooling}. Clearly, the hydrodynamic approach as a macroscopic approximation of the underlying microscopic kinetic theory degrades in far-off-equilibrium situations. This might not have been expected from the evolutions of the normalized shear viscous stress shown in panel (c) which shows discrepancies between microscopic and macroscopic approaches that are of similar size for both sets of initial conditions. On the other hand, the evolution of the normalized bulk viscous pressure shown in panel (d) also shows much larger discrepancies between ``hydro'' and ``KT'' for the far-off-equilibrium initial conditions, in particular in the amplitude of its oscillations around zero at early times. Both features are shared by both weakly and strongly coupled fluids, at least qualitatively.

\begin{figure}[h!]
\centering
 \includegraphics[width=0.8\linewidth]{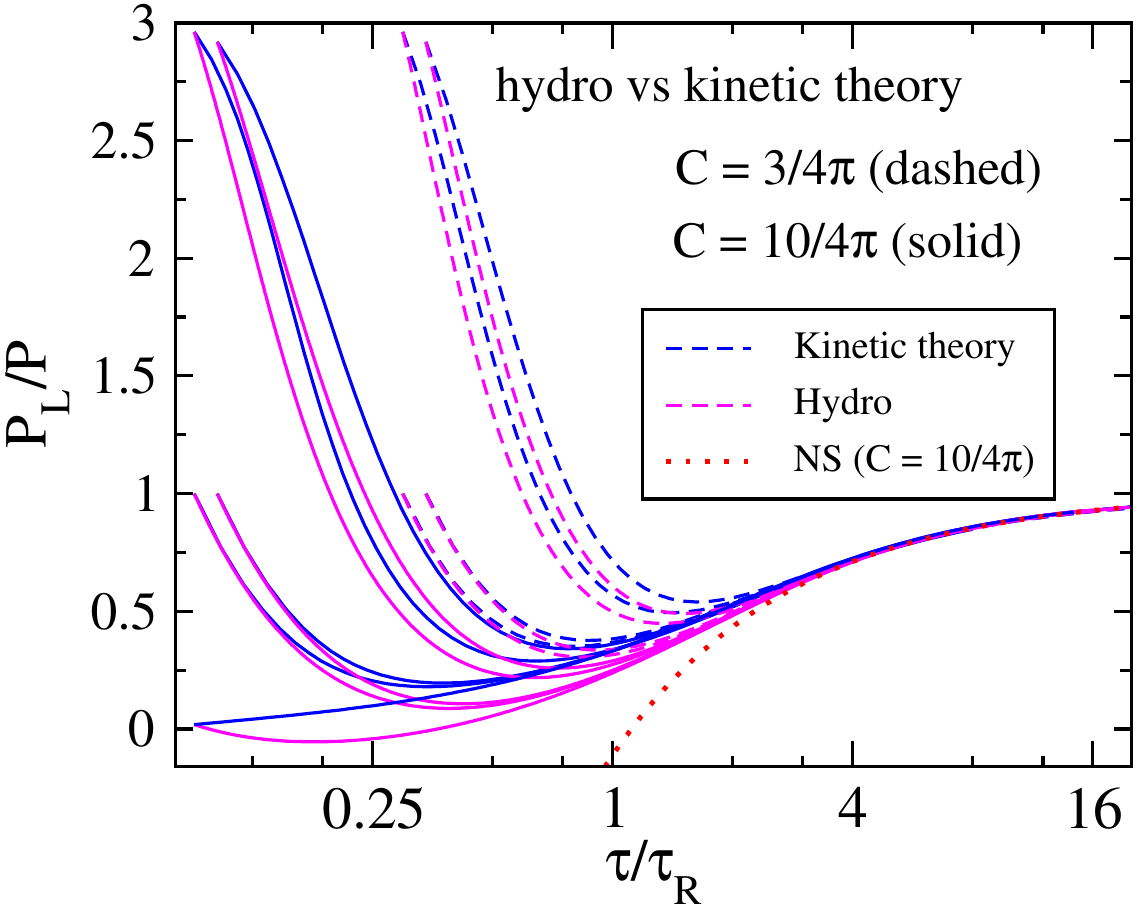}
 \vspace{-3mm}
 \caption{%
    Evolution of the scaled longitudinal pressure in kinetic theory (blue curves) and second-order hydrodynamics (magenta curves). The kinetic theory trajectories are identical to those in Fig.~\ref{F12}.
	\label{F14}
	}
\end{figure} 

Figure~\ref{F14} illustrates the observation made in Refs.~\cite{Jaiswal:2021uvv, Chattopadhyay:2021ive} that the failure of the (standard) second-order viscous hydrodynamic approximation is most severe at early times when the expansion rate is large and the system is far away from equilibrium. During this stage the expansion is dominated by free-streaming and the exact kinetic evolution (blue) of the longitudinal pressure is ruled by a ``far-from-equilibrium attractor'' that is not reproduced in the hydrodynamic approximation (magenta) \cite{Chattopadhyay:2021ive}. The discrepancies are particularly large for far-from-equilibrium initial condition ($(P/P_L)_0\gg 1$ or $(P/P_L)_0\ll 1$).\footnote{%
    In the latter case the hydrodynamic approximation even produces a negative longitudinal pressure at early times, violating a fundamental kinetic theory limit for the system at hand \cite{Chattopadhyay:2021ive}.} 
Readers interested in a more detailed discussion are referred to Refs.~\cite{Jaiswal:2021uvv, Chattopadhyay:2021ive}. 

\bibliographystyle{elsarticle-num}
\bibliography{references}

\end{document}